\documentstyle[12pt,epsfig]{article}

\newcommand{\lsim}{\raisebox{-0.13cm}{~\shortstack{$<$ \\[-0.07cm] $\sim$}}~}
\newcommand{\gsim}{\raisebox{-0.13cm}{~\shortstack{$>$ \\[-0.07cm] $\sim$}}~}

\newcommand{\ra}{\rightarrow}
\newcommand{\ee}{e^+e^-}
\newcommand{\tb}{\tan \beta}
\newcommand{\s}{\smallskip}
\newcommand{\nn}{\noindent}
\newcommand{\non}{\nonumber}
\newcommand{\beq}{\begin{eqnarray}}
\newcommand{\eeq}{\end{eqnarray}}

\catcode`@=11
\def\citer{\@ifnextchar
[{\@tempswatrue\@citexr}{\@tempswafalse\@citexr[]}}
\def\@citexr[#1]#2{\if@filesw\immediate\write\@auxout{\string\citation{#2}}\fi
  \def\@citea{}\@cite{\@for\@citeb:=#2\do
    {\@citea\def\@citea{--\penalty\@m}\@ifundefined
       {b@\@citeb}{{\bf ?}\@warning
       {Citation `\@citeb' on page \thepage \space undefined}}%
\hbox{\csname b@\@citeb\endcsname}}}{#1}}
\catcode`@=12

\begin{document}

\vspace*{.1cm} 
\baselineskip=17pt

\begin{flushright}
PM/02--10\\
May 2002\\
\end{flushright}

\vspace*{0.9cm}

\begin{center}

{\large\sc {\bf The MSSM Higgs bosons in the intense--coupling regime}}

\vspace{0.7cm}

{\sc Edward BOOS}$^{1,2}$ , {\sc  Abdelhak DJOUADI}$^3$ , 
\vspace*{2mm} 

{\sc Margarete M\"UHLLEITNER}$^3$ and {\sc Alexander VOLOGDIN}$^1$ 

\vspace*{5mm}

$^1$ Institute for Nuclear Physics, Moscow State University, \\
119899 Moscow, Russia. 
\vspace*{2mm} 

$^2$ Deutsches Elektronen--Synchrotron, DESY, \\ 
D--22603 Hamburg, Germany. 
\vspace*{2mm} 

$^3$ Laboratoire de Physique Math\'ematique et Th\'eorique, UMR5825--CNRS,\\
Universit\'e de Montpellier II, F--34095 Montpellier Cedex 5, France. 
\end{center} 

\vspace*{1cm} 

\begin{abstract}

\nn We perform a comprehensive study of the Higgs sector of the Minimal
Supersymmetric extension of the Standard Model in the case where all Higgs
bosons are rather light, with masses of ${\cal O}(100~{\rm GeV})$, and couple
maximally to electroweak gauge bosons and strongly to standard third generation
fermions, i.e. for large values of the ratio of the vacuum expectation values
of the two Higgs doublet fields, $\tb$. We first summarize the main
phenomenological features of this ``intense--coupling" scenario and discuss the
available constraints from direct searches of Higgs bosons at LEP2 and the
Tevatron as well as the indirect constraints from precision measurements such
as the $\rho$ parameter, the $Zb\bar{b}$ vertex, the muon $(g-2)$ and the decay
$b\to s \gamma$.  We then analyze the decay branching fractions of the neutral
Higgs particles in this regime and their production cross sections at the
upcoming colliders, the Tevatron Run II, the  LHC, a 500 GeV $e^+e^-$ linear
collider (in the $\ee$ and $\gamma \gamma$ options) as well as at a $\mu^+
\mu^-$ collider.

\end{abstract}

\newpage 

\subsection*{1. Introduction}

A firm prediction of the Minimal Supersymmetric Standard Model (MSSM)
\cite{MSSM} is that, among the five scalar particles which are present in the
extended Higgs sector \cite{HHG}, i.e.\, two CP--even $h$ and $H$, a
pseudoscalar $A$ and two charged $H^\pm$ bosons, at least the lightest Higgs
boson $h$ must have a mass below $130$ GeV when radiative corrections are taken
into account \citer{RC1,RC3}. If the minimal version of Supersymmetry is 
indeed realized in Nature, this particle should be therefore accessible at the
next generation of high--energy colliders, the high--luminosity Tevatron
\cite{Tevatron}, the LHC \cite{LHC,Houches} and a future $\ee$ linear collider
\cite{LC,TESLA}.  \s

In the decoupling regime \cite{Decoup}, that is when the Higgs bosons $H,A$ and
$H^\pm$ are very heavy implying that they are almost degenerate in mass, $M_A
\sim M_H \sim M_{H^\pm}$, the lightest Higgs particle $h$ will have the
properties of the Standard Model (SM) Higgs boson. The experimental search for
this particle will therefore be straightforward and there is now little doubt,
in view of the detailed phenomenological and experimental analyses performed in
the recent years, that it will not escape detection and that at least some of
its fundamental properties can be pinned down. Unfortunately, the other MSSM
Higgs bosons will have masses of ${\cal O}(1~{\rm TeV})$, and will therefore be
too heavy to be accessible directly, implying that one would not be able to
distinguish between the MSSM and the SM Higgs sectors. \s

An opposite and much more interesting situation would be the one where the mass
of the pseudoscalar $A$ boson is not much larger than the maximal value allowed
for $M_h$. In this case, the three neutral Higgs bosons $h,H$ and $A$ and the
charged Higgs particles will have comparable masses, $M_h \sim M_H \sim M_A
\lsim 130$ GeV and $M_{H^\pm} \lsim 150$ GeV.  A particularly interesting
scenario is when the parameter $\tb$, the ratio of the vacuum expectation
values of the two Higgs doublet fields which break the electroweak symmetry in
the MSSM, is large\footnote{Values $\tb \gsim 3$--10, depending on the mixing
in the scalar top sector, are required to maximize the $h$ boson mass and to
evade the experimental constraint from LEP2 searches \citer{LEPevid,LEPAh}, as
will be discussed later. Very large values of $\tb \sim m_t/m_b \sim {\cal O}
(50)$ are favored if one requires Yukawa coupling unification at the GUT scale;
see Ref.~\cite{Yukawa}. In the constrained MSSM or minimal Supergravity model
\cite{mSUGRA}, large values of $\tb$ lead naturally to rather light $A,H$ and
$H^\pm$ bosons; see e.g. Ref.~\cite{benchmark}.}, leading to a CP--even Higgs
boson $\Phi_A$ [either $h$ or $H$] which is almost degenerate in mass with the
pseudoscalar $A$ boson.  As a consequence of the SUSY constraints on the Higgs
sector, this CP--even $\Phi_A$ boson will have almost the same couplings as the
$A$ boson, and therefore will couple strongly to the third generation
$b$--quarks and $\tau$ leptons [since the couplings are proportional to $\tb$],
while the couplings of the $Z$--bosons to $\Phi_A A$ pairs [and the
corresponding couplings of the $W$ bosons to $\Phi_A H^\pm$ pairs] will be
maximal, i.e. not suppressed by (sine or cosine of) mixing angle factors.  The
other CP--even Higgs particle, that we will denote by $\Phi_H$, will have
almost the couplings of the SM Higgs boson, i.e. its couplings to weak gauge
bosons and to top quarks are not strongly suppressed by mixing angle factors
and are therefore almost maximal. \s

This scenario, with all MSSM Higgs bosons being light with almost maximal 
couplings to gauge bosons and strong couplings to third generation fermions, 
will be called hereafter, the ``intense--coupling" regime.  

In this non--decoupling scenario, all three neutral Higgs bosons, as well as
the charged Higgs particle, would be accessible at the next generation of
experiments. The experimental searches for the neutral Higgs bosons in this
case will be slightly more involved and much more interesting than in the
decoupling regime, since a plethora of production and decay processes, with
rates which can be violently different from the ones of the SM Higgs particles,
have to be considered and studied in detail in order to detect individually all
the three particles and to determine their basic properties. In addition, two
features might render the situation somewhat more delicate: the $h,H,A$ boson
masses can be very close to each other, implying that backgrounds for a given
Higgs boson signal would come from another Higgs signal and the total decay
widths of some of the Higgs bosons can be rather large  [in particular for very
large values of $\tb$] implying broader signals.  \s

The search for the charged Higgs particle would be straightforward. Indeed,
since in the MSSM the $A$ and $H^\pm$ boson masses are related by $M_{H^\pm}^2
\sim M_A^2 +M_W^2$, $M_{H^\pm}$ will be smaller than $\sim 150$ GeV for $M_A
\lsim 130$ GeV, implying that $H^\pm$  can always be produced in top quark
decays, $t \to H^+ b$, and can easily be detected at hadron 
\cite{Tevatron,LHC} or $\ee$ colliders \cite{LC,TESLA}. \s

The purpose of this paper is to perform a comprehensive study of the MSSM Higgs
sector in the intense--coupling regime and the implications of the near mass
degeneracy $M_h \sim M_H \sim M_A$, for the search of the Higgs bosons at the
Tevatron, the LHC, a 500 GeV $\ee$ collider and a $\mu^+ \mu^-$
collider\footnote{Several articles in the literature have dealt with some
aspects of the scenario discussed here, and in particular, with the consequences
of large $\tb$ values for MSSM Higgs boson production at future colliders
\cite{hightbpapers1,hightbpapers2}.  The conclusions of these papers overlap
with the ones we obtain here.}.  We will first discuss the parameterization of
the Higgs sector and derive rather simple, accurate and useful expressions for
the radiative corrections to the Higgs masses and couplings which are valid in
the region 90 GeV $\lsim M_A \lsim 130$ GeV for moderate ($\sim 10)$ and large
($\sim 30$--50) $\tb$ values.  We will then analyze in this
``intense--coupling" regime, the various decay branching ratios of the $h,H$
and $A$ particles and collect all their production cross sections at these
machines. We will not only analyze the main production channels which allow to
detect these particles, but also sub--leading processes which would allow for
the measurement of some of their fundamental properties. The various
experimental signatures for the Higgs particles will be discussed. \s

Before that, we will  discuss in detail the various constraints on this
intense--coupling regime with high $\tb$ values: the direct constraints from
the Tevatron and LEP2 searches for MSSM Higgs bosons [and the implications of
the possible $\sim 2 \sigma$ evidence \cite{LEPevid} for a SM--like Higgs boson
with a mass around 115 GeV] as well as all the indirect constraints from
precision measurements \cite{PDG,LEP1}, i.e. the $W$ boson mass, the effective
electroweak mixing angle $\sin^2\theta_W$ and the $Z$ boson decays into
$\bar{b}b$ final states, the radiative decay $b \to s\gamma$ \cite{bsg} and the
anomalous magnetic moment of the muon, $(g-2)_\mu$ \cite{gm2}. \s 

The paper is organized as follows. In the next section, we will present the
physical set--up and summarize our parameterization of the MSSM Higgs sector,
and display the Higgs boson masses and couplings in the intense--coupling
regime.  In section 3, we discuss the implications of the LEP2 and Tevatron
direct Higgs searches and the indirect constraints from high precision
measurements on this scenario. The various decay branching fractions of the
neutral Higgs bosons will be analyzed in section 4.  The production cross
sections of the neutral Higgs bosons at the Tevatron, the LHC, a 500 GeV $\ee$
linear collider and a $\mu^+ \mu^-$ collider will be given in section 5,
together with a discussion of the various processes and signatures which can be
used. A conclusion will be given in section 6.

\subsection*{2. The MSSM Higgs sector and the intense--coupling regime} 

\subsubsection*{2.1 Radiative corrections in the Higgs sector}

In the MSSM, besides the four Higgs boson masses, two mixing angles define the
properties of the scalar particles and their interactions with gauge bosons and
fermions: the ratio of the two vacuum expectation values $\tb = v_2/v_1$ of the
Higgs doublets and a mixing angle $\alpha$ in the neutral CP--even sector.
Supersymmetry leads to several relations among these parameters and, in fact,
only two of them, taken for convenience to be $M_A$ and $\tb$, are independent.
These relations impose, at the tree--level, a strong hierarchical structure on
the mass spectrum [e.g.~$M_h<M_Z$] some of which are, however, broken by
radiative corrections. The leading part of these corrections grows as $m_t^4$
and logarithmically with the common stop mass $M_S$ \cite{RC1}. Including the
mixing in the stop and sbottom sectors and all the subleading terms, the full
expressions of the radiative corrections become rather involved 
\cite{RC2,RC3}. \s

To illustrate the qualitative behavior of the Higgs boson masses and couplings,
we will use simple analytic expressions [comparable in simplicity to those
obtained in \cite{RC1} for the leading $m_t^4$ piece] where the stop mixing and
some of the subleading terms are incorporated. In many cases, these expressions
give rather good approximations for the neutral Higgs boson masses [at the
percent level] and couplings [at the ten percent level] compared to what is
obtained with the full set of corrections in the renormalization group improved
approach of Ref.~\cite{RC2}. This is sufficient for our purpose in the present
section, since our aim is simply to understand qualitatively the
phenomenological implications of almost degenerate Higgs bosons with enhanced
Yukawa couplings.  \s

For our numerical analysis throughout the paper, we will use the program {\tt
HDECAY} \cite{hdecay}, which calculates the Higgs spectrum and decay widths in
the MSSM. The complete radiative corrections due to top/bottom quark and squark
loops within the effective potential approach, leading NLO QCD corrections
[through renormalization group improvement] and the full mixing in the stop and
sbottom sectors are incorporated using the analytical expressions of
Ref.~\cite{RC2}. [Note also that we will include the leading one--loop and
two--loop corrections to the Higgs self--couplings.] However, even in this
case, our treatment of the radiative corrections in the Higgs sector will not
be complete, since two sets of potentially large corrections are presently not 
included in the program: \s

-- The SUSY--QCD and dominant Yukawa radiative corrections to the bottom and
top quark Yukawa couplings which, in many cases, can be rather important.  In
particular, the corrections to the bottom Yukawa couplings can be very large
for high values of $\tb$ and the higgsino mass parameter $\mu$ \cite{Yukawacr}.
The net effect of these corrections can be viewed as an increase or decrease of
the bottom quark mass, depending on the sign of $\mu$. The phenomenological
impact of these corrections has been already discussed in the literature, see
e.g.\ Ref.~\cite{hightbpapers1,Yukawacr}, and we have little to add. \s

-- Recently, the two--loop ${\cal O}(\alpha_t^2)$ and ${\cal O}(\alpha_s
\alpha_b)$ Yukawa corrections to the Higgs boson masses have been calculated
\cite{Pietro}. The former set of corrections can increase the lightest $h$
boson mass by several GeV in some areas of the MSSM parameter space, while the
later corrections can be sizeable for large $\tb$ values. \s 

We have nevertheless verified that the numerical results that we obtain for the
MSSM Higgs spectrum, and in particular for the lightest $h$ boson mass, are
rather close to those obtained from the complete results of the Feynman
diagrammatic approach implemented in the program {\tt FeynHiggs}
\cite{herrors}, which include some of the Yukawa corrections of
Ref.~\cite{Pietro}.  The difference, of the order of a few GeV, is of the same
size (within a factor of two) as the expected theoretical error on the
determination of the Higgs boson masses.  However, as already pointed out, this
difference will not be of utmost importance for our analysis, and our general
conclusions will not be altered in a significant way.  

\subsubsection*{2.2 The Higgs boson masses}

Before discussing the Higgs boson masses in the intense--coupling regime, 
i.e. when the $A$ boson is rather light and $\tb$ is very large, let us first 
present the rather simple analytical expression which approximate the Higgs 
boson masses. We first define the quantity $\epsilon$, which parameterizes the 
main radiative correction,
\beq
\epsilon = \frac{3 G_{F} \overline{m}_t^4}{\sqrt{2} \pi^2\sin^2\beta} \,
\bigg[ t + \frac{X_t}{2} \bigg] - \frac{3 G_{F}}{\sqrt{2} \pi^2} 
\bigg[ \frac{\overline{m}_t^2 M_Z^2 t}{2}  +  \frac{2 \alpha_S}{\pi}
\frac{\overline{m}_t^4}{\sin^2\beta} (X_t t + t^2)  \bigg] 
\eeq
with
\beq
t = \log \bigg (\frac{M_S^2}{m_t^2} \bigg) \ \ , \ \ 
X_t = \frac{2 A_t^2}{M_S^2} \bigg(1-  \frac{A_t^2}{12M_S^2} \bigg) \non  
\hspace*{1.5cm}
\eeq
where $\overline{m}_t$ is the running ${\rm \overline{MS}}$ top quark mass to
account for the leading QCD corrections, $A_t$ is the stop trilinear coupling,
and $\alpha_S$ the strong coupling constant. The Higgs boson masses can be
approximated to an accuracy of the order of a few percent compared to the 
complete result [in particular in the case where the splitting between the 
two stop masses, and to a lesser extent sbottom masses, is not very large], as 
functions of $M_A$, $\tan\beta$ and $\epsilon$ with the expression
\begin{eqnarray}
M_{h,H}^2 &=& \frac{1}{2} (M_A^2+ M_Z^2+\epsilon) \left[ 1 \mp 
\sqrt{1- 4 \frac{ M_Z^2 M_A^2 \cos^2 2\beta +\epsilon ( M_A^2 \sin^2\beta +
 M_Z^2 \cos^2\beta)} {(M_A^2+ M_Z^2+\epsilon)^2} } \right] 
\end{eqnarray}
This radiative correction pushes the maximum value of the lightest $h$ boson
mass upwards from $M_Z$ by several tens of GeV: in the so--called maximal
mixing scenario $\tilde{A}_t= A_t - \mu/\tan\beta = \sqrt{6} M_S$ and with
$M_A$ and $M_S$ of about 1 TeV, one obtains an upper mass bound, $M_{h} \lsim
130$ GeV for $\tb \gg1$, a bound that is comparable to the one obtained
including the full set of corrections in the RG improved approach \cite{RC2}.
\s  

In the case of the mixing angle $\alpha$ of the CP--even Higgs sector,  the 
correction eq.~(1) is not accurate enough and one needs to introduce another 
correction which involves the ratio $\mu/M_S$, where $\mu$ is the higgsino 
mass parameter. Defining the parameter $\epsilon'$ to be
\beq
\epsilon' = \frac{G_F \overline{m}_t^4}{2\sqrt{2}\pi^2 \sin^2\beta} 
\frac{\mu}{M_S} \bigg[ \frac{A_t^3}{M_S^3} - \frac{6 A_t}{M_S} \bigg]
\bigg[ 1- \frac{4 \alpha_S t}{\pi} \bigg]
\eeq
the mixing angle $\alpha$ is given in terms of $M_A, \tb, \epsilon$ and
$\epsilon'$, by 
\beq
\tan 2\alpha = \tan 2\beta \,  \frac{M_A^2 + M_Z^2 -\epsilon' /\sin 2\beta}
{M_A^2 - M_Z^2 + \epsilon /\cos 2 \beta } \ , 
\hspace*{1cm}- \frac{\pi}{2} \leq \alpha \leq 0 \ .
\label{alpha}
\eeq
Here again, the accuracy of the formula, compared to the case where the full 
corrections are taken into account, is at the percent level if the splitting 
between the stop (and to a lesser extent, sbottom) masses is not too large. \s

[In the case of the charged Higgs boson mass, one can also derive a very simple 
expression for the radiative corrections which gives a result
that is accurate at the percent level,
\beq
M_{H^\pm}=\sqrt{M_A^2+M_W^2-\epsilon_+} \ \ {\rm with} \ \ \epsilon_+ = 
\frac{3 G_F M_W^2 t}{4 \sqrt{2} \pi^2} \bigg[ \frac{\overline{m}_t^2}
{\sin^2\beta} + \frac{\overline{m}_b^2}{\cos^2\beta}
\bigg]
\eeq
where $\overline{m}_b$ is the running $\overline{\rm MS}$ $b$ mass at the 
scale of the top quark mass.]\s 

Let us now discuss in some details the neutral Higgs boson masses in the
intense--coupling regime, taking into account the dominant radiative
corrections $\epsilon$ and $\epsilon'$. In fact, this regime can be defined as
the one where the two CP--even Higgs bosons $h$ and $H$ are almost degenerate
in mass, $M_h \simeq M_H$. We will therefore first concentrate on the Higgs
sector in the limit $M_h=M_H$. \s

Solving the equation (2) for $M_H^2- M_h^2=0$,  which is a second order 
polynomial equation in the variable $M_A^2$, 
\beq 
M_A^4+2 M_A^2 [M_Z^2 (1-2\cos^2 2\beta)+ \epsilon
\cos2\beta ] +M_Z^4 +\epsilon^2 - 2M_Z^2 \epsilon \cos 2\beta =0 
\eeq
one obtains a discriminant $\Delta'= -\sin^2 2\beta (2 M_Z^2 \cos 2\beta-
\epsilon)^2 \leq 0$. The only way for the solution to be real is therefore to
have either $\sin2\beta =0$ or $\epsilon = 2M_Z^2 \cos 2\beta$. The last
possibility gives $M_A^2=-M_Z^2$ which has to be rejected, while the former
possibility gives $M_A^2=M_Z^2 +\epsilon$ with $\beta= \frac{\pi}{2}$. In fact,
this solution or critical mass corresponds to the maximal value allowed for 
$M_h$ and the minimal value that $M_H$ can take,
\beq 
M_{C}= M_h^{\rm max} = M_H^{\rm min} = \sqrt{M_Z^2+\epsilon}
\eeq
In addition, in the large $\tb$ regime, eq.~(2) for the masses of the CP--even
Higgs bosons simplifies to 
\beq
M^2_{h,H}= \frac{1}{2} (M_A^2+M_Z^2+\epsilon \mp |M_A^2-M_Z^2-\epsilon|) 
\eeq
which means that
\beq
M_A \ge M_{C} &\Rightarrow& M_H=M_A \quad {\rm and} \quad M_h=M_{C} 
\non \\
M_A \le M_{C} &\Rightarrow& M_h=M_A \quad {\rm and} \quad M_H=M_{C}  
\eeq
and therefore the $A$ boson is always degenerate in mass with one of the
CP--even Higgs bosons, and if the masses of the latter are equal, one has 
$M_H=M_h=M_A = M_C$. \s

\subsubsection*{2.3 The Higgs boson couplings}

We turn now to the couplings of the Higgs bosons, which determine to a large 
extent, the production cross sections and the decay widths. The pseudoscalar 
Higgs boson has couplings to isospin down (up) type fermions that are
(inversely) proportional to $\tb$ and, because of CP--invariance, it has no 
tree--level couplings to gauge bosons. In the case of the CP--even $h$ and $H$ 
bosons, the mixing angle $\alpha$ enters in addition. 
The couplings of the CP--even Higgs bosons to down (up) type fermions are 
enhanced (suppressed) compared to the SM Higgs couplings for values $\tb> 1$; 
the couplings to gauge bosons are suppressed by $\sin(\beta-\alpha)$ or
$\cos(\beta-\alpha)$ factors; see Table 1. In fact, the $h$ and $H$ bosons 
share the couplings of the SM Higgs to the gauge bosons; also, the squared sum 
of the couplings to fermions do not depend on $\alpha$ and obey the sum rules: 
\beq
g_{hdd}^2+g_{Hdd}^2= 1/\cos^2\beta \ , \ 
g_{huu}^2+g_{Huu}^2= 1/\sin^2\beta \ , \ 
g_{hVV}^2+g_{HVV}^2= 1 
\eeq
Developing the formulae for $\sin 2\alpha$ and $\cos 2\alpha$ for large 
values of $\tb$, $\tb \gg 1$, 
\beq
\sin 2\alpha = \sin 2\beta \,  \frac{M_A^2 + M_Z^2 -\epsilon' /\sin 2\beta}
{M_h^2-M_H^2} \ , \ 
\cos 2\alpha = \cos 2\beta \,  \frac{M_A^2 - M_Z^2 +\epsilon /\cos 2\beta}
{M_h^2-M_H^2} 
\eeq
one obtains depending on whether $M_A$ is, respectively, above or below $M_{C}$:
\beq
M_A> M_C: && \quad \cos\alpha \approx \sin\beta \approx 1 
\quad , \quad \quad  \ \, \sin\alpha  \approx f_1 -f_2/\tan\beta
\non \\
M_A< M_C: && \quad \sin\alpha \approx  -\sin\beta 
\approx -1 \quad , \quad \cos\alpha  \approx -f_1 +f_2/\tan\beta \non \\
&& {\rm with} \ \ f_1 = \frac{\epsilon' /2}{M_H^2-M_h^2} \ \ , \quad f_2 = 
\frac{M_A^2+M_Z^2}{M_H^2-M_h^2}
\eeq
The couplings of the neutral Higgs bosons are displayed in Table 1 together
with their limiting values for $\tb \gg 1$ and $M_A \geq M_{C}$ (upper values)
or $M_A < M_{C}$ (lower values). The accuracy is of the order of $10\%$ for
$A_t, A_b$ and $\mu$ of the order of 1 TeV and improves  in the case of zero
mixing.  One can see that, not only in the decoupling limit $M_A \to \infty$
but also in the intense--coupling regime with $\tb \gg 1$, the CP--even Higgs
boson with a mass $M_{C}$ has, up to a sign,  SM--like Higgs couplings to gauge
bosons and up--type quarks. For down--type fermions, the situation is more
complicated: in general the coupling is larger than the SM Higgs coupling, but
in some cases, when all terms add up to zero [i.e. when $f_2 \sim f_1 \tb$ in
our approximation], the coupling becomes very small and even vanishes.  The
other Higgs boson, with a mass $M_A$, has couplings of the same order as the
$A$ boson, i.e. it has rather small couplings to gauge bosons and couples to
up-type and down--type fermions proportionally to, respectively, $\mp 1/\tb$
and $\tb$ [in the case of small $f_1$]. \s 

In the intense--coupling regime, we will therefore define a SM--like Higgs 
boson $\Phi_H$ and a pseudoscalar--like Higgs boson $\Phi_A$ as follows: 
\beq
&& \Phi_H \equiv h \ \ {\rm if} \ \ M_A>M_C \ \ {\rm and} \ \ \Phi_H \equiv 
H \ \ {\rm if} \ M_A< M_C \non \\
&& \Phi_A \equiv h \ \ {\rm if} \ \ M_A<M_C \ \ {\rm and} \ \ \Phi_A \equiv 
H \ \ {\rm if} \ M_A>M_C 
\eeq

\begin{table}[htbp]
\renewcommand{\arraystretch}{1.8}
\begin{center}
\vspace*{-7mm}
\begin{tabular}{|c|c|c|c|c|} \hline
$\  \Phi  \ $ &$ g_{\Phi \bar{u}u} $      & $ g_{\Phi \bar{d} d} $ &
$g_{ \Phi VV} $ \\ \hline
$h$ &
$\cos\alpha/\sin\beta^{\ \ \rightarrow 1}_{\ \ \ra -f_1+f_2/\tb}$ & 
$-\sin\alpha/\cos\beta^{\ \ \rightarrow -f_1 \tan\beta +f_2}_{\ \ \ra \tb}$ & 
$\sin(\beta-\alpha)^{\ \ \rightarrow 1}_{\ \ \ra -f_1 + (1+f_2)/\tan\beta}$ 
\\ \hline
$H$ &
$\sin\alpha/\sin\beta^{\ \  \rightarrow f_1-f_2/\tb}_{\ \ \ra -1}$ & 
$\cos\alpha/\cos\beta^{\ \ \rightarrow \tb}_{ \ \ \ra -f_1 \tan\beta +f_2}$ & 
$\cos(\beta-\alpha)^{\ \  \rightarrow f_1 + (1-f_2)/\tan\beta}_{\ \ \ra -1}$ 
\\ \hline
$A$  & \ $\; 1/ \tb \; $\ & \ $ \; \tb \; $ \   & \ $ \; 0 \; $ \ \\ \hline
\end{tabular}
\end{center}
\vspace*{-5mm}
\caption[]{\it Neutral Higgs boson couplings to fermions and gauge bosons 
in the MSSM normalized to the SM Higgs boson couplings, and their limits for 
$\tb \gg 1$ for $M_A \geq M_C$ (upper values) and  $M_A < M_C$ (lower values).}
\end{table}

An additional set of couplings that we will need in this study are the
trilinear couplings among the neutral Higgs bosons. These couplings will be
defined in units of $\lambda_0= M_Z^2/v$ and are renormalized not only
indirectly by the re\-nor\-ma\-li\-za\-tion of the angle $\alpha$ but also
directly by additive terms proportional to the parameter $\epsilon$. In the
$\epsilon$ approximation [which here, gives only the magnitude of the 
Higgs self--couplings, i.e. a few ten percent in general but in some cases, 
more than a factor of two] and keeping only $1/\tb$ terms, one has: 
\beq
\lambda_{hhh} &=& 3 \cos2\alpha \sin (\beta+\alpha) 
+ 3 \frac{\epsilon}{M_Z^2} \frac{\cos \alpha}{\sin\beta} \cos^2\alpha  \non\\ 
&\to&  \left\{ \begin{array}{l} 3M_C^2/M_Z^2 \\ 
 3[f_1+(1-f_2)/\tan\beta]-3 \frac{\epsilon}{M_Z^2}
 f_1^2 [f_1 -3 f_2/\tan\beta] \end{array} \right.  \non \\
\lambda_{Hhh} &=& 2\sin2 \alpha \sin (\beta+\alpha) -\cos 2\alpha \cos(\beta
+ \alpha) + 3 \frac{\epsilon}{M_Z^2} \frac{\sin \alpha}{\sin\beta}
\cos^2\alpha \non\\
& \to &  \left\{ \begin{array}{l} 5[f_1-f_2/\tan\beta] -1/\tan\beta+
3 \frac{\epsilon}{M_Z^2}[f_1-f_2/\tan\beta] \\ 
-4f_1 [f_1 + (1-2 f_2) /\tan\beta ] +1 -
3 \frac{\epsilon}{M_Z^2} f_1 [f_1- 2f_2/\tan\beta] \end{array} \right. \non \\
\lambda_{HHh} &=& -2 \sin 2\alpha \cos (\beta+\alpha) - \cos 2\alpha \sin(\beta
+ \alpha) + 3 \frac{\epsilon}{M_Z^2} \frac{\cos \alpha}{\sin\beta}
\sin^2\alpha \non\\
& \to & \left\{ \begin{array}{l} 
4 f_1 [f_1 -(1+2f_2) /\tan\beta]-1+
3 \frac{\epsilon}{M_Z^2} f_1 [f_1-2f_2/\tan\beta] \\ 
-5[f_1-f_2/\tan\beta]-1/\tan\beta-
3 \frac{\epsilon}{M_Z^2}[f_1-f_2/\tan\beta] \end{array} \right. \non \\
\lambda_{HHH} &=& 3 \cos 2\alpha \cos (\beta+\alpha) 
+ 3 \frac{\epsilon}{M_Z^2} \frac{\sin \alpha}{\sin\beta} \sin^2 \alpha
\non\\
& \to & \left\{ \begin{array}{l} -3[f_1-(1+f_2)/\tan\beta]+
3 \frac{\epsilon}{M_Z^2}f_1^2 [f_1-3 f_2/\tan\beta] \\ -3M_C^2/M_Z^2 
\end{array} \right. \non \\
\lambda_{hAA} &=& \cos 2\beta \sin(\beta+ \alpha)+ 
\frac{\epsilon}{M_Z^2} \frac{\cos \alpha}{\sin\beta} \cos^2\beta 
\non\\
&\to & \left\{ \begin{array}{l} -1 
\\ \phantom{-}[f_1+(1-f_2)/\tan\beta]
\end{array} \right. \non \\
\lambda_{HAA} &=& - \cos 2\beta \cos(\beta+ \alpha) + 
\frac{\epsilon}{M_Z^2} \frac{\sin \alpha}{\sin\beta} \cos^2\beta
\non\\
& \to & \left\{ \begin{array}{l} -[f_1-(1+f_2)/\tan\beta] \\ 
+ 1 \end{array} \right.
\label{coup}
\eeq
Again, the upper (lower) values are the limits for $\tb \gg 1$ with $M_A$
larger (smaller) than $M_{C}$. The SM trilinear self--coupling
$3M_{H^0}^2/M_Z^2$ is approached in the case of $h$ in the decoupling limit,
and in the case of $H$ (up to a sign) in the intense--coupling limit. Without
radiative corrections, one of these Higgs bosons would have $A$--like couplings
while the other would decouple for large $\tb$ values. \s 

If 90 GeV $\lsim M_A \lsim 130$ GeV, one would have also a rather light $H^\pm$
boson with a mass 120 GeV $ \lsim M_{H^\pm} \lsim 150$ GeV. Its coupling to 
fermions is a parity violating mixture of scalar and pseudoscalar currents, 
$g_{H^+ u \bar{d}}  \propto (1-\gamma_5) m_u/\tb+(1+\gamma_5) m_d \tb$. As in 
the case of the $A$ boson, the couplings to down (up) type fermions are 
(inversely) proportional to $\tb$. Only the couplings to the $(t,b)$ and 
$(\tau,\nu_\tau)$ isodoublets are relevant in most cases.
We will not discuss the properties of this particle further in this paper. \s

Finally, note that the couplings of the CP--even $h$ and $H$ bosons to $ZA$ and
$W^+H^-$ pairs are proportional to $\cos(\beta-\alpha)$ and
$\sin(\beta-\alpha)$, respectively, while the $W^+H^-A$ coupling is not
suppressed by these factors. This means again, that for large $\tb$ values the
CP--even $\Phi_A$ boson with a mass close to $M_A$ couples maximaly to $ZA$
and $W^\mp H^\pm$ states while the SM--like $\Phi_H$ boson has small couplings.

\subsubsection*{2.4 Inputs for the numerical analysis}

As mentioned previously, our numerical analyses all through this paper will be
based on the program {\tt HDECAY} for the evaluation of the masses, couplings
and decay branching ratios. We will use the running mass of the pseudoscalar
Higgs boson $A$, together with $\tb$ and the relevant MSSM parameters which
enter in the radiative corrections [$A_t, \mu$ and the soft SUSY breaking third
generation squark masses, $m_{\tilde{Q}_L}= m_{\tilde{U}_R} =m_{\tilde{D}_R}=
M_S$] as inputs, to obtain the masses of the two CP--even $h$ and $H$ bosons
and the charged $H^\pm$ particles. The pole bottom and top quark masses will be
fixed to $m_b=4.25$ GeV and $m_t=175$ GeV. In most of the analysis, we will use
two values of $\tb$, a moderate and a large one, $\tb=10$ and 30, and deal with
the maximal mixing scenario with $A_t \simeq \sqrt{6} M_S$ with $M_S$ fixed to
1 TeV. For the Higgs boson masses however, we will also discuss the no--mixing
and typical mixing scenarii with, respectively, $A_t=0$ and $A_t=M_S$, and give
some illustrations for a small and very large value of $\tb$, 5 and 50. \s

Because of the approximations discussed in section 2.1 [in particular because
the SUSY radiative corrections to the Yukawa couplings are not incorporated in
the present version of {\tt HDECAY}] the other SUSY parameters play only a
minor role and we will fix them to $A_b=1$ TeV and $\mu=M_2=350$ GeV for the
discussion of the Higgs boson masses, couplings and decay branching ratios and
$A_b=\mu=M_2=1$ TeV for the production cross sections. We have verified that
within our approximation, this difference does not lead to significant
changes\footnote{However, we recall again, that if the SUSY corrections to the
$b$--quark Yukawa coupling are included, this might not be the case for large
$\tb$ values \cite{hightbpapers1,Yukawacr}. This is particularly true in the
discussion of the branching ratios: while in our case, BR$(\Phi \to
b\bar{b})$/BR$(\Phi \to \tau^+ \tau^-)$ for instance is simply given by $
\simeq 3\bar{m}_b^2(M_\Phi^2) |^{\rm SM}/m_\tau^2$, it can increase or decrease
for large values of $\tb$ depending on the sign of $\mu$.}.  

\begin{figure}[htbp]
\vspace*{-5.5cm}
\hspace*{-1.5cm}
\mbox{\psfig{figure=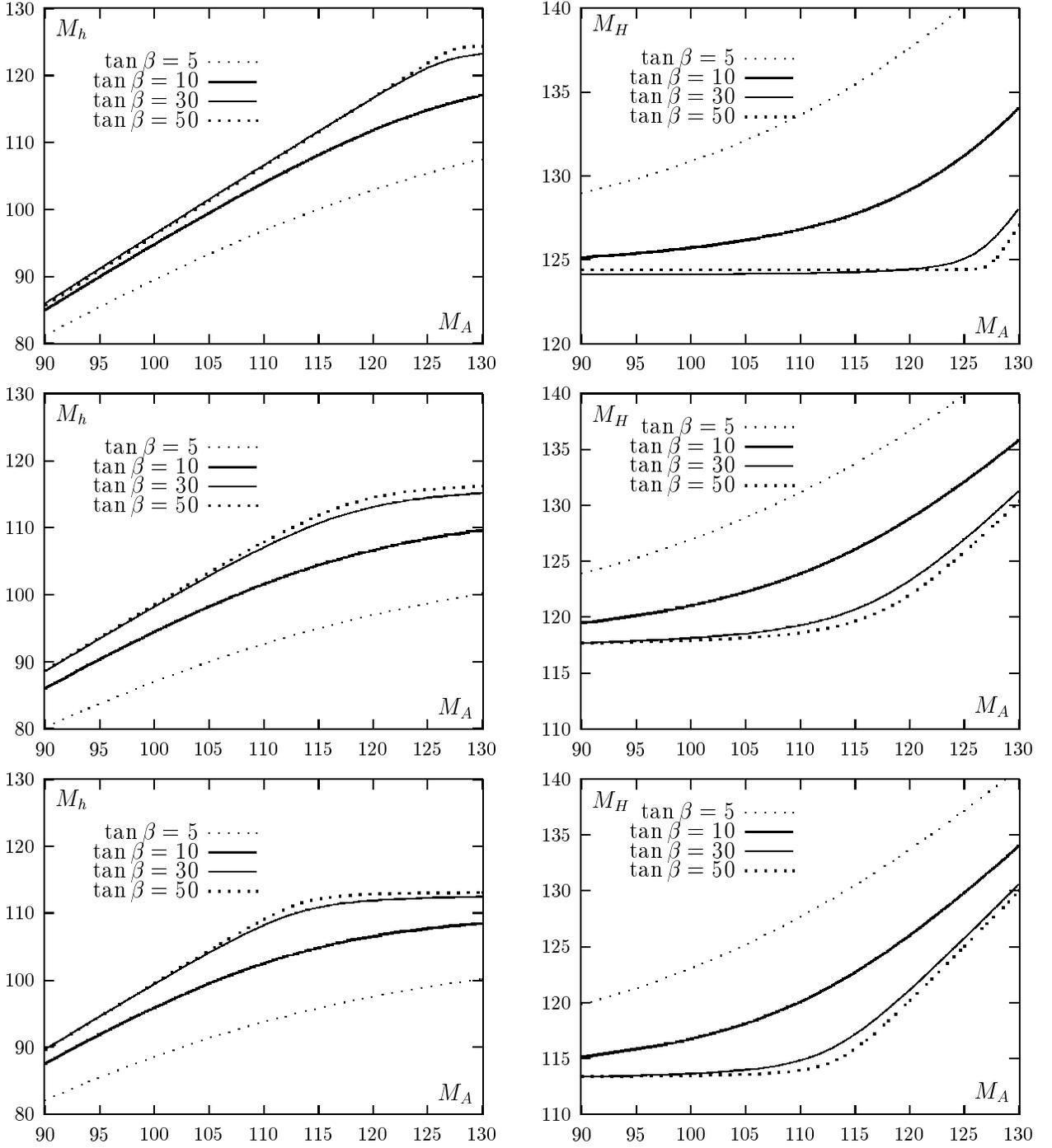,width=19cm}}
\vspace*{-4.2cm}
\caption[]{\it The masses of the lightest and heavier CP--even Higgs bosons 
$h$ and $H$ [in GeV] as a function of $M_A$ for four values of $tan \beta=5,10,
30$ and 50 in the case of maximal mixing (upper), typical mixing (central) 
and no--mixing (lower).}
\label{figmass}
\end{figure}

The neutral Higgs boson masses that we obtain are shown in Fig.~1 for a mass of
the $A$ boson varying from 90 to 130 GeV for four representative values of
$\tb$ from small to very large, $\tb=5,10,30$ and 50, in the three scenarii of
stop mixing discussed above. The normalized Higgs boson couplings to fermions
and gauge bosons are shown in Fig.~2 for $\tb=10,30$ and 50 in the scenario 
with large mixing, $A_t=2.6$ TeV.

\begin{figure}[htbp]
\vspace*{-5.5cm}
\hspace*{-2.3cm}
\mbox{\psfig{figure=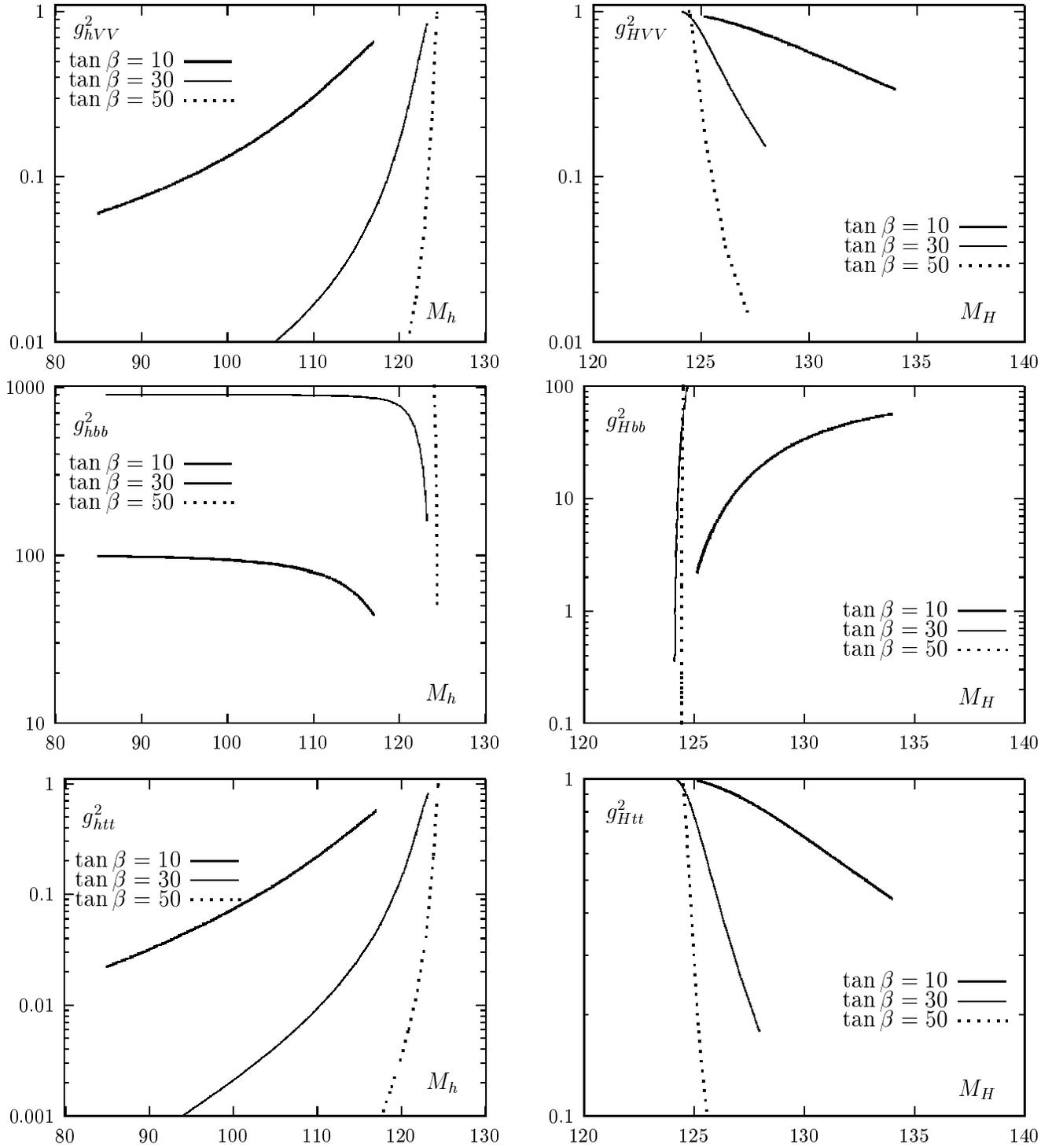,width=20cm}}
\vspace*{-4.5cm}
\caption[]{\it The couplings of the CP--even Higgs bosons $h$ and $H$ 
[normalised to the SM Higgs couplings] as a function of the masses for 
three values of $tan \beta=10, 30$ and 50,
in the maximal mixing scenario.}
\label{figcoup}
\end{figure}

\subsection*{3. Experimental Constraints on the intense--coupling scenario}

\subsubsection*{3.1 Direct constraints from LEP2 and Tevatron searches}

The search for the Higgs bosons was the main motivation for extending the LEP2
energy up to $\sqrt{s}\simeq 209$ GeV \cite{LEPevid}. In the SM, a lower bound
$M_{H^0} > 114.1$ GeV has been set at the 95\% confidence level, by
investigating the Higgs--strahlung process, $\ee \to ZH^0$ \cite{LEPH0}. In the
MSSM, this bound is valid for the lightest CP--even Higgs particle $h$ if its
coupling to the $Z$ boson is SM--like, i.e. if $g^2_{ZZh}/g^2_{ZZH^0} \equiv
\sin^2(\beta- \alpha) \simeq 1$ [almost in the decoupling regime] or in
the less likely case of the heavier $H$ particle if $g^2_{ZZH}/g^2_{ZZH^0}
\equiv \cos^2 (\beta- \alpha) \simeq 1$ [i.e. in the non--decoupling regime
with a rather light $M_H$]; see e.g. Ref.~\cite{CapitalH}. \s

A complementary information is obtained from the search of Higgs bosons in the
associated production processes $\ee \to Ah$ where the 95\% confidence level
limits, $M_h>91.0$ GeV and $M_A >91.9$ GeV \cite{LEPAh}, on the $h$ and $A$
masses have been set.  This bound is obtained in the limit where the coupling
of the $Z$ boson to $hA$ pairs is maximal, $g^2_{ZhA}/g^2_{ZZH^0} \equiv
\cos^2(\beta- \alpha) \simeq 1$, i.e. in the non--decoupling regime for large
values of $\tb$. This limit is lower than the one from the Higgs--strahlung
process, due to the less distinctive signal and the $\beta^3$ suppression near
threshold for spin--zero particle pair production. Note that for small $M_A$
and large $\tb$ values, $M_H$ becomes small enough, cf.~Fig.~1, in the
no--mixing scenario to allow for the possibility of the process $\ee \to HA$,
which is suppressed by $\sin^2(\beta-\alpha)$. \s

Deriving a precise bound on $M_h$ for arbitrary values of $M_A$ and $\tb$ [i.e.
not only in the decoupling limit or for $\tb\gg 1$] and hence, for all values
of the angle $\alpha$, is more complicated since one has to combine results
from two different production channels, which have different kinematical
behavior, cross sections, backgrounds, etc.. However, some exclusion plots for
$\sin^2 (\beta-\alpha)$ versus $M_h$ from the Higgs--strahlung process [and
which can be used to constrain the mass of the $H$ boson if
$\sin^2(\beta-\alpha)$ is replaced by $\cos^2(\beta-\alpha)$] and $\cos^2
(\beta-\alpha)$ versus $M_A+M_h$ [with $M_h \sim M_A$] from the pair production
process, have been given in Refs.~\cite{LEPH0} and \cite{LEPAh},
respectively\footnote{The exclusion plot $\cos^2(\beta-\alpha)$ versus $M_h+
M_A$ has been obtained with the assumption that the total widths of the Higgs
bosons are rather small, which is not the case for large values of $\tb \gsim
30$. We will assume here, that one can nevertheless extrapolate these limits up
to values $\tb \simeq 50$.}.  We have fitted these exclusion contours and
delineated the regions allowed by the LEP2 data up to $\sqrt{s}=209$ GeV in the
$[M_A, \tb]$ and $[M_h, \tb]$ planes in the no--mixing, typical mixing and
maximal mixing scenarii [the other SUSY parameters are fixed as in section 2]. 
The domains allowed by the LEP2 constraints\footnote{These domains agree rather
well with those derived by the LEP collaborations in their combined analyses 
\cite{LEPH0,LEPAh}. This means that although we use different programs for 
the implementation of the radiative corrections [{\tt HDECAY} with an 
improved RG Higgs potential versus a complete implementation of the two--loop 
radiative corrections in the Feynamn diagramatic approach], the obtained Higgs 
boson spectrum is the same with a very good approximation, ${\cal O}(2$ GeV). 
This also means that our fitting procedure of the LEP2 contraints on the Higgs 
boson masses versus the mixing angle is rather good.} are shown in Fig.~3. \s

\begin{figure}[htbp]
\vspace*{-5mm}
\begin{center}
\epsfig{figure=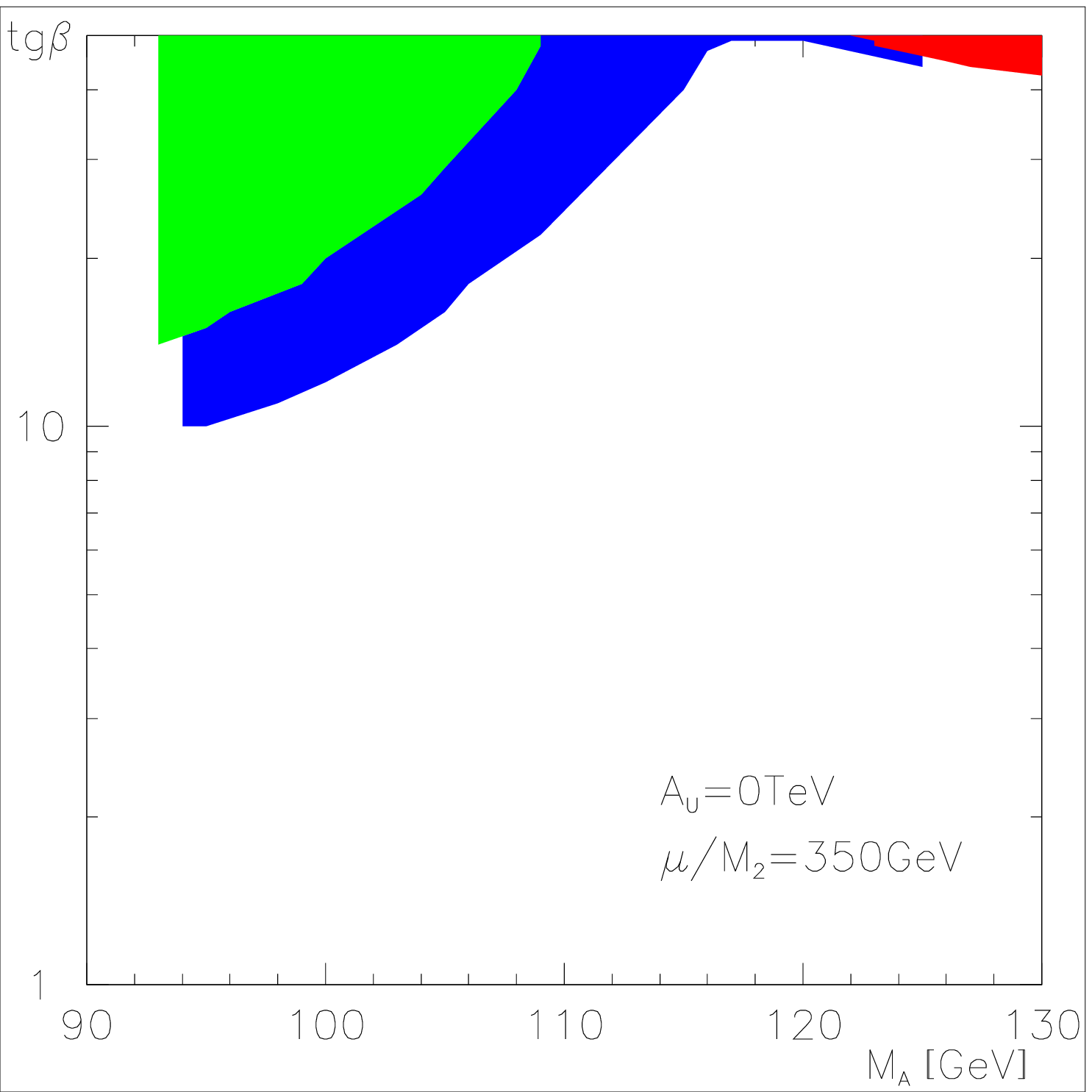,bbllx=3,bblly=3,bburx=410,bbury=420,height=6.5cm,width=7.5cm,clip=}
\hspace{1cm}
\epsfig{figure=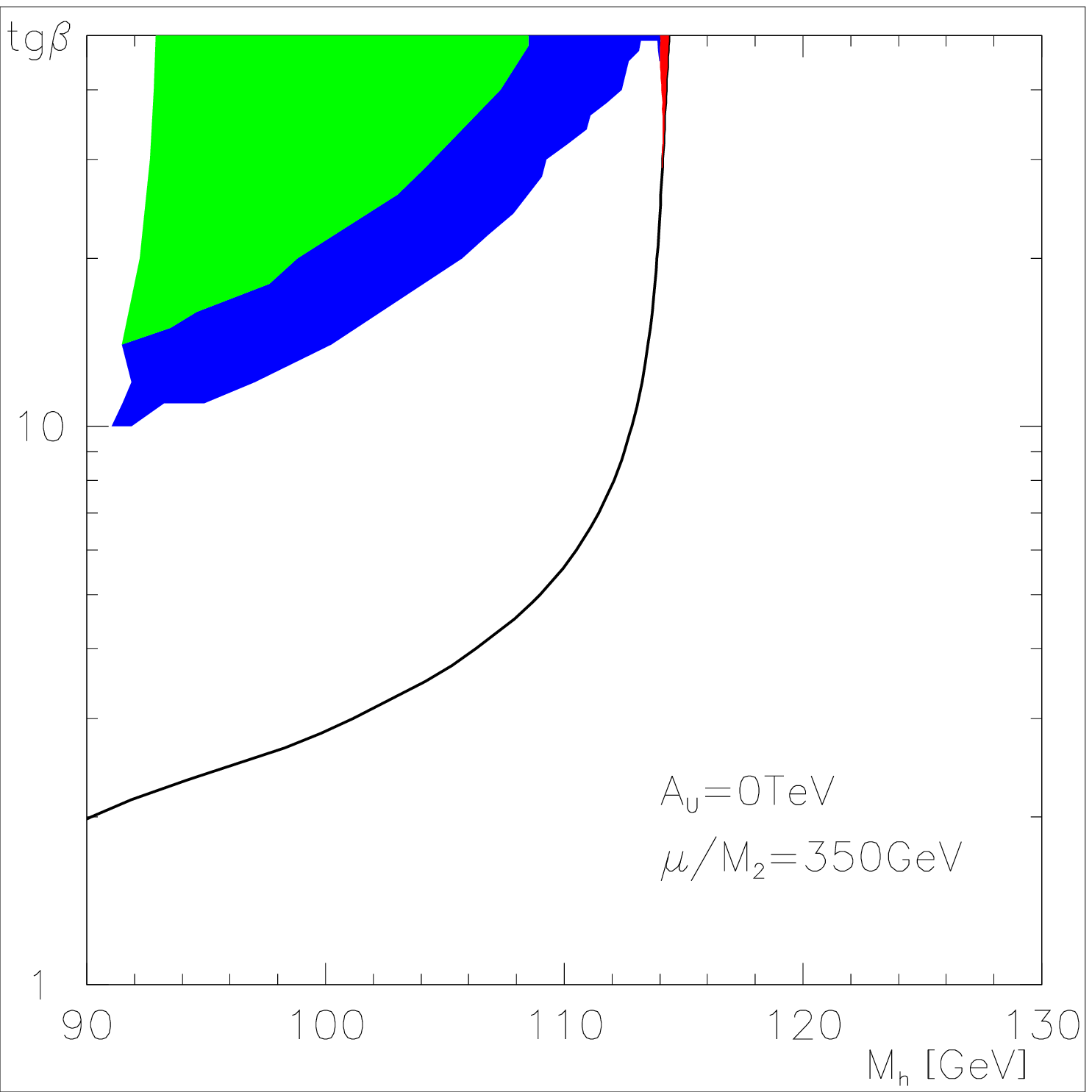,bbllx=3,bblly=3,bburx=410,bbury=420,height=6.5cm,width=7.5cm,clip=}
\\[0.1cm]
\epsfig{figure=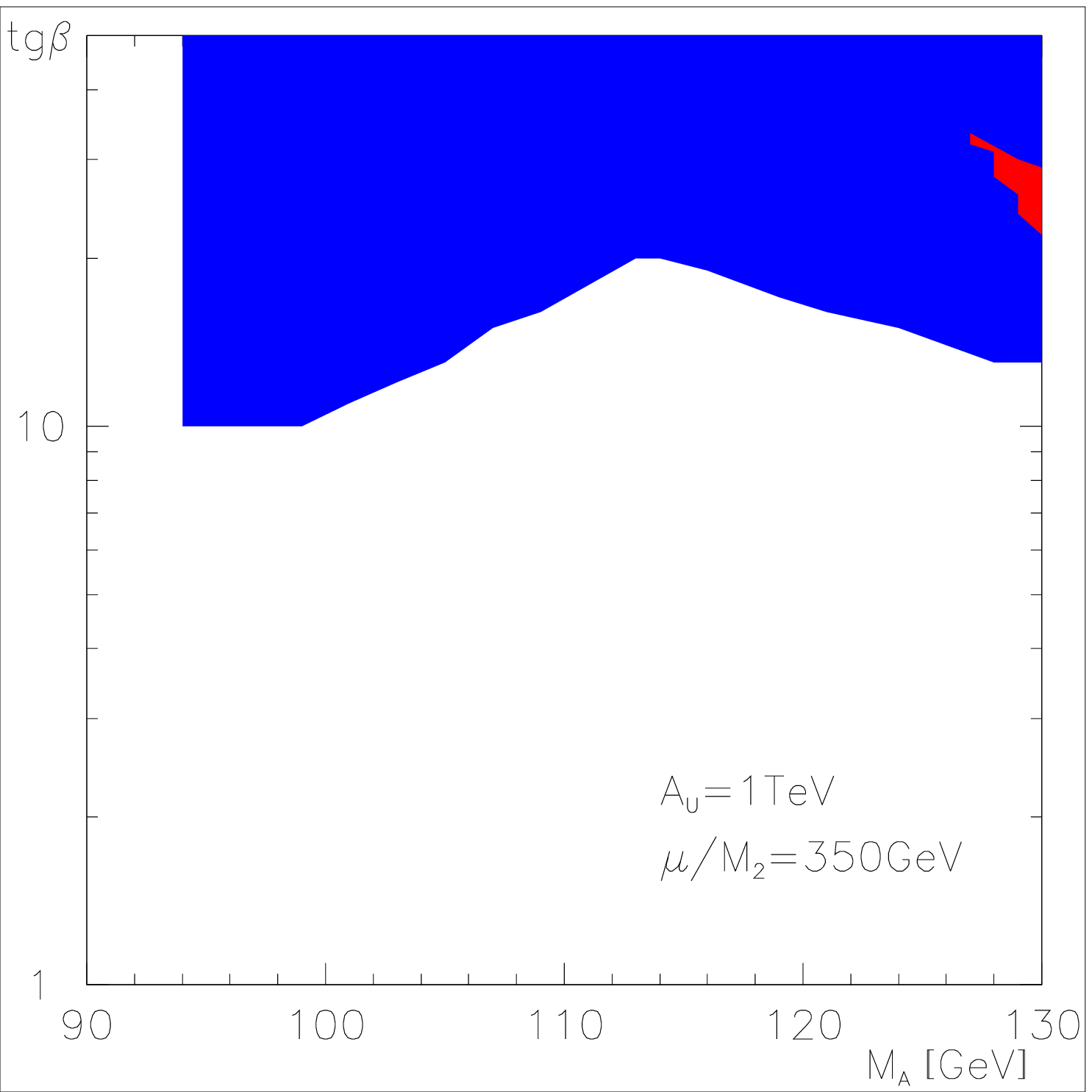,bbllx=3,bblly=3,bburx=410,bbury=420,height=6.5cm,width=7.5cm,clip=}
\hspace{1cm}
\epsfig{figure=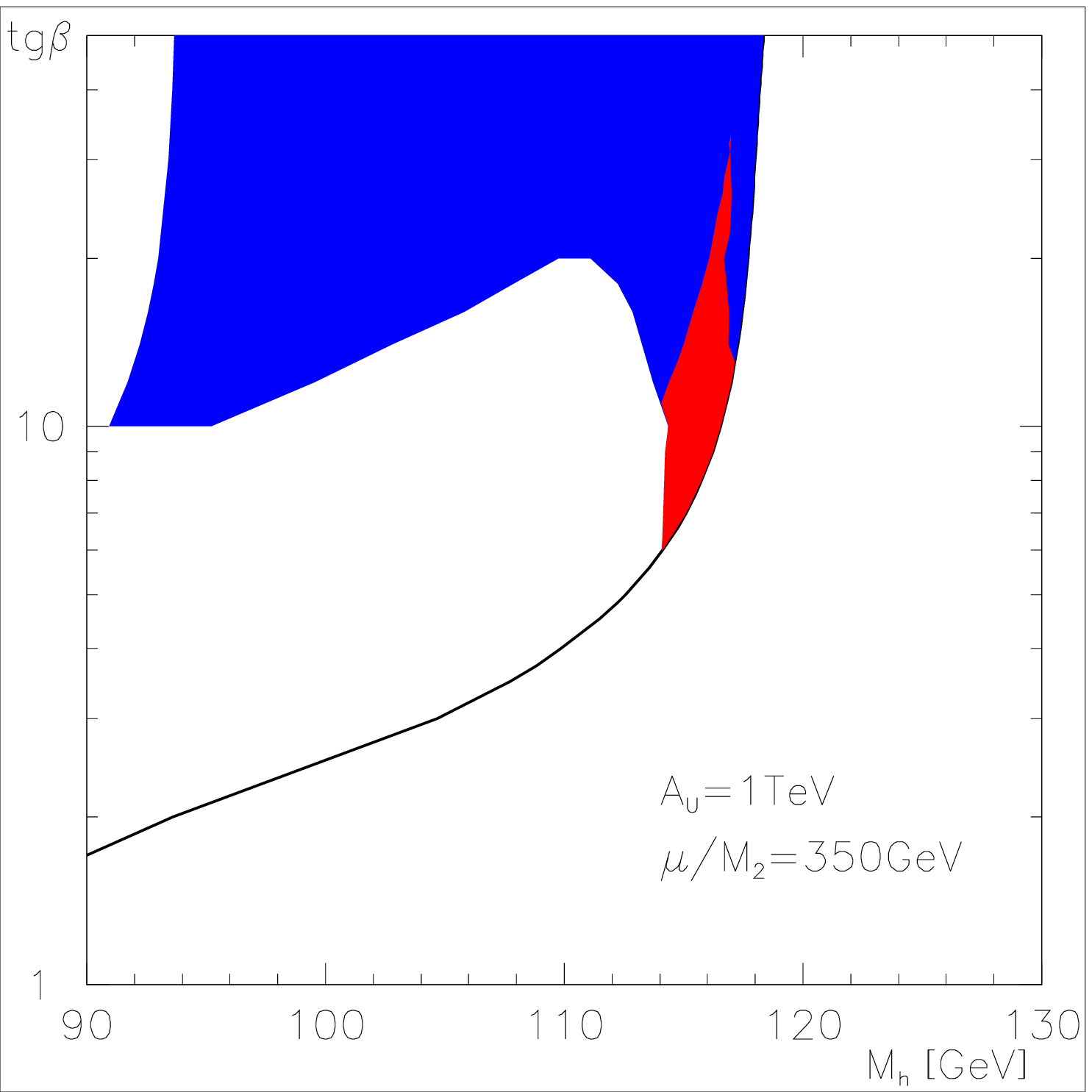,bbllx=3,bblly=3,bburx=410,bbury=420,height=6.5cm,width=7.5cm,clip=}
\\[0.1cm]
\epsfig{figure=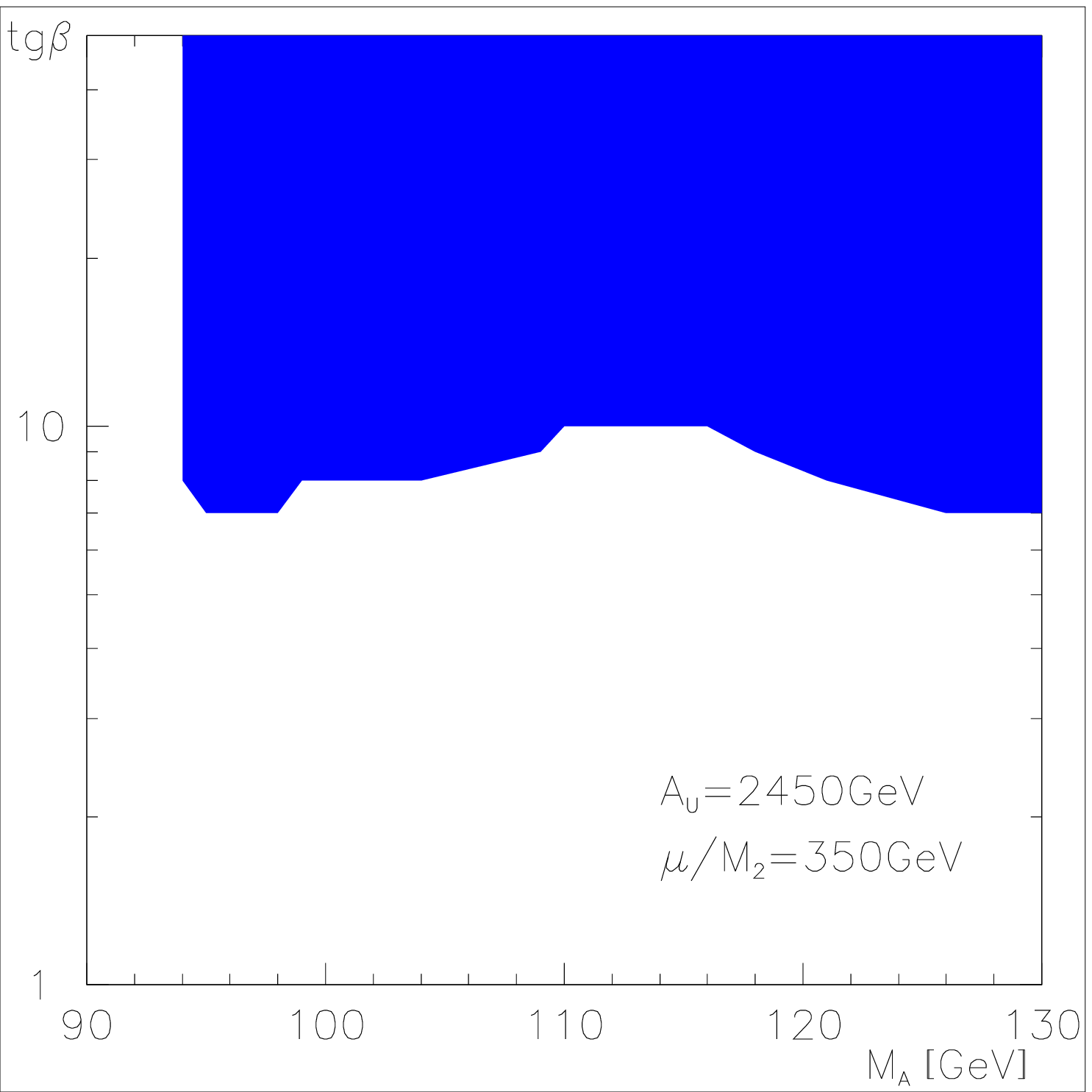,bbllx=3,bblly=3,bburx=410,bbury=420,height=6.5cm,width=7.5cm,clip=}
\hspace{1cm}
\epsfig{figure=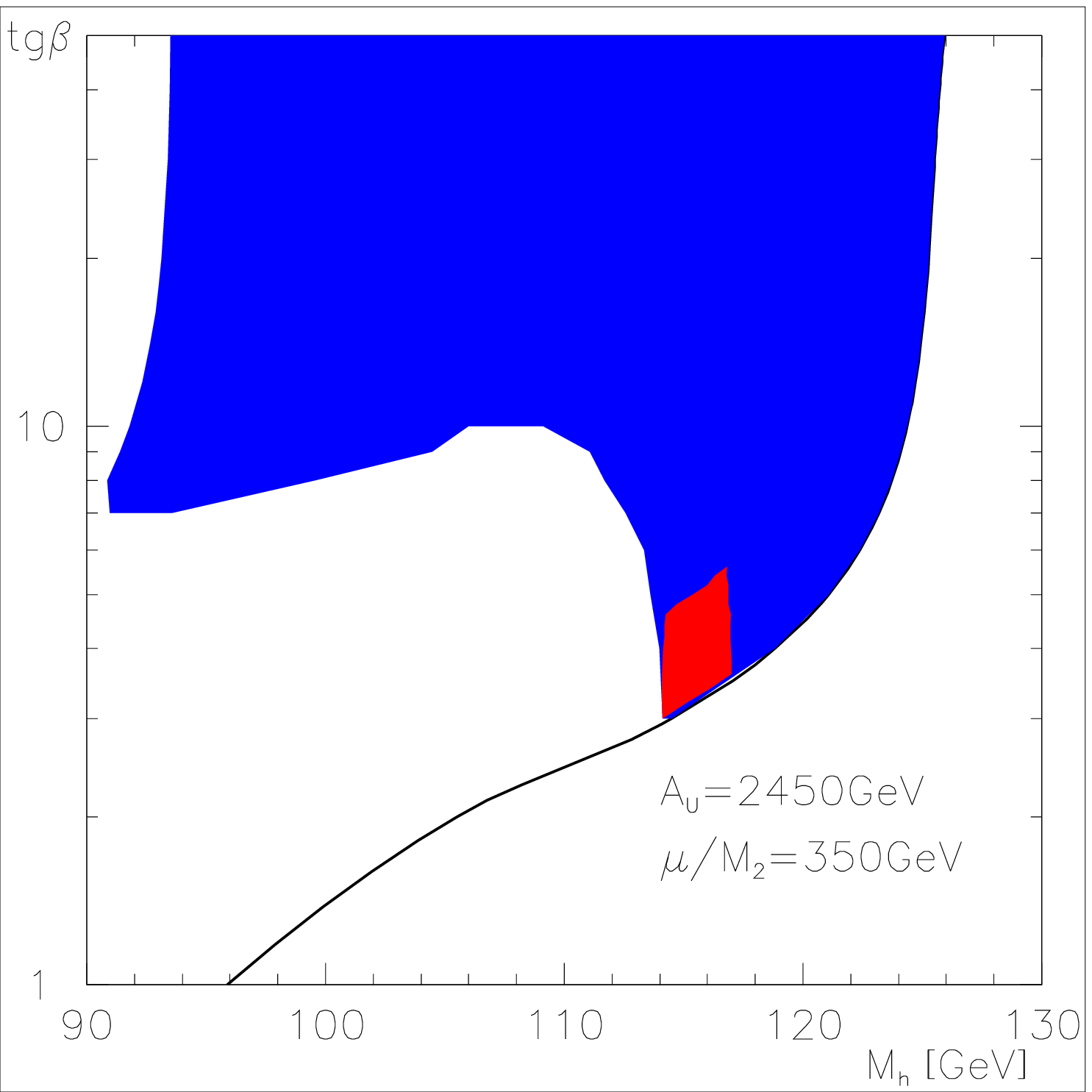,bbllx=3,bblly=3,bburx=410,bbury=420,height=6.5cm,width=7.5cm,clip=}
\caption{\it The allowed regions for $M_A$ [left] and $M_h$ [right] from LEP2
searches as a function of $\tan \beta$ (colored regions, or black, dark grey
and light grey regions) in the case of maximal (bottom), typical (central) and
no--mixing (up). The red (or dark grey) regions indicate where
$114$~GeV$<M_h<117$~GeV and $\sin^2(\beta-\alpha)>0.9$ and the green (or light
grey) regions indicate where $114$~GeV$<M_H<117$~GeV and
$\cos^2(\beta-\alpha)>0.9$.}
\end{center}
\end{figure}

In these figures the colored (blue, green and red) areas correspond to the
domains of the parameter space which are still allowed by the LEP2 searches. As
can be seen, the allowed regions are different for the three scenarii of mixing
in the stop sector. \s

In the no--mixing case, values $\tan \beta \lsim 10$ are excluded for
any value of the pseudoscalar mass $M_A \gsim 90$ GeV. For a light $A$ boson,
$M_A \lsim 100$ GeV and $\tan \beta \gsim 10$, the masses $M_h \sim M_A$ pass the
bound ($\gsim$ 91 GeV) from the Higgs pair production processs, while $H$ is
light enough ($M_H \lsim 115$ GeV) to be probed in the  Higgs--strahlung
process.  For increasing $M_A$, this situation holds only for larger $\tan
\beta$: for instance, for $M_A \gsim 115$ GeV, one needs values $\tan \beta
\gsim 40$ to cope with the experimental bounds [i.e. to make $H$ light enough 
to be produced]. 
In the typical mixing case, values $\tb \lsim 10$ are excluded for
$M_A$ close to 90 or 130 GeV. For intermediate $M_A$ values, one needs larger
$\tan \beta$ values to evade the constraint from the Higgs--strahlung process
since the $g_{hZZ}$ coupling is increasing while $M_h$ is relatively small,
leading to a sizeable $\sigma (\ee \to hZ)$. [Note that here, there 
is a strong interplay between the variation of $\sin^2(\beta-\alpha)$ and $M_h$
in the phase space in $\sigma (\ee \to hZ)$, which explains the bump around 
$M_A \sim 115$ GeV.] For large $M_A$, $\tb$ values close to 5 are still 
allowed since $M_h$ can be larger than 114 GeV. 
Finally, in the maximal mixing case, values $\tb \gsim 7$ are excluded for
90 GeV $\lsim M_A \lsim 130$ GeV. However, for large $M_A$, only values $\tb 
\gsim 3$ are excluded since here, the large stop mixing increases the maximal 
value of $M_h$ to the level where it exceeds the discovery limit, $M_h \gsim 
114$ GeV. \s

We turn now to the implications of the $2.1 \sigma$ evidence for a SM--like
Higgs boson with a mass $M_{\Phi_H}= 115.6$ GeV, seen by
the LEP collaborations in the Higgs--strahlung process \cite{LEPevid}. In view
of the theoretical and experimental uncertainties, this result will be
interpreted as favoring the mass range $114 \ {\rm GeV} \lsim M_{\Phi_H} \lsim
117 \ {\rm GeV}$. Furthermore, since this Higgs boson should be almost
SM--like, we will impose the constraint $g^2_{ZZ\Phi_H}/g_{ZZH^0}^2 \geq 0.9$
so that the cross section for the Higgs--strahlung process, $\sigma(\ee \to
\Phi_H Z)$, is maximized. Of course, at the same time [and in particular in the
case where the ``observed" Higgs boson is $H$] the experimental constraints
from the pair production process, $M_A, M_h \gsim 91$ GeV, should be taken 
into account. \s

The green areas show the regions where the ``observed" Higgs boson is the
heavier CP--even $H$ particle. This occurs only for no mixing with  values
$\tan \beta \gsim 15$ and $M_A \lsim 110$ GeV. In this case, $M_H \sim 115$ GeV
and has SM--like couplings, while $M_h$ is still larger than 91 GeV. For large
stop mixing, the $H$ boson is always heavier than 117 GeV and
cannot be observed. The red areas correspond to the regions where the
``observed" Higgs boson is in turn the lighter $h$ particle, while $M_H$ is
heavier than 117 GeV. These regions are larger for the typical mixing scenario:
in the no--mixing case, it is unlikely that the $h$ boson mass exceeds 114 GeV
except for very large $\tan \beta$, while in the maximal mixing scenario, the
$h$ boson is usually heavier than 117 GeV, except for very low $\tan \beta$.  

\newpage

Finally, let us briefly discuss the constraints on the intense--coupling
scenario from the Higgs boson searches at the Tevatron. The first important
constraint is coming from decays of the top quarks into $b$--quarks and charged
Higgs bosons, $t \to bH^+$. For $\tb$ values around unity [which are ruled out
as discussed above] and for large values of $\tb$, the $H^- t\bar{b}$ coupling,
$g_{H^-tb} \propto m_t/\tb + m_b \tb$, is very strong. The branching ratio
BR($t \to bH^+$) becomes then comparable and even larger than the branching
ratio of the standard mode, $t \to bW^+$, which allows the detection of
top quarks at the Tevatron. A search for this top decay mode, with the
$H^-$ boson decaying into $\tau \nu_\tau$ final states, by the CDF and D0
Collaborations \cite{charged} allows to place stringent limits on the value of
$\tb$ for charged Higgs boson masses below $\sim 150$ GeV: for $M_{H^\pm}\simeq
120$ GeV [i.e. $M_A \simeq 90$ GeV] one has $\tb \lsim 50$--60, while the bound
becomes very weak for $M_{H^\pm} \simeq 150$ GeV [i.e $M_A \simeq 130$ GeV]
where one has $\tb \lsim 100$ [i.e. beyond the limit where the $b$--quark
Yukawa coupling is perturbative]. \s

Furthermore, constraints on the MSSM Higgs sector at the Tevatron can be
derived \cite{Manuel} by exploiting the data on $\tau^+ \tau^- +$ 2 jets used
by the CDF Collaboration \cite{CDF} to place limits on third generation
leptoquarks. Indeed, in the associated production of the pseudoscalar $A$ boson
with $b\bar{b}$ pairs, $pp \to q\bar{q}/gg \to b\bar{b} A$, with the $A$ boson
decaying into $\tau^+ \tau^-$ final states, the cross section is proportional
to $\tan^2\beta$ and can be very large for $\tb \gg 1$. In addition, since in
this regime, one of the CP--even Higgs bosons is always degenerate in mass with
the $A$ boson and has couplings to $b\bar{b}$ [and $\tau^+\tau^-$] pairs which
are also proportional to $\tb$, the rate for $b\bar{b} \tau^+ \tau^-$ final
states due to Higgs bosons in this type of process is multiplied by a factor 2.
Since there is no evidence for non--SM contributions in this final state as 
analyzed by the CDF Collaboration, one can place bounds in the plane spanned 
by $\tb$ and $M_A$. The upper bound is $\tb \lsim 80$ for $M_A \gsim 
90$ GeV, and is weaker for higher values of $M_A$, $\tb \lsim 100$ for $M_A 
\sim 130$ GeV \cite{Manuel}. Therefore, this constraint is less severe than 
the one due to top decays into charged Higgs bosons \cite{charged}. \s 

There are two other processes for MSSM Higgs boson production which could be
relevant at the Tevatron as will be discussed later. The gluon--gluon fusion
mechanism, $gg \to \Phi$ with $\Phi \equiv h,H$ or $A$, can have large cross
sections in particular for large $\tb$ values, but the backgrounds in the main
decay channel $\Phi \to b\bar{b}$ are too large while for the cleaner $\gamma
\gamma$ decays, $\sigma \times {\rm BR}(\Phi \to \gamma \gamma)$ is too small.
The associated production of the CP--even Higgs bosons $h$ or $H$ with a $W$ or
$Z$ boson, has cross sections which are too small, in particular if the gauge
bosons are required to decay leptonically.  These processes will therefore be
useful only when an integrated luminosity of a few femtobarn will be
accumulated, i.e. only at the Run II of the machine [see section 5].  \s

To summarize, the constraints on the Higgs sector from Tevatron data are not 
very strong and values of $\tb$ slightly above $\tb \sim 50$ [and below] are 
still allowed experimentally.

\subsubsection*{3.2 Indirect constraints from precision data}

Indirect constraints on the parameters of the MSSM Higgs sector, in particular
on $M_A$ and $\tb$, come from the high--precision data: the measurements of the
$\rho$ parameter from $M_W$ and $\sin^2\theta_W$, the decays $Z \to b \bar{b}$,
the muon anomalous magnetic moment $(g_\mu-2)$ and the radiative decay $b \to
s\gamma$. In discussing these constraints, we will neglect the contributions of
the SUSY particles which are assumed to be heavy [$\mu=M_2=350$ GeV and $M_S=1$
TeV with $A_t=\sqrt{6}M_S$]. For an analysis of these SUSY contributions in a
constrained MSSM, see Ref.~\cite{benchmark} for instance. For completeness, we
will also give the analytical expressions of the Higgs boson loop contributions
to the observables in the intense--coupling regime. \bigskip

\nn {\bf a) The $\rho$ parameter} \s

Loop contributions of the MSSM Higgs bosons can alter the values of the
electroweak observables which have been precisely measured at LEP1, the SLC and
the Tevatron \cite{PDG,LEP1}. The dominant contributions to the weak
observables, in particular the $W$ boson mass and the effective mixing angle
$s_W^2 \equiv \sin^2\theta_W$, enter via a deviation from unity of the $\rho$
parameter which, in terms of $W$ and $Z$ boson self--energies at zero momentum
transfer, is defined by $\rho=(1-\Delta \rho)^{-1}$ with $\Delta \rho=
\Pi_{ZZ}(0)/M_Z^2 -\Pi_{WW}(0)/M_W^2$. Precision measurements constrain the
contribution of New Physics to be $\Delta \rho^{\rm NP} \lsim 1.1 \cdot
10^{-3}$ at the $1\sigma$ level \cite{PDG}. \s

In the intense-coupling regime, the contributions of the MSSM Higgs sector to 
the $\rho$ parameter is given by [note that the same expression holds exactly 
in the decoupling regime]:
\beq
\Delta \rho^{\rm Higgs}= - \frac{G_F M_W^2}{8 \sqrt{2}\pi^2} \bigg[ 3 f_1 \bigg(
\frac{M_C^2}{M_Z^2} \bigg) + 2 f_2 \bigg( \frac{M_{H^\pm}^2}{M_W^2},
\frac{M_{A}^2}{M_W^2} \bigg) \bigg]
\eeq
with the two functions $f_1$ and $f_2$ given by \cite{deltarho} 
\beq
f_1(x) = x \left[ \frac{\ln c_W^2 - \ln x}{c_W^2-x} + \frac{\ln x}{ c_W^2
(1-x)} \right] \ \ , \ 
f_2(x_1,x_2) = \frac{x_1 x_2}{x_1-x_2} \ln \frac{x_2}{x_1} +
\frac{1}{2} (x_1+x_2) 
\eeq
with $c_W^2=1-s_W^2$. The first contribution, $\Delta \rho^{\rm Higgs}_{\rm SM}
= - 3G_F M_W^2/(8 \sqrt{2}\pi^2) f_1(M_C^2/M_Z^2)$, is simply the one of the SM
Higgs boson with a mass $M_{H^0}=M_C$, which is close to the Higgs mass, ${\cal
O}(100$ GeV), favored by the global fits of the electroweak data \cite{LEP1}.
The second contribution, proportional to $f_2$, is the genuine MSSM
contribution.  The contribution of loops involving CP--even and CP--odd Higgs
boson, $\sim -f_2(M_A^2/M_W^2, M_A^2/M_W^2)$ in the bracket, vanishes since
$f_2(x,x)=0$. In fact, only loops which involve particles which have a large
mass splitting contribute significantly to $\Delta \rho$. This is the reason
why the genuine MSSM contribution, $ \Delta \rho^{\rm Higgs}_{\rm non-SM}= -
G_F M_W^2/(4 \sqrt{2}\pi^2) f_2(M_{H^\pm}^2/M_W^2, M_{A}^2/M_W^2)$, shown in
Table 2, is always extremely small: the mass difference between the $H^\pm$ and
$A$ bosons is not large enough. In turn, the SM--like Higgs contribution has a
size which is comparable to the experimental error for $M_{H^0}=M_C \sim 127$ 
GeV [we use $\tb=50$].

\begin{table}[htbp]
\renewcommand{\arraystretch}{1.6}
\begin{center}
\vspace*{-5mm}
\begin{tabular}{|c||c|c|c|} \hline
$M_A~[{\rm GeV}]$ &  90  & 110 & 130 \\ \hline
$\Delta^{\rm Higgs}_{\rm non-SM}$ & $ -0.53\cdot 10^{-4}$ & $-0.38 \cdot 
10^{-4}$ &  $ -0.28 \cdot 10^{-4}$ \\ \hline
$\Delta^{\rm Higgs}_{\rm SM}$ & $ 4.7\cdot 10^{-4}$ & $4.7 \cdot 10^{-4}$ 
&  $ 4.7\cdot 10^{-4}$ \\ \hline
\end{tabular}
\end{center}
\vspace*{-2mm}
\caption[]{\it The genuine and the SM--like contributions of the MSSM Higgs 
sector to the $\rho$ parameter in the intense--coupling regime with $\tb=50$. }
\end{table}

\nn {\bf b) The g--2 of the muon} \s

The precise measurement of the anomalous magnetic moment of the muon recently
performed at BNL, $a_\mu \equiv g_\mu-2 = 11 659 202 (20) \cdot 10^{-10}$
\cite{gm2} where we have added the statistical and systematical errors, can
provide very stringent tests of models of New Physics. In the MSSM, the Higgs
sector will contribute to $a_\mu$ through loops involving the neutral Higgs
bosons $h,H$ and $A$ with muons and loops involving the charged Higgs bosons
$H^\pm$ with neutrinos. The contributions are sizeable only for large values of
$\tb$ for which the $\Phi \mu^+ \mu^-$ and $H^+ \mu \nu_\mu$ couplings are
enhanced; for a recent analysis see Ref.~\cite{Haber}. \s

Taking into account only the leading, $\propto \tan^2\beta$, contributions [i.e.
neglecting the contribution of the SM--like CP--even Higgs boson $\Phi_H$] and
working in the intense--coupling regime, one obtains for the MSSM Higgs sector
contribution to $a_\mu$ [again, the expression holds also in the decoupling 
limit in this approximation]
\beq
a_\mu^{\rm Higgs} \simeq \frac{G_F m_\mu^2}{24 \pi^2 \sqrt{2}} \tan^2 \beta 
\bigg[ 4 \frac{m_\mu^2}{M_A^2} - \frac{m_\mu^2}{M_{H^\pm}^2} \bigg] 
\eeq
This contribution is given in Table 3 for $\tb=50$ for $M_A=90,110$ and 130 GeV,
and one can see that it is far too small compared to the experimental error. 
[Here, the approximated result which is displayed, is about 10 to 20 \% smaller 
or larger than the exact result].

\begin{table}[htbp]
\renewcommand{\arraystretch}{1.6}
\begin{center}
\vspace*{-5mm}
\begin{tabular}{|c||c|c|c|} \hline
$M_A~[{\rm GeV}]$ &  90  & 110 & 130 \\ \hline
$a_\mu^{\rm Higgs}$ & $ 4.5\cdot 10^{-12}$ & $2.9 \cdot 10^{-12}$ 
&  $ 2.0 \cdot 10^{-12}$ \\ \hline
\end{tabular}
\end{center}
\vspace*{-2mm}
\caption[]{\it Contribution of the MSSM Higgs sector to $a_\mu$ for $\tb=50$.} 
\end{table}

\nn {\bf c) The Zbb vertex} \s

An observable where the MSSM Higgs sector can have sizeable effects is the $Z$
boson decay into $b\bar{b}$ final states. The neutral Higgs particles $h,H,A$
as well as the charged $H^\pm$ bosons can be exchanged in the $Z b\bar{b}$
vertex \cite{Zbb} and for large values of $\tb$, for which the Higgs boson
couplings to $b$ quarks are strongly enhanced, they can alter significantly the
values of the partial decay width $\Gamma(Z \to b\bar{b})$ [or equivalently the
ratio $R_b = \Gamma(Z \to b\bar{b})/\Gamma(Z \to {\rm hadrons})$ which is
measured with a better accuracy] and the forward--backward asymmetry $A_{FB}^b$
[or the polarization asymmetry $A^b_{LR}$]. In terms of the left-- and
right--handed $Zf\bar{f}$ couplings,  $g_{L/R}^f= I_f^3 -e_f s_W^2$, and
neglecting the $b$--quark mass for simplicity, they are given by: 
\beq
\Gamma (Z \to b\bar{b}) \simeq \frac{G_F M_Z^3}{\sqrt{2}\pi} 
[(g_L^b)^2+(g_R^b)^2]  \ \ ,  \ \ 
A_{FB}^b \simeq \frac{3}{4} \frac{(g_L^e)^2-(g_R^e)^2}{(g_L^e)^2+(g_R^e)^2}
\frac{(g_L^b)^2-(g_R^b)^2}{(g_L^b)^2+(g_R^b)^2}
\eeq
In the limit $\tb \gg 1$, the MSSM neutral (N) and charged (C) Higgs boson 
contributions to the left-- and right--handed bottom quark couplings are given 
by 
\beq
&& \hspace*{2cm} \delta g_{R/L}^b = \delta  g_{R/L}^b|_N + \delta g_{R/L}^b|_C
\non \\
\delta g_{R/L}^b|_N &=& \mp \bigg( \frac{g m_b \tb}{ \sqrt{2}M_W} \bigg)^2 
\bigg[ C_2 (m_b,M_A,M_A) \pm g_{L/R}^b C_1( M_A, m_b, m_b) \bigg] \non \\
\delta g_{R}^b|_C &=& - \bigg( \frac{g m_b \tb}{ \sqrt{2}M_W} \bigg)^2 
\bigg[ \bigg( 2s_W^2 - 1 \bigg) C_2 (m_t,M_{H^\pm},M_{H^\pm}) + 
g_{L}^t C_1( M_{H^\pm}, m_t, m_t) \non \\ &&
- g_{R}^t C_0  (M_{H^\pm}, m_t, m_t) \bigg]   \ \ \ , \ \ \  \delta g_{L}^b|_C 
= 0
\eeq
where the functions $C_{1,2}$ are given in terms of the 
Passarino--Veltman two-- and three--point functions [the latter evaluated at 
$q^2=M_Z^2$ with $m_b \simeq 0$] by: 
\beq
C_2(m_1,m_2,m_2)&=&C_{24}(m_1,m_2,m_2) + \frac{1}{2} B_1(m_1,m_2) \non \\
C_1(m_1,m_2,m_2)&=&-\frac{1}{2}+2 C_{24}(m_1,m_2,m_2) -M_Z^2 C_{23} 
(m_1,m_2,m_2)+ B_1(m_1,m_2) 
\eeq
The latest experimental values of $R_b$ and the forward backward asymmetry
are \cite{LEP1}: $R_b = 0.21653 \pm 0.00069$ and $A_{FB}^b=0.099 \pm 0.002$
[note that they deviate by, respectively, $+1.1 \sigma$ and $-2.4 \sigma$ from
the predicted values in the SM]. This means that the virtual effects of the 
MSSM Higgs bosons should be, in relative size, of the order of $0.3\%$ in 
$R_b$ and $2\%$ in $A_{FB}^b$ to be detectable. This is far to be the case
even for $\tb$ values close to 50 and for $M_A \gsim 90$ GeV as can be 
seen from Table 4 where the approximate contributions [which are very close, at 
most a few percent,  to the exact contributions] are displayed.  \s

\begin{table}[htbp]
\renewcommand{\arraystretch}{1.7}
\begin{center}
\vspace*{-5mm}
\begin{tabular}{|c||c|c|c|} \hline
$M_A~[{\rm GeV}]$ &  90  & 110 & 130 \\ \hline
$\Delta R_b/R_b $ & $ -5.0\cdot 10^{-4}$ & $-7.5 \cdot 10^{-4}$ 
&  $ -8.3\cdot 10^{-4}$ \\ \hline
$\Delta A_{\rm FB}^b/A_{\rm FB}^b$ & $ 2.5\cdot 10^{-3}$ & $2.5 \cdot 
10^{-3}$ &  $ 2.4\cdot 10^{-3}$ \\ \hline
\end{tabular}
\end{center}
\vspace*{-1mm}
\caption[]{\it Relative contribution of the MSSM Higgs sector to the two
observables $R_b=\Gamma(Z\to b\bar{b})/\Gamma (Z \to {\rm hadrons})$ and 
$A_{\rm FB}^b$ for $\tb=50$.} 
\end{table}

\bigskip
\nn {\bf d) The $b \to s \gamma$ decay} \s

Finally, in the radiative and flavor changing decay $b \to s\gamma$, in
addition to the SM contribution built--up by $W$ boson and top quark loops,
loops of charged Higgs bosons and top quarks can significantly contribute in
the MSSM together with SUSY particle loops \cite{bsg0,bsg2}.  Since SM and MSSM
Higgs contributions appear at the same order of perturbation theory, the
measurement of the inclusive branching ratio of the $B \ra X_s \gamma$ given by
the CLEO and Belle Collaborations \cite{bsg} is a very powerful tool for
constraining $\tb$. We use the value \cite{PDG}
\beq
{\rm BR}(b \ra s \gamma) = (3.37 \pm 0.37 \pm 0.34 \pm 0.24^{+0.35}_{-0.16}
\pm 0.38) \cdot 10^{-4}
\eeq
where the errors are, respectively, the statistical error, the systematical
error, the error from model dependence, the one due to the extrapolation from
the data to take into account the full range of photon energies,  and finally
an estimate of the theory uncertainty. We allow the branching ratio to vary 
within the conservative range: $ 2 \times 10^{-4} \leq {\rm BR}(b \ra s 
\gamma) \leq 5 \times 10^{-4}$. \s

For the calculation of the MSSM Higgs boson and SUSY particle contributions, we
will use the most up--to--date determination of the $b\to s \gamma$ decay rate
\cite{bsg2}, where all known perturbative and non--perturbative effects are
included\footnote{We thank the authors of Ref.~\cite{bsg2}, in particular
P.~Gambino, for providing us with their FORTRAN code for the calculation of
BR($b \to s\gamma$) at NLO.}. This includes all the possibly large 
contributions which can occur at NLO, such as terms $\propto \tan \beta$ [in 
particular in $m_b$], and/or terms containing logarithms of $M_S/M_W$.
For the input parameters we will use the values given in Refs.~\cite{bsg2},
except for the cut--off on the photon energy, $E_{\gamma} > (1-\delta)m_b/2$ in
the bremsstrahlung process $b \to s\gamma g$, which we fix to $\delta=0.9$. 
For the SUSY spectrum, we will work in the scenario discussed in section 2,
i.e. a common squark mass $M_S= 1$ TeV, the maximal mixing scenario $A_t
=\sqrt{6} M_S$ and $A_b=1$ TeV, $\mu=M_2=350$ GeV. \s

\begin{table}[htbp]
\renewcommand{\arraystretch}{1.5}
\begin{center}
\vspace*{-5mm}
\begin{tabular}{|c||c|c|c|} \hline
$M_A~[{\rm GeV}]$ &  90  & 110 & 130 \\ \hline
$\tan \beta=10$ & $ 5.2\cdot 10^{-4}$ & $5.0 \cdot 10^{-4}$ 
&  $ 4.8 \cdot 10^{-4}$ \\ \hline
$\tan \beta=30$ & $ 3.7\cdot 10^{-4}$ & $3.5 \cdot 
10^{-4}$ &  $ 3.3\cdot 10^{-4}$ \\ \hline
$\tan \beta=50$ & $ 2.5\cdot 10^{-4}$ & $2.3 \cdot 
10^{-4}$ &  $ 2.2\cdot 10^{-4}$ \\ \hline
\end{tabular}
\end{center}
\vspace*{-4mm}
\caption[]{\it BR($b \to s\gamma)$ in the MSSM for $\tb=10,30,50$ 
and $M_A=90,110,130$ GeV.}
\end{table}

Values of BR($b \to s\gamma$) in the MSSM are displayed for three choices 
of $\tb$ and $M_A$ in Table 5, and one can see that they are in most cases
within the allowed range $2 \times 10^{-4} \leq {\rm BR}(b \ra s \gamma) \leq 5 
\times 10^{-4}$. An exception is when $\tb \sim 10$ and the $A$ boson 
is light, $M_A \lsim 110$ GeV, where the branching ratio slightly exceeds the 
maximal value; however, we have checked that for other values of the SUSY 
parameters  a reasonable BR($b \to s\gamma)$ can be accommodated also in this 
case. 

\subsection*{4. Branching fractions and total decay widths} 

The  branching ratios and the total decay widths of the three neutral Higgs
bosons \cite{Decays}, again evaluated with the program {\tt HDECAY} with the
set of input parameters discussed in section 2.4, are shown in Figs.~4--5 and 
6 as functions of their masses for $M_A$ varying between 90 and 130 GeV, for two
values of $\tb$, 10 and 30. \s

In the case of the $A$ boson, the situation is rather simple since in the
entire mass range shown, the pattern is the same: BR$(A \to \bar{b}b$) and
BR($A\to \tau^+ \tau^-$) are approximately 90\% and 10\%, respectively, while
the gluonic branching fraction is slightly above $10^{-3}$ and BR$(A \to \mu^+
\mu^-) \sim 5\times 10^{-4}$. The branching ratios for the other decay modes,
including the $A \to \gamma \gamma$ decay, are below the level of $10^{-5}$ for
$\tb \gsim 10$. This is due to the strong enhancement (suppression) of the $A$
couplings to isospin $-\frac{1}{2} (+\frac{1}{2})$ fermions [the $Agg$ coupling
is mediated here mainly by the $b$--loop contribution]. Some of the final
states which are important for larger $M_A$ values [such as the decay $A \to
hZ$] are strongly suppressed by phase space in addition to the coupling
suppression for large $\tb$ values. \s 

The pattern for the CP--even Higgs bosons is similar to that of the $A$ boson
if their masses are close to $M_A$ and $\tb \gg1$. The exception is when $h$
and $H$ have masses very close to $M_{\rm C}$. In this limit, they are SM--like
and decay into charm quarks and gluons with rates similar to the one for
$\tau^+ \tau^-$ [$\sim$ a few $\%$] and in the high mass range, $M_{H^0} \sim
125$ GeV, into $WW$ ($ \sim 10 \%$) and $ZZ$ (a few $\%$) pairs with one of the
gauge  bosons being virtual.  However, this limit is not reached in practice,
in particular for the lightest $h$ boson with $\tb \lsim 50$. \s

\begin{figure}[htbp]
\vspace*{-5.5cm}
\hspace*{-2.cm}
\mbox{\psfig{figure=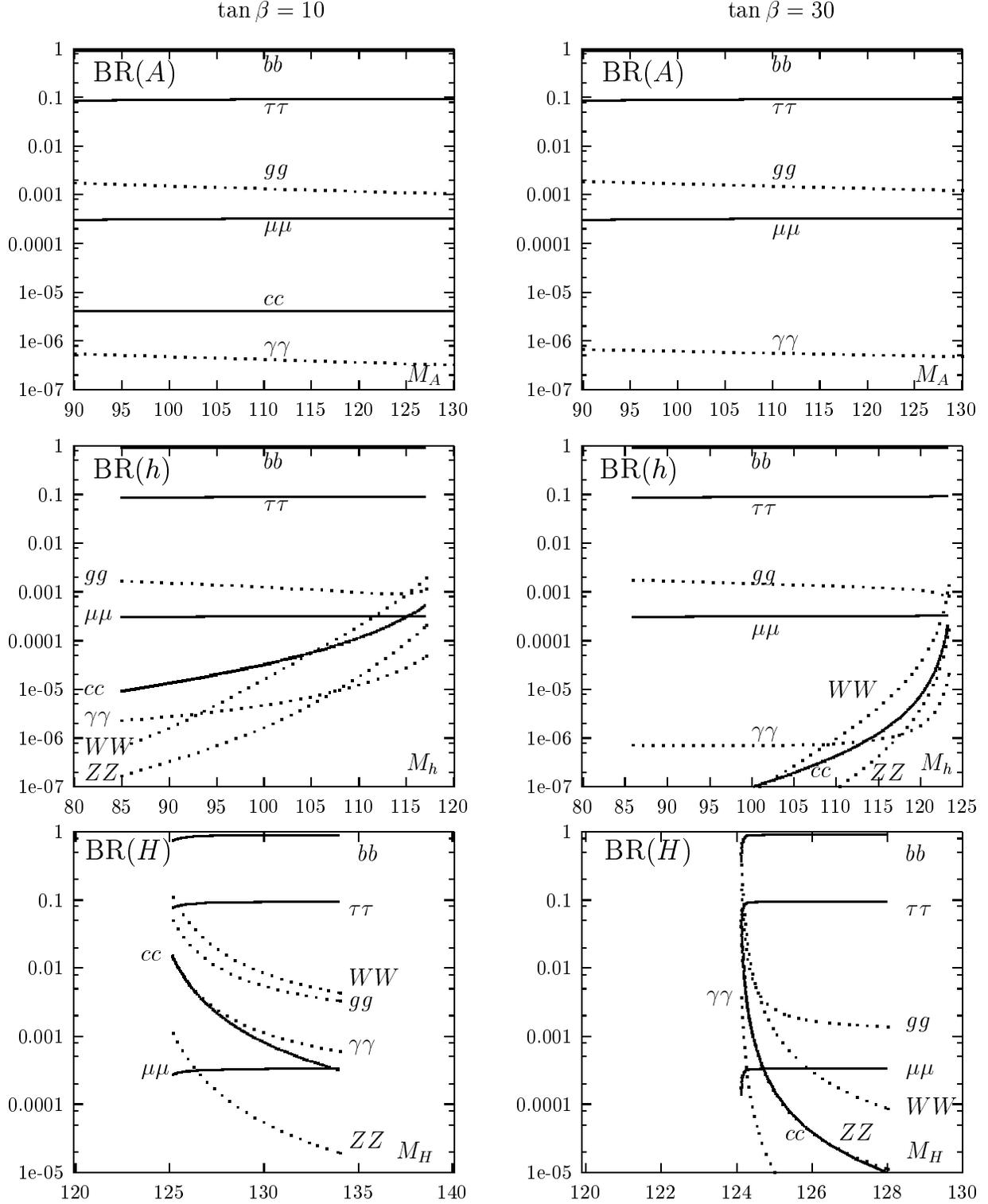,width=20cm}}
\vspace*{-4.cm}
\label{figbr}
\caption[]{\it The branching ratios of the neutral Higgs bosons $h,H$ and $A$
as a function of their masses for two values of $tan \beta=10$ (left panel)
and  30 (right panel).}
\end{figure}
\begin{figure}[htbp]
\vspace*{-2.3cm}
\begin{center}
\epsfig{figure=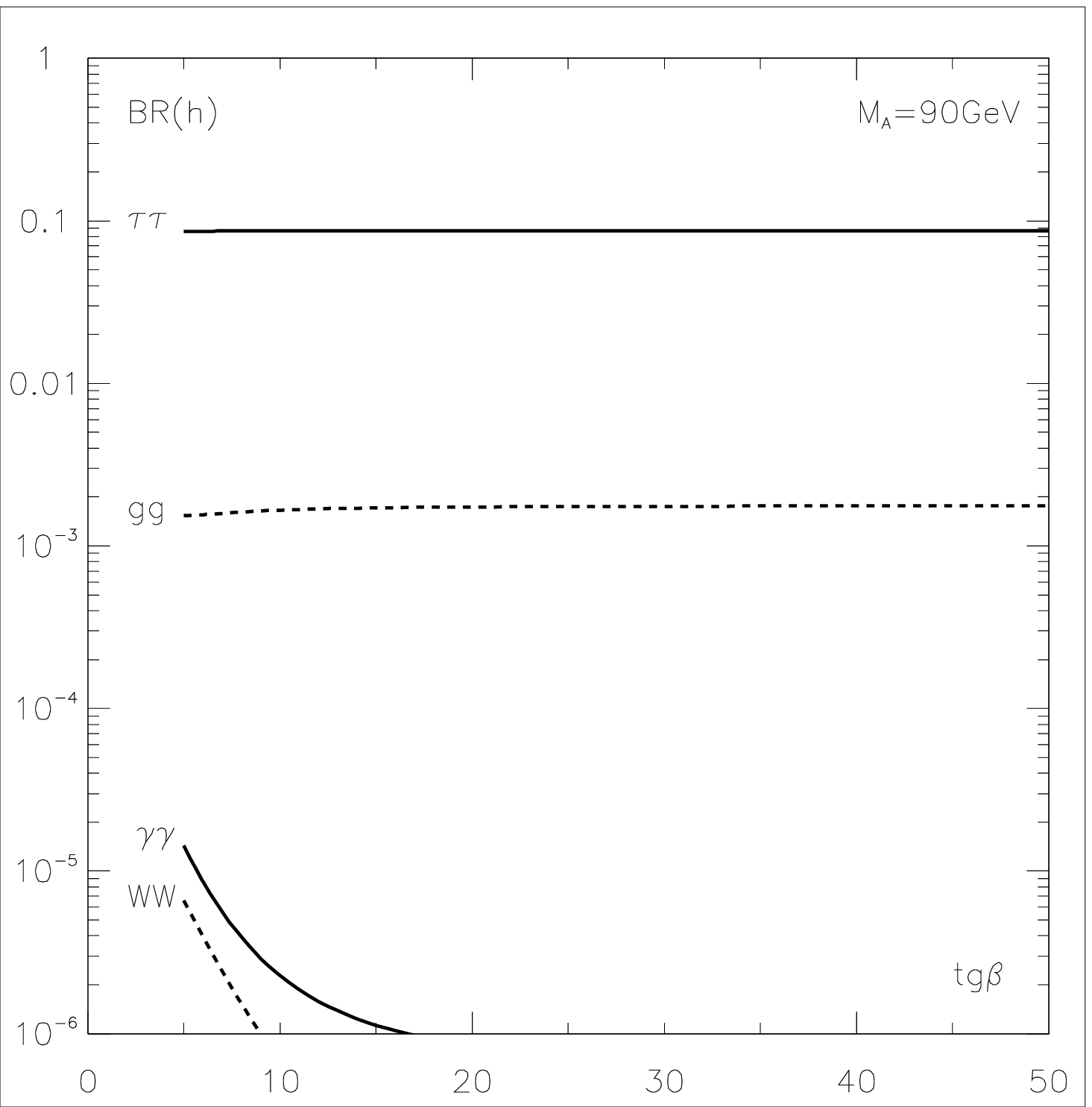,bbllx=3,bblly=3,bburx=410,bbury=420,width=7.cm,height=5cm,clip=}
\hspace{1cm}
\epsfig{figure=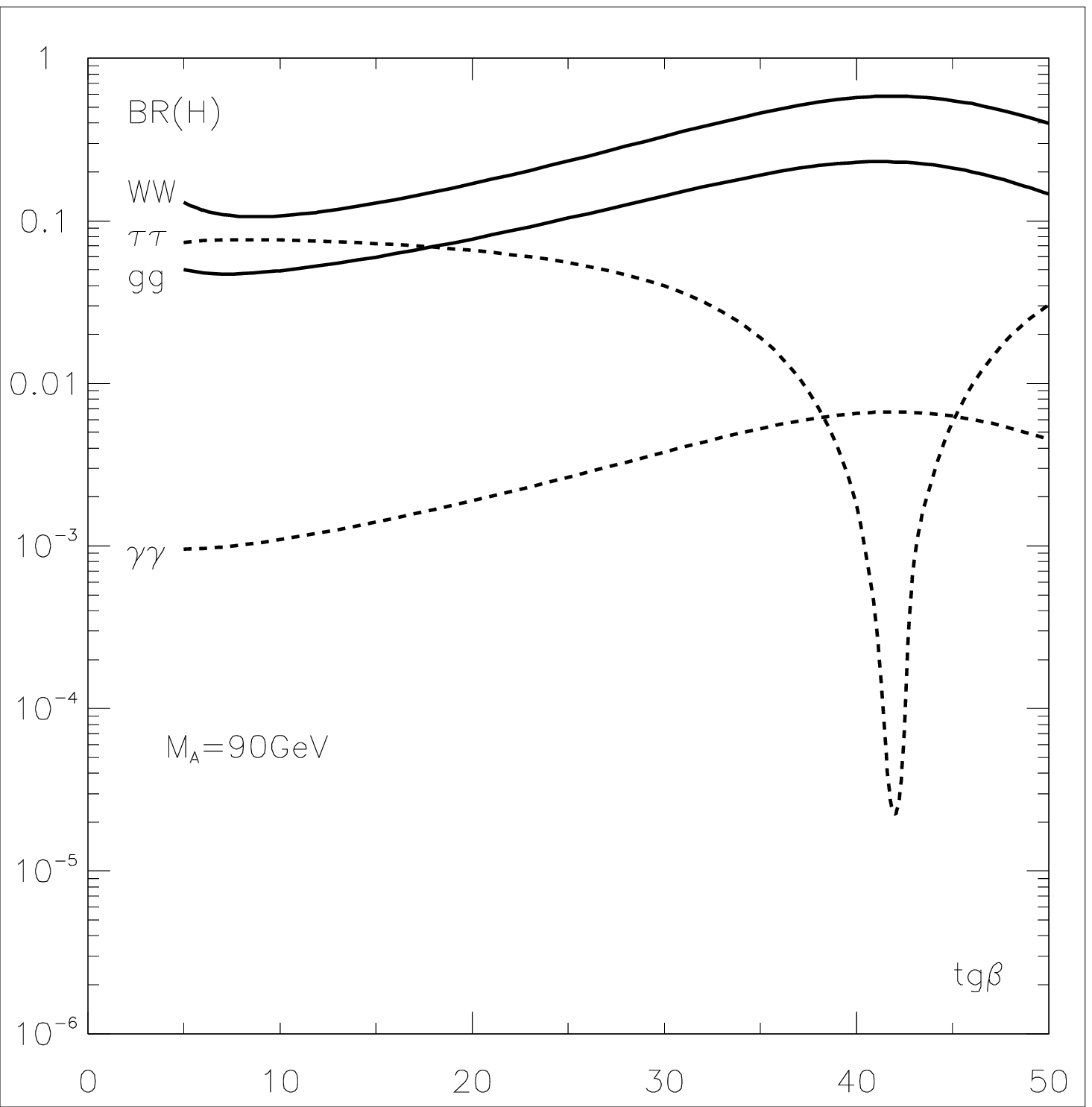,bbllx=3,bblly=3,bburx=410,bbury=420,width=7.cm,height=5cm,clip=}
\\[0.2cm]
\epsfig{figure=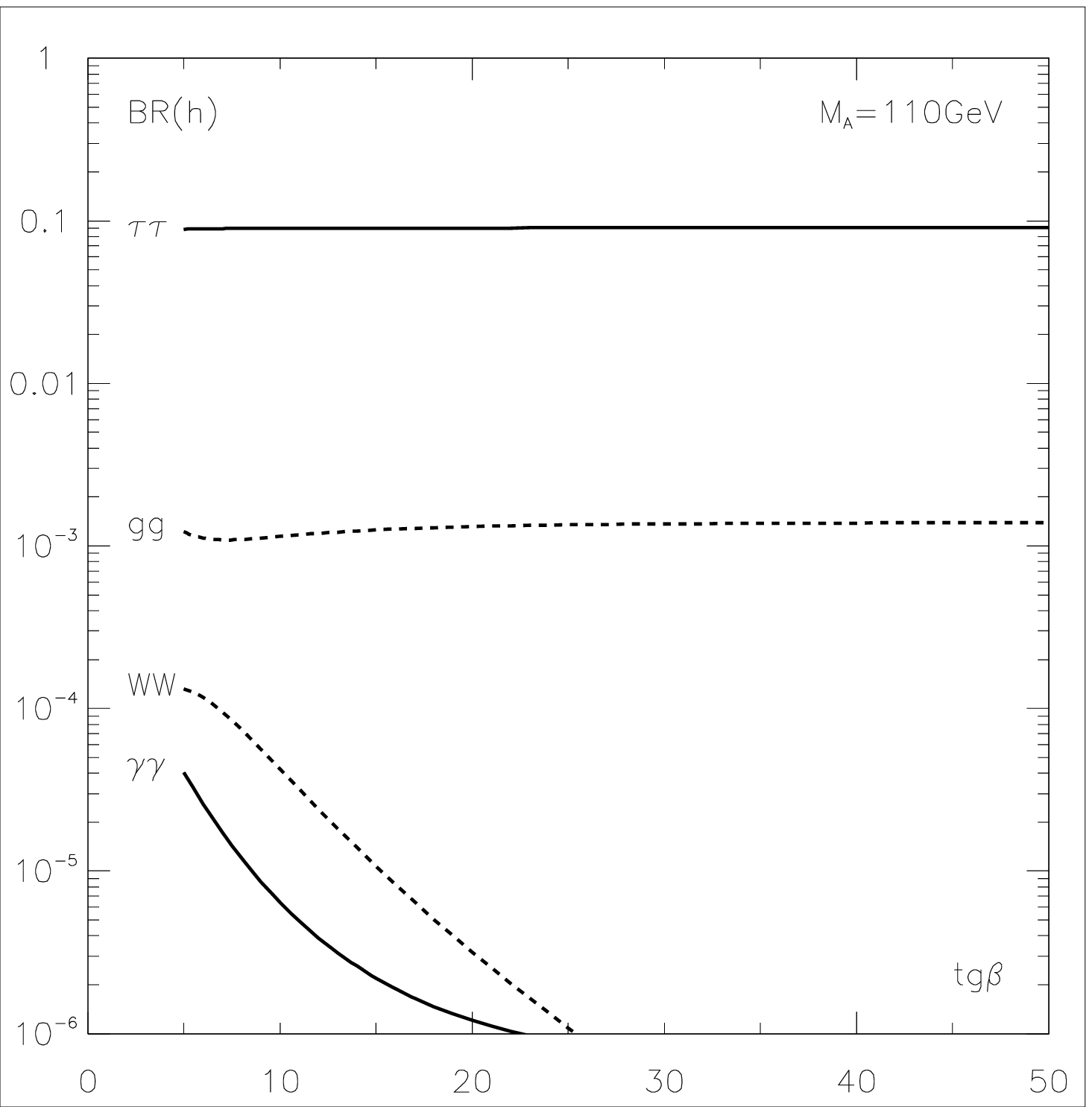,bbllx=3,bblly=3,bburx=410,bbury=420,width=7.cm,height=5cm,clip=}
\hspace{1cm}
\epsfig{figure=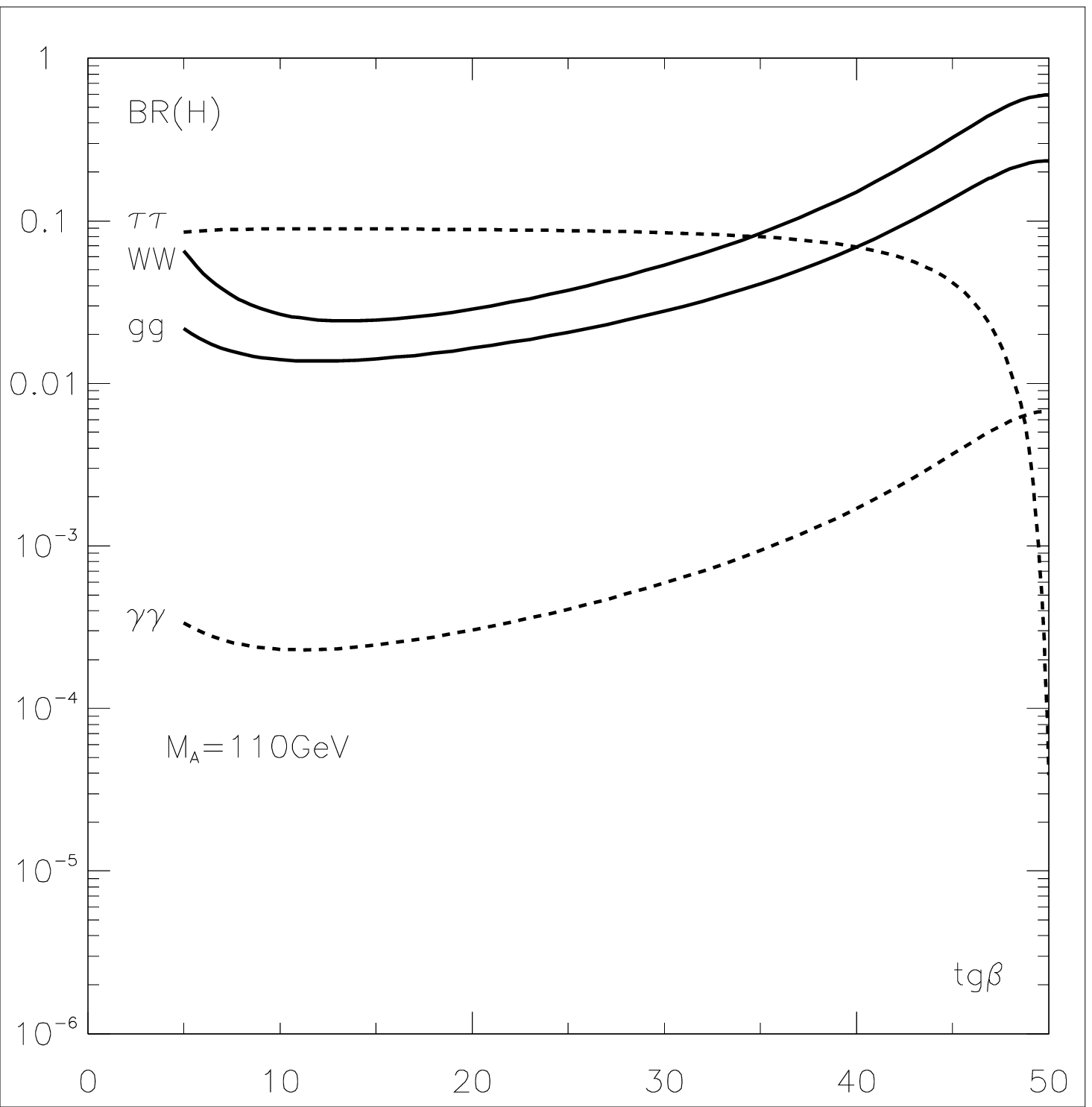,bbllx=3,bblly=3,bburx=410,bbury=420,width=7.cm,height=5cm,clip=}
\\[0.2cm]
\epsfig{figure=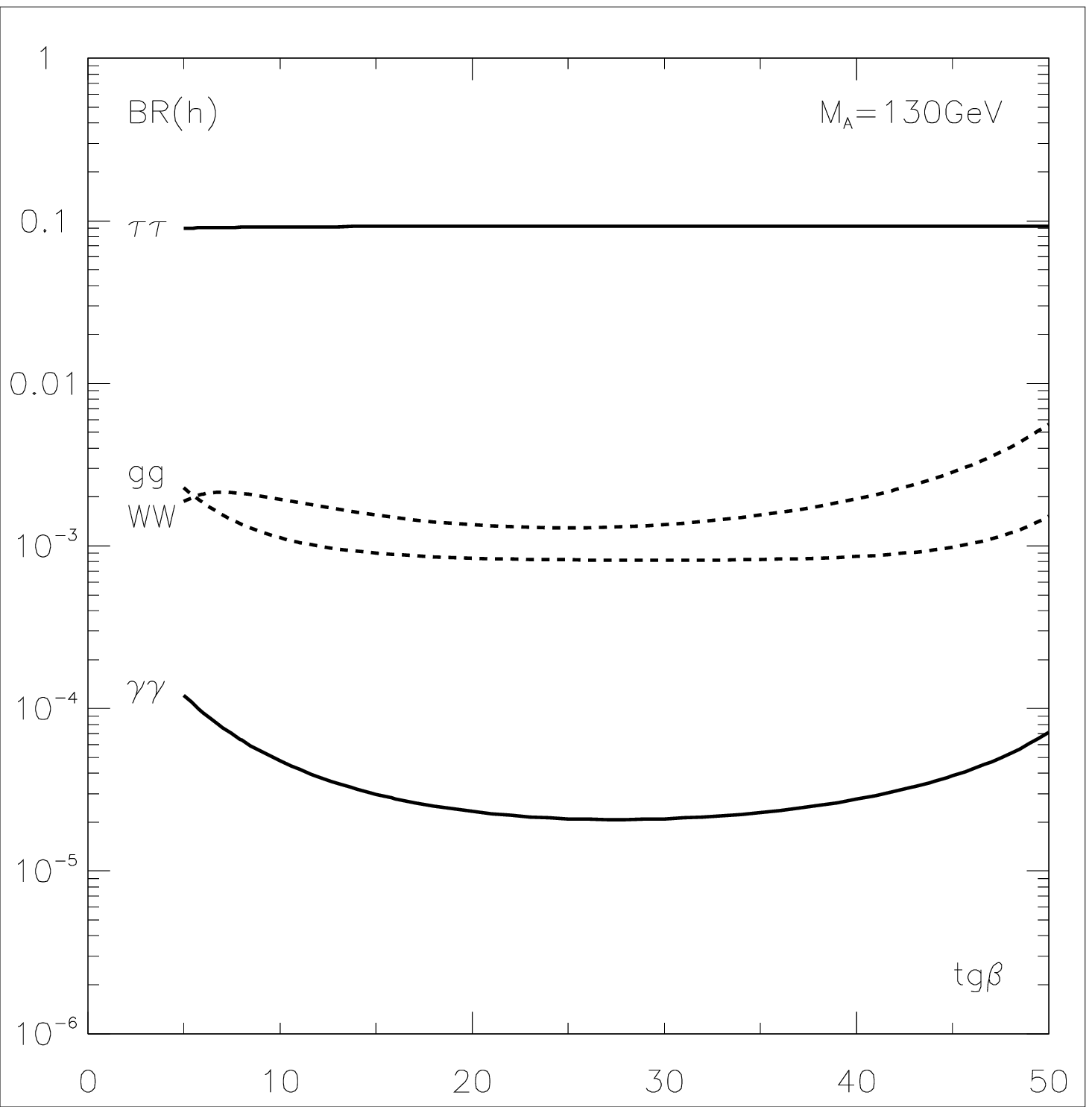,bbllx=3,bblly=3,bburx=410,bbury=420,width=7.cm,height=5cm,clip=}
\hspace{1cm}
\epsfig{figure=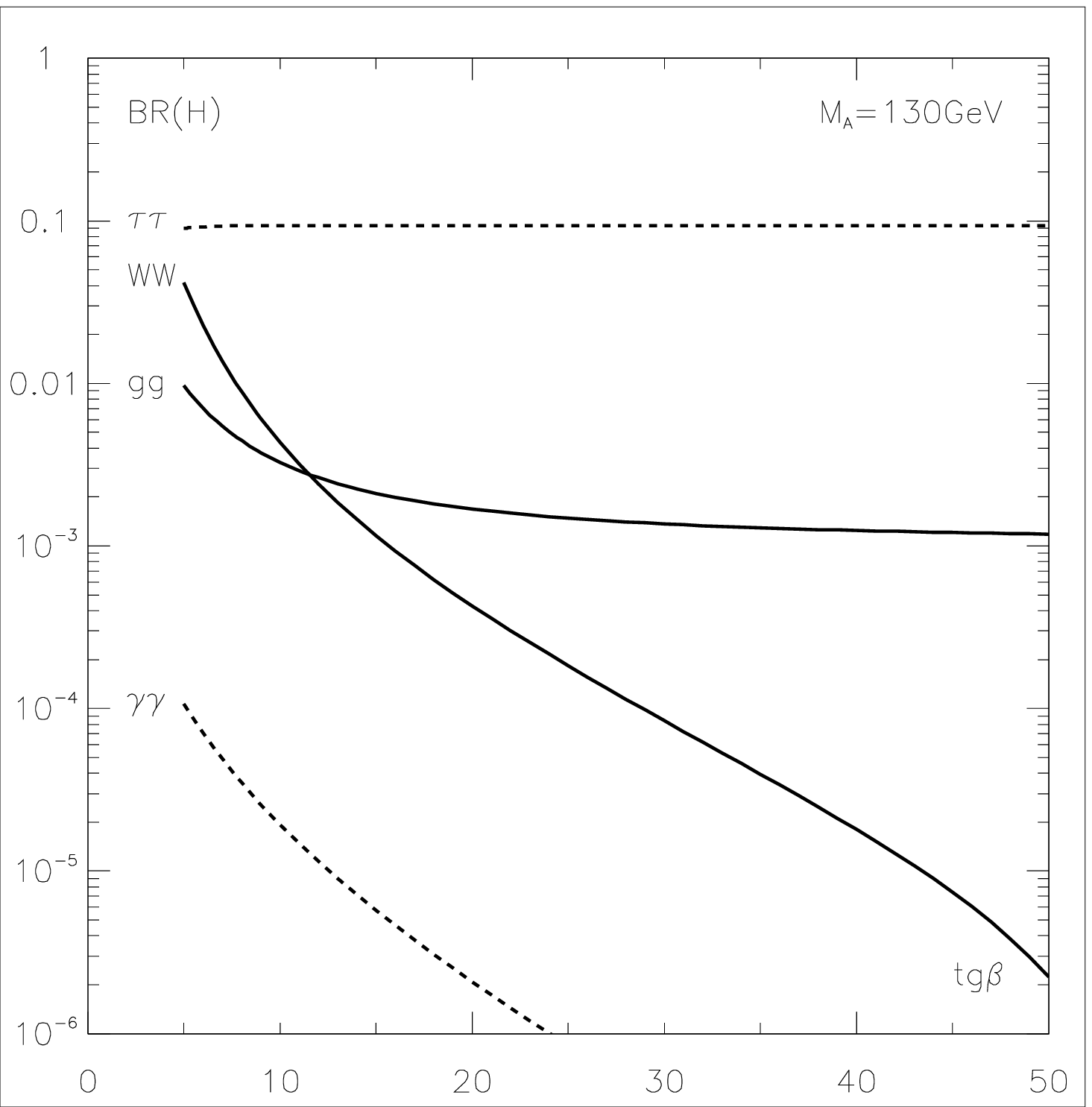,bbllx=3,bblly=3,bburx=410,bbury=420,width=7.cm,height=5cm,clip=}
\\[0.2cm]
\epsfig{figure=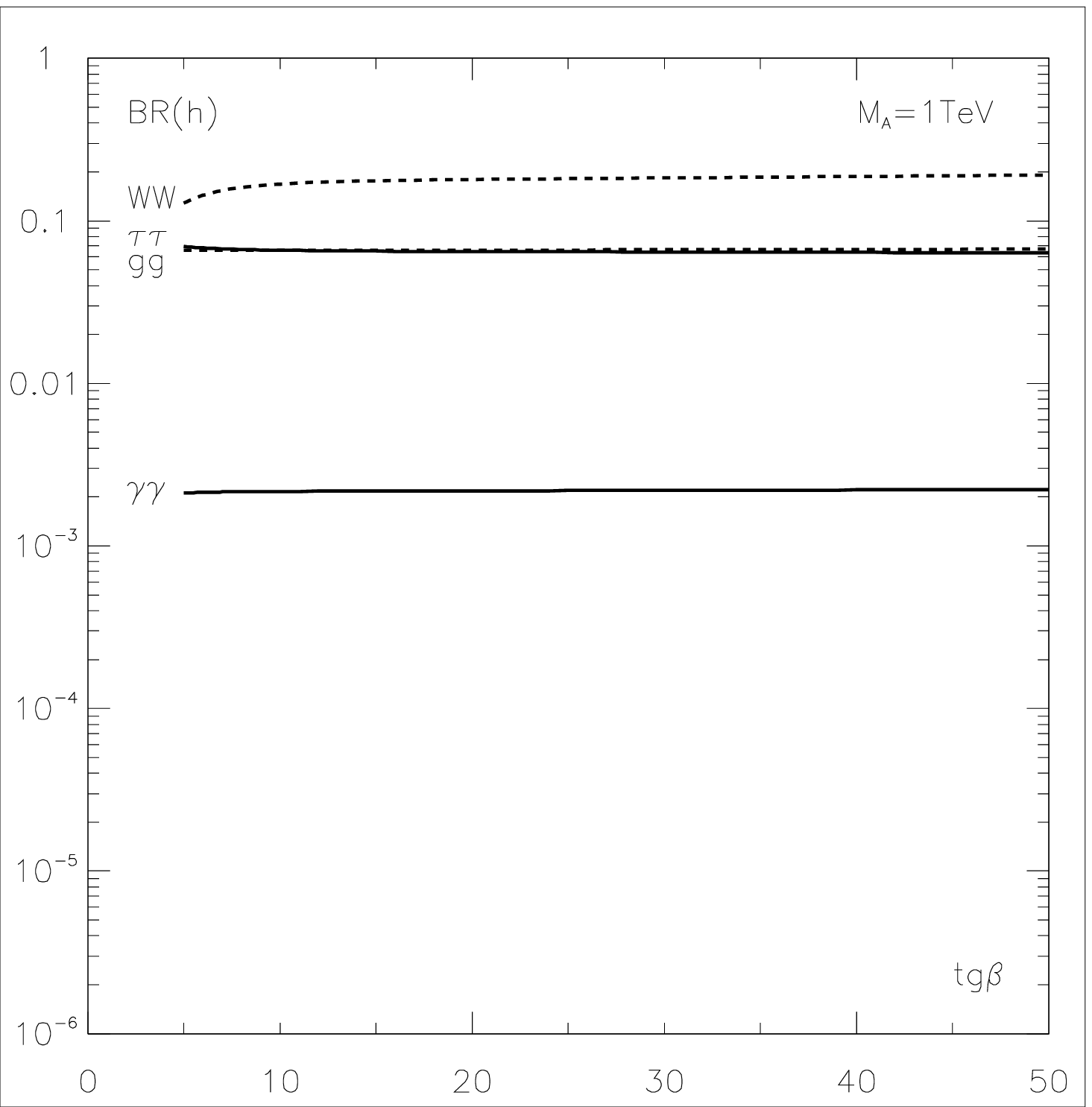,bbllx=3,bblly=3,bburx=410,bbury=420,width=7.cm,height=5cm,clip=}
\hspace{1cm}
\epsfig{figure=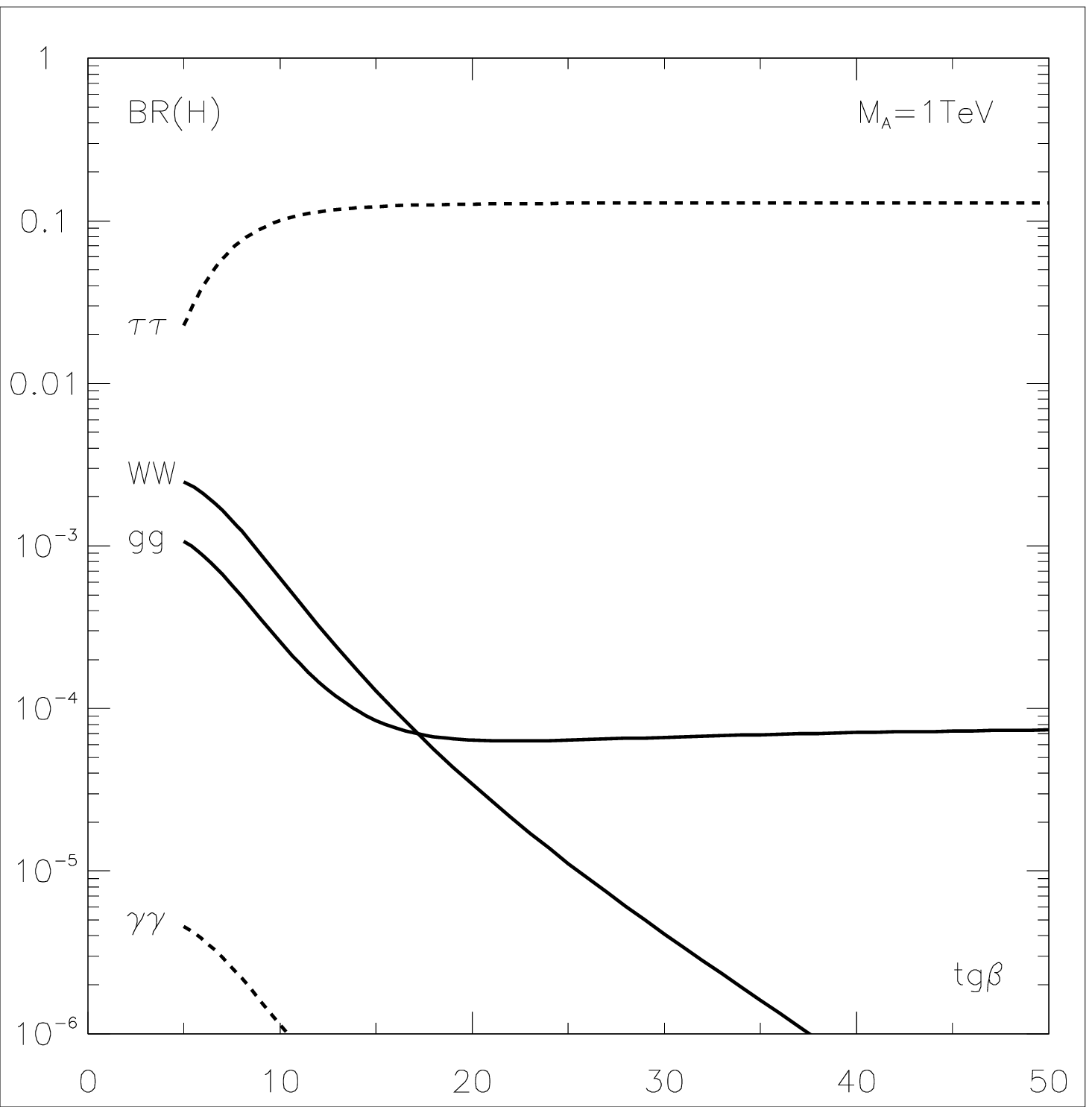,bbllx=3,bblly=3,bburx=410,bbury=420,width=7.cm,height=4.8cm,clip=}
\\
\caption{\it The branching ratios of the $h$ (left panel) and $H$ (right panel) 
bosons into $\gamma\gamma$, $gg$, $WW$, $\tau^+\tau^-$ final states as a 
function of $\tan\beta$ for $M_A=90, 110, 130$~GeV and $1$~TeV.}
\end{center}
\vspace*{-1.2cm}
\end{figure}

Indeed, as can be seen in Fig.~4, all decays of the $h$ boson other than
$b\bar{b}$ and $\tau^+ \tau^-$, are below the level of $10^{-3}$ for $\tb=10$
and 30. In particular, we have BR($h \to \gamma \gamma) \lsim 10^{-4}$ and
BR($h \to WW) \lsim 10^{-2}$ even for $M_h \sim 120$ GeV, i.e. at least one
order of magnitude smaller than in the case of the SM Higgs boson. This is
a consequence of the fact that the $h$ boson coupling to gauge bosons (and to
up--type fermions) does not reach the SM limit while the couplings to
$b$--quarks are still enhanced even for $M_h \sim M_C$; see Fig.~2
for the couplings. The $gg$ branching fraction is in turn almost constant,
BR($h \to gg) \sim 10^{-3}$, since the $h \to gg$ decay width is dominantly
mediated by the $b$--quark loop, but is suppressed by a factor of more than 20
compared to the SM case. \s

In the case of the heavier $H$ boson, the trend depends strongly on the value
of $M_A$. For $M_H \gsim 125$ GeV, the branching ratios into the $\gamma\gamma,
WW$ and $gg$ final states are much smaller than for the SM Higgs boson: for
$\tb$ values larger than $\sim 10$, they are at least two orders of magnitude
smaller. Near the critical mass $M_C$, the branching fractions of the $H$ boson
are in general closer to those of the  SM Higgs particle. This is
particularly true for $\tan \beta$ values smaller than $\sim 30$, where the
branching fractions into $WW, gg$ and $\gamma \gamma$ ($\tau^+\tau^-$) final
states are only slightly, less than a factor of two, smaller (larger) than for
the SM Higgs particle. However, for larger values of $\tb$ [not shown in
Fig.~4a] a new feature occurs: the $H$ boson coupling to down--type fermions
which is proportional to $\tb$ [in the approximation discussed in Section 2,
$g_{Hdd} \sim -f_1 \tb+ f_2$ for $M_A <M_C$] becomes strongly suppressed and 
at some stage [close to $\tb \sim f_2/f_1$ in our approximation] vanishes 
[note that this is not the case for up--type fermions and gauge bosons]. \s

All these features can be seen more explicitly in Figs.~5, where the branching 
fractions of  the $h$ and $H$ bosons into the $\tau^+\tau^-, WW, gg$ and 
$\gamma \gamma$ final states are displayed as functions of $\tb$ for three 
values of the $A$ mass in the intense--coupling regime $M_A=90,
110$ and 130 GeV as well as in the decoupling limit with $M_A=1$ TeV. In 
particular, one can see that the branching fractions BR($H \to \tau^+ \tau^+)$,
and hence BR$(H \to b\bar{b})$, drop at a given value of $\tb$ [around $\tb 
\sim 40$ for $M_A=90$ GeV and $\tb \sim 50$ for $M_A=110$ GeV], therefore 
enhancing the rates of the $WW, gg$ and $\gamma \gamma$ decay modes. This
is the only situation where the branching ratios for these decays are 
larger than in the case of the SM Higgs boson\footnote{This ``pathological"
situation, where the couplings to down--type fermions are strongly suppressed,
is known in the case of the the lightest $h$ boson and is discussed in several
places in the literature; see for instance Ref.~\cite{pathos}.}. \s
 
Finally, the total decay widths of the three neutral Higgs bosons are shown in
Fig.~6.  For $h$ and $H$ when they have masses close to $M_C$, the total decay
widths are comparable to those of the SM Higgs boson, i.e. they are rather
small. For $\tb \gsim 10$, the $A$ boson and one of the CP--even Higgs bosons,
i.e. $\Phi_A$, decay most of the time into $b$ quarks and $\tau$ leptons and
the total decay widths are to a good approximation given by: $\Gamma_{\rm tot}
\simeq G_FM_{A,\Phi_A}/(4 \sqrt{2} \pi) \times \tan^2\beta (m_\tau^2 +3
\overline{m}_b^2)$ and are therefore rather large. For $\tb=50$ and $M_A \sim
130$ GeV, the total widths of the three neutral Higgs bosons are $\Gamma_{\rm
tot}(A) \simeq 7.2$ GeV, $\Gamma_{\rm tot}(H) \simeq 6.96$ GeV (with $M_H=
127.6$ GeV) and $\Gamma_{\rm  tot}(h) \simeq 0.11$ GeV (with $M_h = 123.9$
GeV). 
 
\begin{figure}[htbp]
\vspace*{-4.7cm}
\hspace*{-2.3cm}
\mbox{\psfig{figure=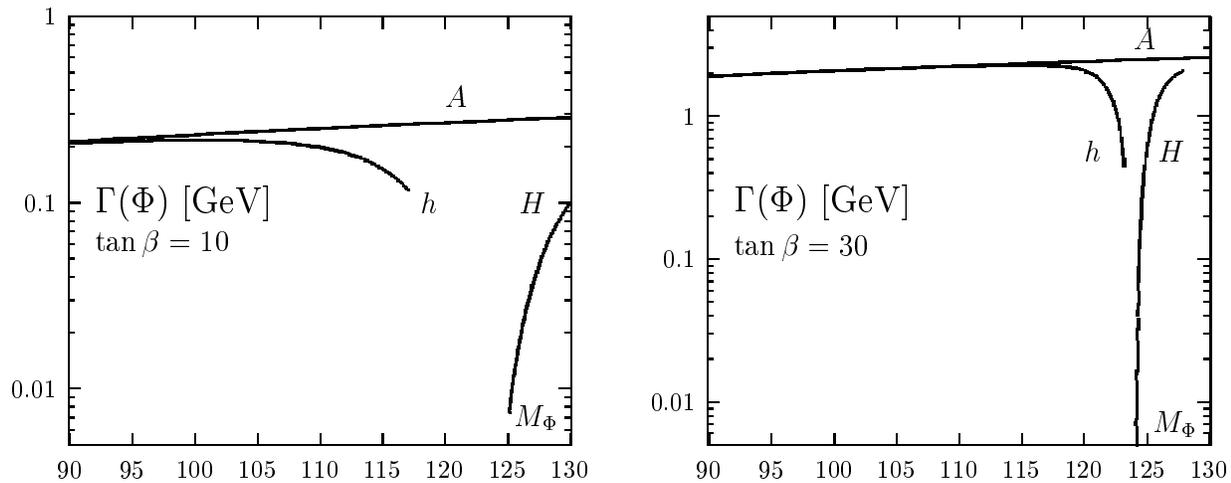,width=20cm}}
\vspace*{-17.7cm}
\label{figwid}
\caption[]{\it The total decay widths of the neutral Higgs bosons $h,H$ and $A$ 
as a function of their masses for two values of $tan \beta=10, 30$.}
\end{figure}


\subsection*{5. Production at Future Colliders}

In this section, we present the production cross sections of the neutral MSSM
Higgs bosons in the intense--coupling regime at future colliders. We will
consider the case of the hadron colliders, the Tevatron Run II and the LHC, a
future $e^+e^-$ linear collider with a c.m. energy of $\sqrt{s}=500$ GeV, in
both the $e^+e^-$ and $\gamma \gamma$ modes [with the photons generated by
back--scattering of laser light] and a $\mu^+ \mu^-$ collider. In all cases, we
will assume the values $\tan \beta=10$ and 30 and a heavy superparticle
spectrum with the soft SUSY--breaking scalar masses set to $M_S=1$ TeV and
trilinear couplings $A_t =2.6$ TeV, $A_b=1$ TeV [we recall the reader that
here, the higgsino and gaugino masses are set to $\mu=M_2=1$ TeV]. Therefore
only the standard processes, i.e.  single or pair production and production in
association with SM particles, will be considered.  We will not discuss
associated Higgs boson production with SUSY particles \cite{HSUSYprod}, Higgs
boson production from the (cascade) decays of SUSY particles \cite{HSUSYdecays}
or the effects of SUSY particle loops in the production processes
\cite{HSUSYloops}.

\subsubsection*{5.1 Production at hadron colliders}

The main production mechanisms of neutral MSSM Higgs bosons at hadron colliders
are the following processes [see Ref.~\cite{ppxsections} and for recent 
reviews Refs.~\cite{HHG,Higgsreview,HiggsTevatron} for instance]:  
\begin{eqnarray}
\begin{array}{lccl}
(a) & \ \ {\rm gluon-gluon~fusion} & \ \ gg  \ \ \ra & h,H,A \nonumber \\
(b) & \ \ {\rm association~with~}\bar{b}b & gg,q\bar{q}\ra & b\bar{b}+h,H,A
\nonumber \\
(c) & \ \ {\rm association~with~}\bar{t}t & gg,q\bar{q}\ra & t\bar{t}+h,H,A
\nonumber \\
(d) & \ \ WW/ZZ~{\rm fusion}       & \ \ VV \  \ra &  h,H \nonumber \\
(e) & \ \ {\rm association~with}~W/Z & \ \ q\bar{q} \ \ \ra & V + h,H \nonumber
\\ \nonumber 
(f) & \ \ {\rm associated~pair~production} & \ \ q\bar{q} \ \ \ra & A + h,H 
\nonumber 
\end{array}
\end{eqnarray}
The single production of the pseudoscalar $A$ boson in the weak boson fusion 
processes or in association with massive gauge bosons, as well as the pair 
production of two CP--even Higgs bosons, do not occur at leading order 
because of CP--invariance. Higgs boson production in association with other 
SM particles has too small cross sections. \s

All the production cross sections are shown in Figs.~7--12 for the LHC with
$\sqrt{s}=14$ TeV and for the upgraded Tevatron with $\sqrt{s}= 2$ TeV. These
cross sections have been obtained using the package {\tt CompHEP}
\cite{Comphep} [see Ref.~\cite{Semenov} for details of the MSSM implementation
in {\tt CompHEP} by means of the {\tt LanHEP} program], with the Higgs sector
adapted from {\tt HDECAY} for consistency with the discussion of the Higgs
properties given previously.  The parton densities are taken from the sets
CTEQ5L and CTEQ5M1 \cite{CTEQ} with the scale set at the Higgs boson mass. The
choice of scale and parton density might lead to a 30\% uncertainty in the
production cross sections. \s

The next--to--leading order QCD corrections are taken into account in the $gg$
fusion processes where they are large, leading to an increase of the production
cross sections by a factor of up to two \cite{QCDHiggs1}. For the other
processes, the QCD radiative corrections are relatively smaller
\cite{QCDHiggs2}: for the associated production with gauge bosons and
associated Higgs pair production, the corrections [which can be inferred from
the Drell--Yan $W/Z$ production] are at the level of $10\%$, while in the case
of the vector boson fusion processes, they are at the level of $30\%$. For the
associated production with heavy quarks, the NLO corrections are only available
in the case of the production of the SM Higgs boson with $t\bar{t}$ pairs
\cite{QCDHiggs3} where they alter the cross section by $\sim 20\%$ if the scale
is chosen properly. To include these corrections and to obtain the
corresponding $K$--factors, we have linked {\tt CompHEP} with the programs {\tt
HIGLU}, {\tt HPAIR}, {\tt HQQ}, {\tt V2HV} and {\tt VV2H} \cite{Spira} which
have all these corrections incorporated. \s

Let us now briefly summarize the main features of these processes, relying for 
the experimental searches on the analyses performed by various collaborations
\citer{Tevatron,Houches}. \bigskip

\noindent {\bf a) Gluon--gluon fusion} \s

This mechanism is mediated by top and bottom quark loops and is therefore
sensitive to both the $b\bar{b}$ and $t\bar{t}$--Higgs Yukawa couplings,
allowing for the measurement of these important parameters [in conjunction with 
other production processes with the same Higgs decay modes] if the uncertainties
from parton densities and higher order QCD corrections are properly
under control. In the large $\tb$ regime, the $b$--loop contributions are
dominant since they are enhanced by $\tan^2\beta$ factors. The top--quark loop
gives a relatively important contribution only in the case of the SM--like
CP--even Higgs boson $\Phi_H$ i.e. $h$ or $H$ for $M_A \geq M_C$ or $M_A \leq
M_C$, respectively. The cross sections are not the same for the pseudoscalar
$A$ boson and the CP--even Higgs boson $\Phi_A$ even for large $\tan\beta$
where the Higgs--$\bar{b}b$ Yukawa couplings are similar: this is due to the
appearance  of different loop induced Higgs--$gg$ form factors for CP--even and
CP--odd particles. The cross sections for the production of the $A$ and $h$
boson decrease quickly for increasing $M_A$ [a factor $\sim 5$ for $M_A$
varying from 90 to 130 GeV]; the cross section for $H$ stays almost constant
since $M_H$ does not vary too much in this range.  \s

At the LHC, for $\tb \sim 10$, the cross sections for the $A$ and $h$ bosons
vary from ${\cal O}(150)$ pb for a low mass $A$ boson, $M_A \sim 90$ GeV, to
${\cal O}(30)$ pb for $M_A \sim 130$ GeV; the cross section for $H$ production
stays at the level of 30 to 50 pb in this mass range. This leads to a huge
number of events, ${\cal O}(10^7)$ per year for an integrated  luminosity of
$\int {\cal L} \sim 100$ fb$^{-1}$ as expected in the high--luminosity option.
For $\tan \beta\sim 30$, the cross sections are one order of magnitude larger
for the $A$ and $\Phi_A$ bosons [while it stays roughly constant for the
$\Phi_H$ boson] as a result of the increase of the Higgs--$b\bar{b}$ Yukawa
coupling.  At the Tevatron, the trend is practically the same as at the LHC
[note that here, the initiated $q\bar{q}$ process plays a more important role
than at high energies due to the reduced gluon luminosity] except that the
cross section is $\sim 30$ times smaller.  Nevertheless, with the integrated 
luminosity of $\int {\cal L} \sim 30$ fb$^{-1}$ expected at the end of the
Run II option, this would lead to at least ${\cal O}(10^4)$ Higgs bosons in 
each of the $gg \to A, \Phi_A$ processes. \s

\begin{figure}[htbp]
\vspace*{-.5cm}
\begin{center}
\epsfig{figure=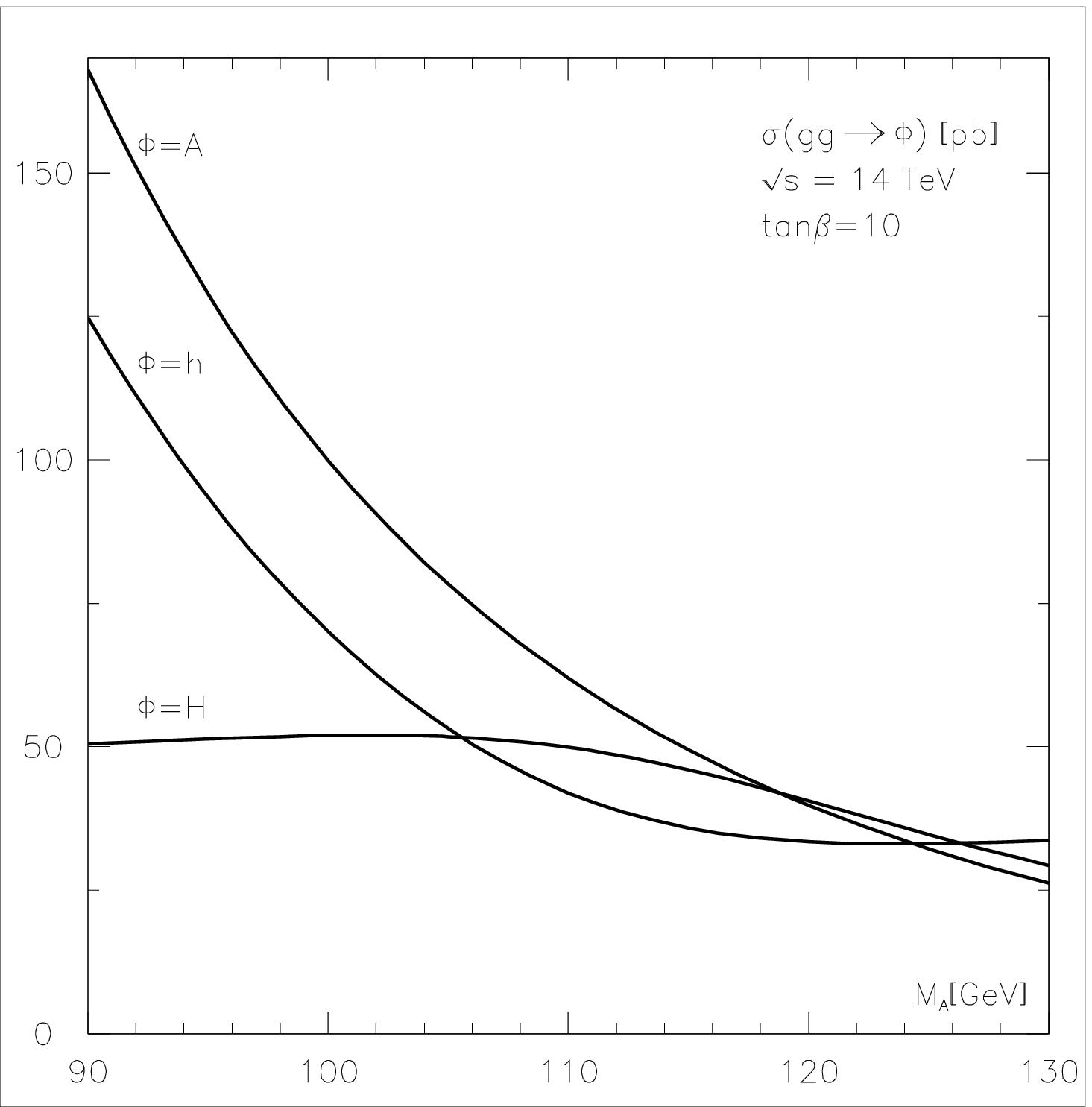,bbllx=3,bblly=3,bburx=418,bbury=420,height=6.5cm,width=7.5cm,clip=}\hspace{1cm}
\epsfig{figure=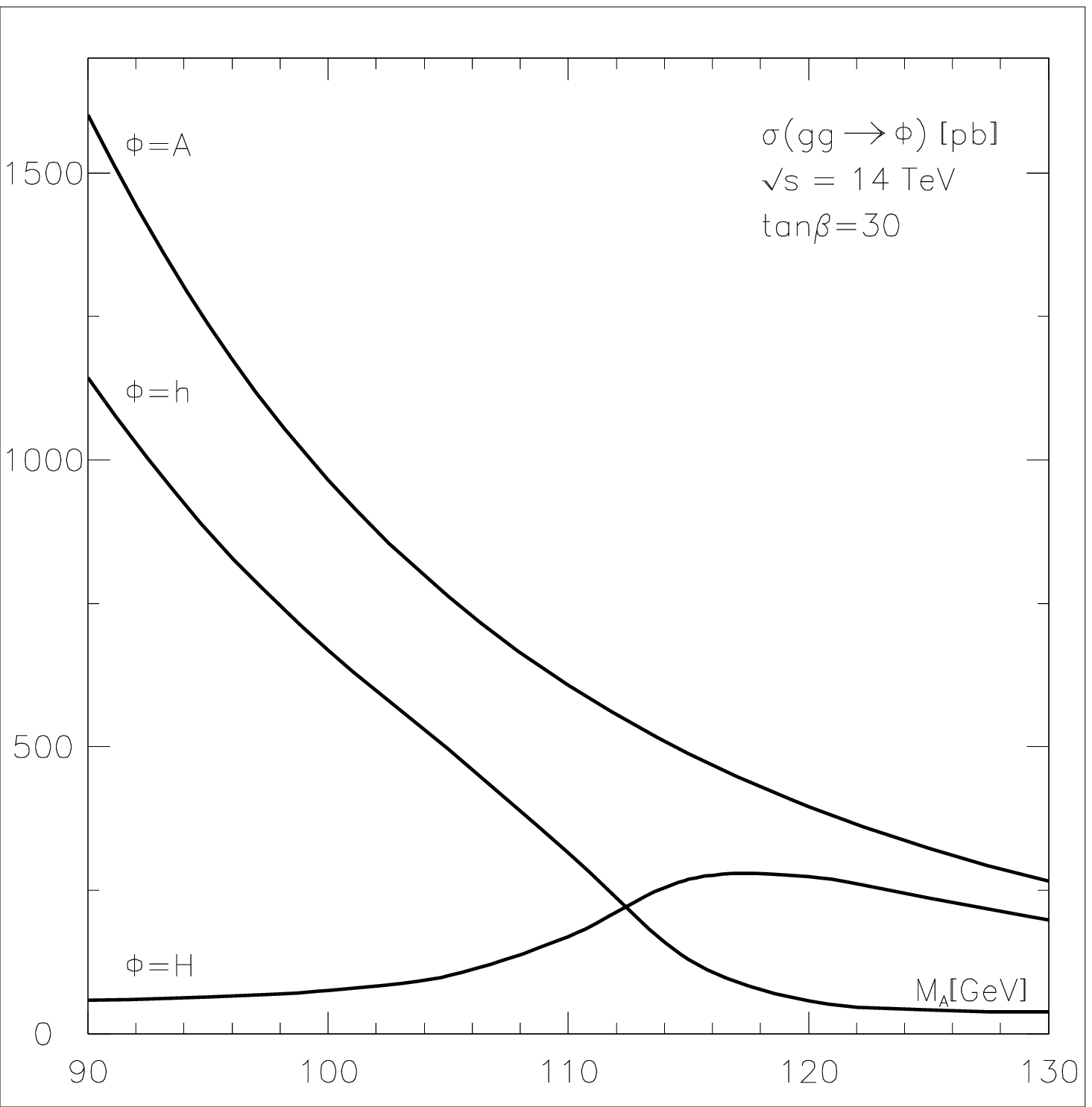,bbllx=3,bblly=3,bburx=418,bbury=420,height=6.5cm,width=7.5cm,clip=}\\[.1cm]
\epsfig{figure=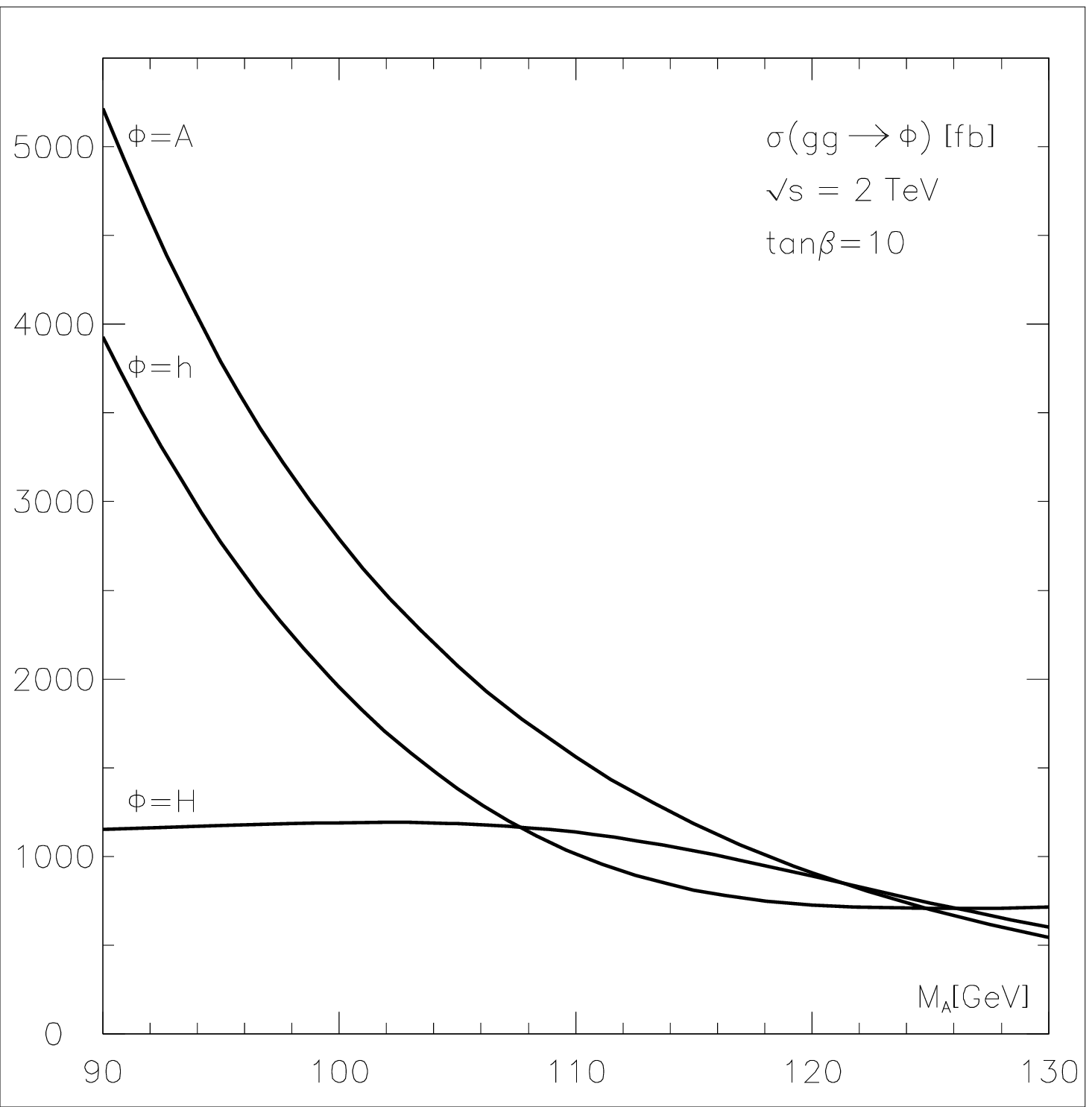,bbllx=3,bblly=3,bburx=418,bbury=420,height=6.5cm,width=7.5cm,clip=}\hspace{1cm}
\epsfig{figure=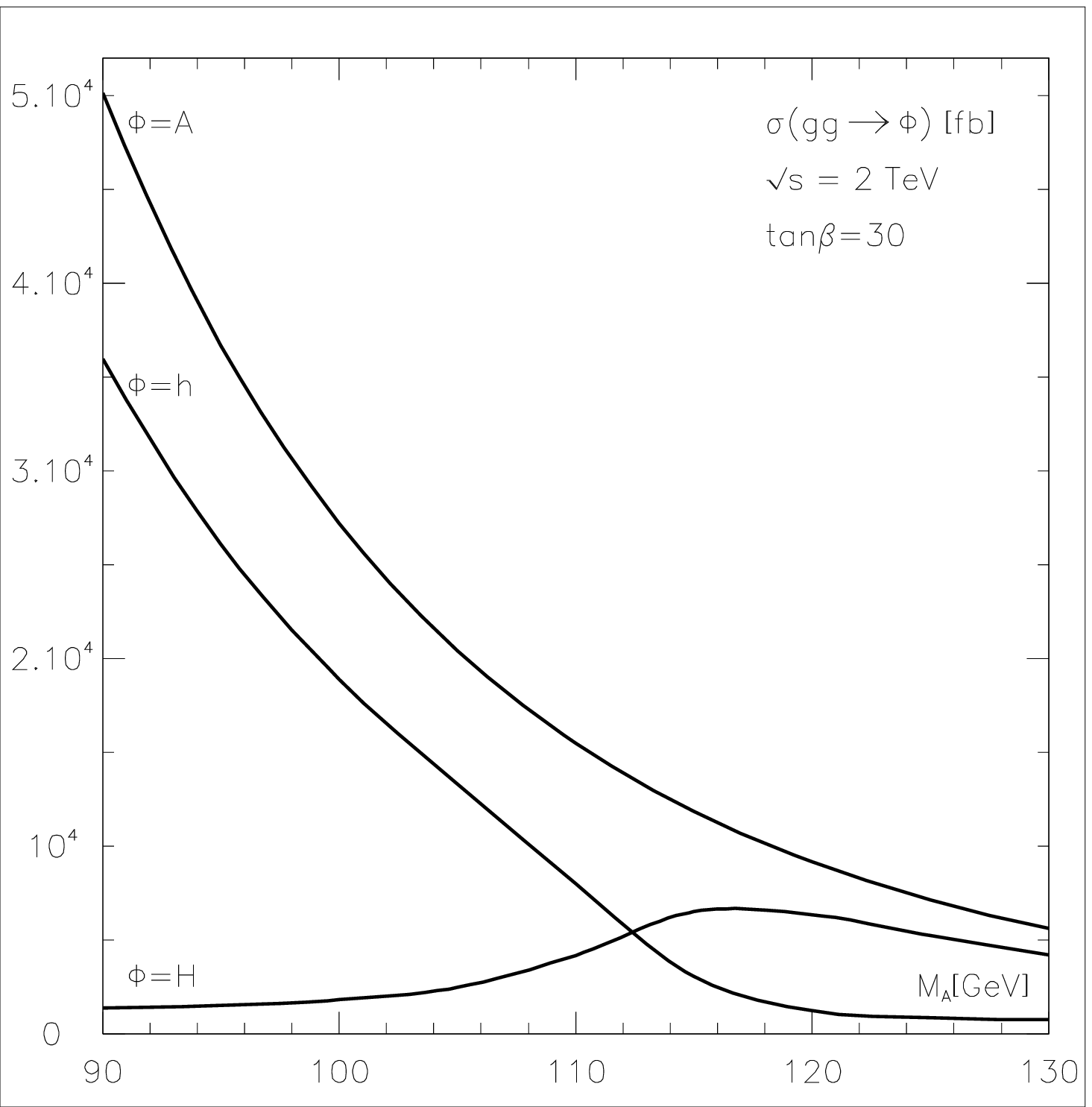,bbllx=3,bblly=3,bburx=418,bbury=420,height=6.5cm,width=7.5cm,clip=}\\[-.1cm]
\caption{\it Production cross sections at the LHC (upper panel) and at the 
Tevatron (lower panel) for the gluon-gluon fusion mechanism as functions of 
$M_A$ for $\tan \beta=10,30$. }
\end{center}
\vspace*{-.5cm}
\end{figure}

Let us now discuss the signatures which would allow for the detection of these
Higgs particles. In all cases, the $b\bar{b}$ final states cannot be used
because of the huge QCD 2--jet background and the lack of a clean leptonic
trigger. At the LHC, the final states which can be probed for the SM--like
Higgs boson $\Phi_H$ would be in principle the $\Phi_H \to \gamma \gamma$ final
state for which the branching ratio can be of order $10^{-3}$. However, this
occurs only if $\Phi_H$ is pure SM--like which, in the scenario discussed in
Fig.~5,  occurs only for the $H$ boson for a rather light $A$ boson and
moderate to large $\tan \beta$ values. This is exemplified in Table 6 where are
shown the cross sections for $h,H$ and $A$ boson production in the $gg$ fusion
mechanism times the Higgs decay branching ratio  into $\gamma \gamma$ final
states, relative to the SM case [with the SM Higgs boson having the same mass
as the $h,H$ or $A$ boson]. In the case of $H$, only for large $\tb$ and
moderate $M_A$ values,  the cross section times branching ratio exceeds the
SM value [with a factor up to 3]. For the pseudoscalar $A$ boson, the very 
small $\gamma \gamma$ branching ratio is not compensated by the large cross 
section.  Therefore, the ratio is much smaller than unity. In the case
of the lighter $h$ boson, we are not yet in the decoupling limit even for
$M_A\sim 130$ GeV and $\tan \beta \sim 50$, and the cross section times
branching ratio $\sigma (gg \to h) \times {\rm BR}(h \to \gamma \gamma$) is at
least two orders of magnitude smaller than in the SM. \s

Thus, there are situations where $\sigma (\Phi) \times {\rm BR}(\Phi \to
\gamma \gamma$) is rather small for the three Higgs bosons $h,H$ and $A$
compared to the SM case, making this channel more difficult to use. In
addition, for the $A$ and $\Phi_A$ particles, the total decay widths are much
larger than in the SM, leading to broader signals and therefore a much larger
$\gamma \gamma$ background. The $\Phi_H \to WW$ and $ZZ$ final states
would be also very difficult to use since in our case $M_{\Phi_H}\sim 130$ GeV 
and the branching ratios are too small even for SM $\Phi_H VV$ couplings.  \s

At the Tevatron, although very difficult, only the $\Phi_H \to \tau^+ \tau^-$
channel might be used as preliminary studies seem to indicate \cite{Hildreth}. 
The branching ratios for the $h,H$ and $A$ boson decays into  $\tau^+ \tau^-$
final states are always of the order of  $10\%$ except in the ``pathological"
situations discussed in section 4, where the $H\tau^+ \tau^-$ coupling is
suppressed.  For the $A$ and $\Phi_A$ bosons, the channels $A, \Phi_A \to
\tau^+ \tau^-$ can be used at the LHC in the high $\tan\beta$ regime, $\tan
\beta\gsim 10$, where the cross sections are large enough. At the Tevatron,
because of the smaller cross sections, this channel is expected to be more
difficult, although the analyses mentioned previously give some hope. \s

\begin{table}[htbp]
\renewcommand{\arraystretch}{1.65}
\begin{center}
\vspace*{-2mm}
\begin{tabular}{|c||c|c||c|c||c|c|} \hline
$\ \tb \ $ & $M_A$ &  $\frac{\sigma (A) \times {\rm
BR}(\gamma \gamma)|_{\rm MSSM} } {\sigma (H^0) \times {\rm BR}(\gamma
\gamma)|_{\rm SM} } $   & $M_h$ & $\frac{\sigma (h) \times {\rm
BR}(\gamma \gamma)|_{\rm MSSM} } {\sigma(H^0) \times {\rm BR}(\gamma
\gamma)|_{\rm SM} } $ & $M_H$ & $\frac{\sigma(H) \times {\rm BR}
(\gamma \gamma) |_{\rm MSSM} } {\sigma (H^0) \times {\rm BR}(\gamma \gamma)
|_{\rm SM} } $ \\ \hline
   & 90  & $1.5\cdot 10^{-3}$ & 84.9 & $7.5\cdot 10^{-3}   $ & 125.1 & $0.6$ \\ 
10 & 110 & $3.7\cdot 10^{-4}$ & 103.9 & $5.6\cdot 10^{-3} $ & 126.8 & $0.15$ \\ 
   & 130 & $1.4\cdot 10^{-4}$& 117.1 & $1.2\cdot 10^{-2} $ & 134.0 & $1.4 \cdot 10^{-2}$
\\ \hline \hline
   & 90  &$1.8\cdot 10^{-2}$ & 85.9 & $2.2\cdot 10^{-2}   $ & 124.0 & $1.9$ \\ 
30 & 110 &$5.0\cdot 10^{-3}$ & 106.6 & $6.7\cdot 10^{-3} $ & 124.2 & $0.36$ \\ 
   & 130 &$2.1\cdot 10^{-3}$ & 123.2 & $1.2\cdot 10^{-2} $ & 128.0 & $1.8\cdot 10^{-3}$
\\ \hline \hline
   & 90  & $5.2\cdot 10^{-2}$ & 85.7 & $5.6\cdot 10^{-2}   $ & 124.3 & $2.0$ \\ 
50 & 110 & $1.4\cdot 10^{-2}$ & 106.5 & $1.5\cdot 10^{-2} $ & 124.4 & $3.2$ \\ 
   & 130 & $6.2\cdot 10^{-3}$ & 124.3 & $2.3\cdot 10^{-2} $ & 127.1 & $4.3 \cdot 10^{-3}$
\\ \hline \hline
\end{tabular}
\end{center}
\vspace*{-1mm}
\caption[]{\it The ratios $\sigma(gg \to \Phi) \times {\rm BR} (\Phi \to \gamma 
\gamma)$ in the MSSM with $\Phi=h,H$ and $A$ relative to the SM case for a 
SM Higgs boson with the same mass, for three values of $\tb= 10, 30$ and 50 
and three values of the pseudoscalar boson mass $M_A=90, 110$ and $130$ GeV.}
\end{table}

\noindent {\bf b) Associated production with bottom quarks} \s

In these processes, the production cross sections for the $A$ and $\Phi_A$
bosons are also strongly enhanced by $\tan^2\beta$ factors. The rates are
similar to those in the $gg$ fusion processes, ${\cal O} (100)$ pb at the LHC and
${\cal O} (1)$ pb at the Tevatron for $M_A\sim 90$--100 GeV and $\tb=10$, and 
are one order of magnitude larger for $\tb=30$. They decrease more slowly with
increasing Higgs boson mass than in the case of $gg$ fusion and the cross
sections are closer in magnitude for the $A$ and $\Phi_A$ bosons [in fact, they
must be equal for equal Yukawa couplings when the final state $b$--quark mass
is neglected, i.e. in the ``chiral" limit].  In turn the cross section for
associated production of the SM--like $\Phi_H$ boson  is much smaller due to a
tiny (not enough enhanced) Yukawa coupling. \s 

\begin{figure}[htbp]
\vspace*{-.6cm}
\begin{center}
\epsfig{figure=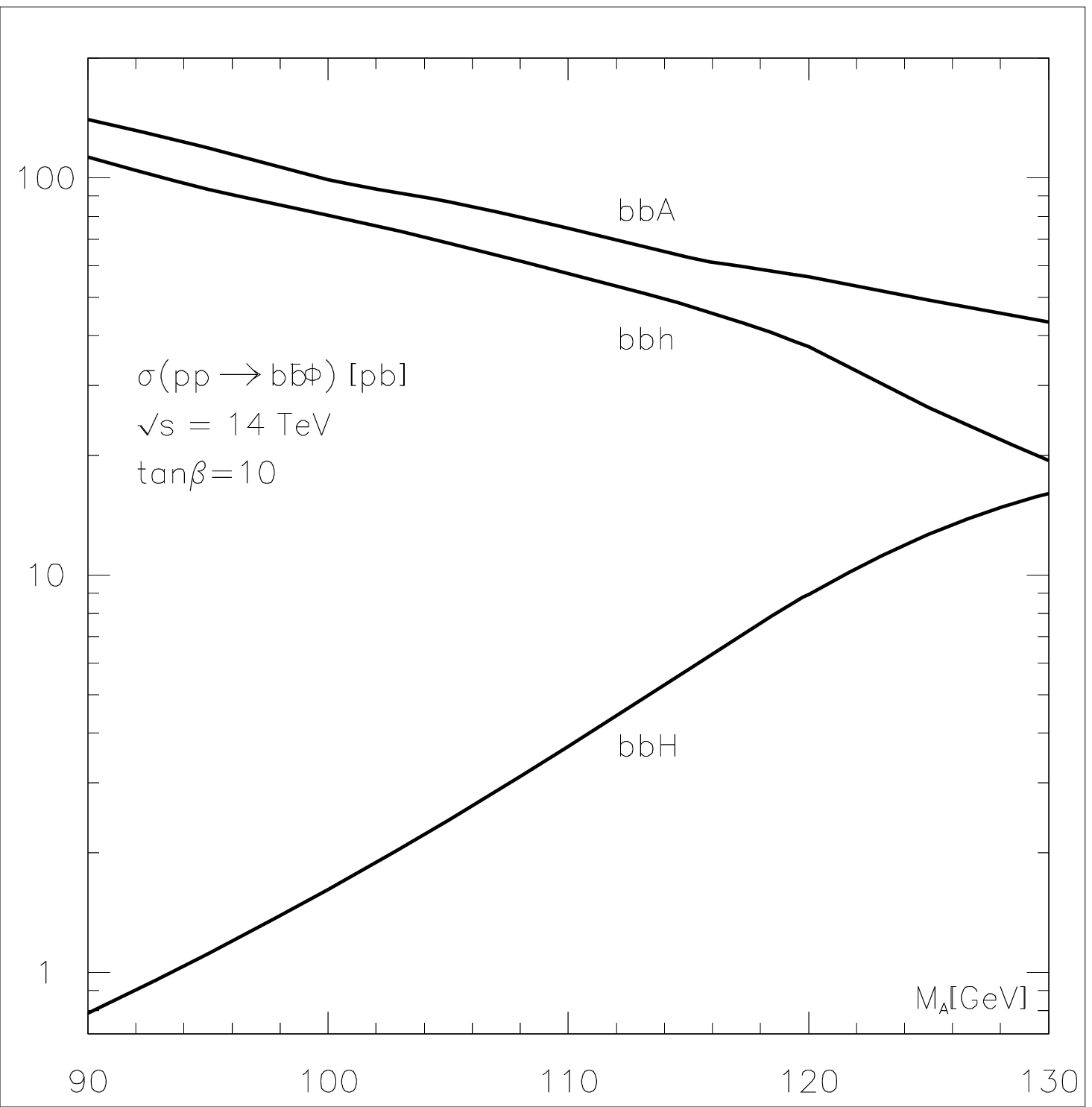,bbllx=3,bblly=3,bburx=418,bbury=420,height=6.5cm,width=7.5cm,clip=}\hspace{1cm}
\epsfig{figure=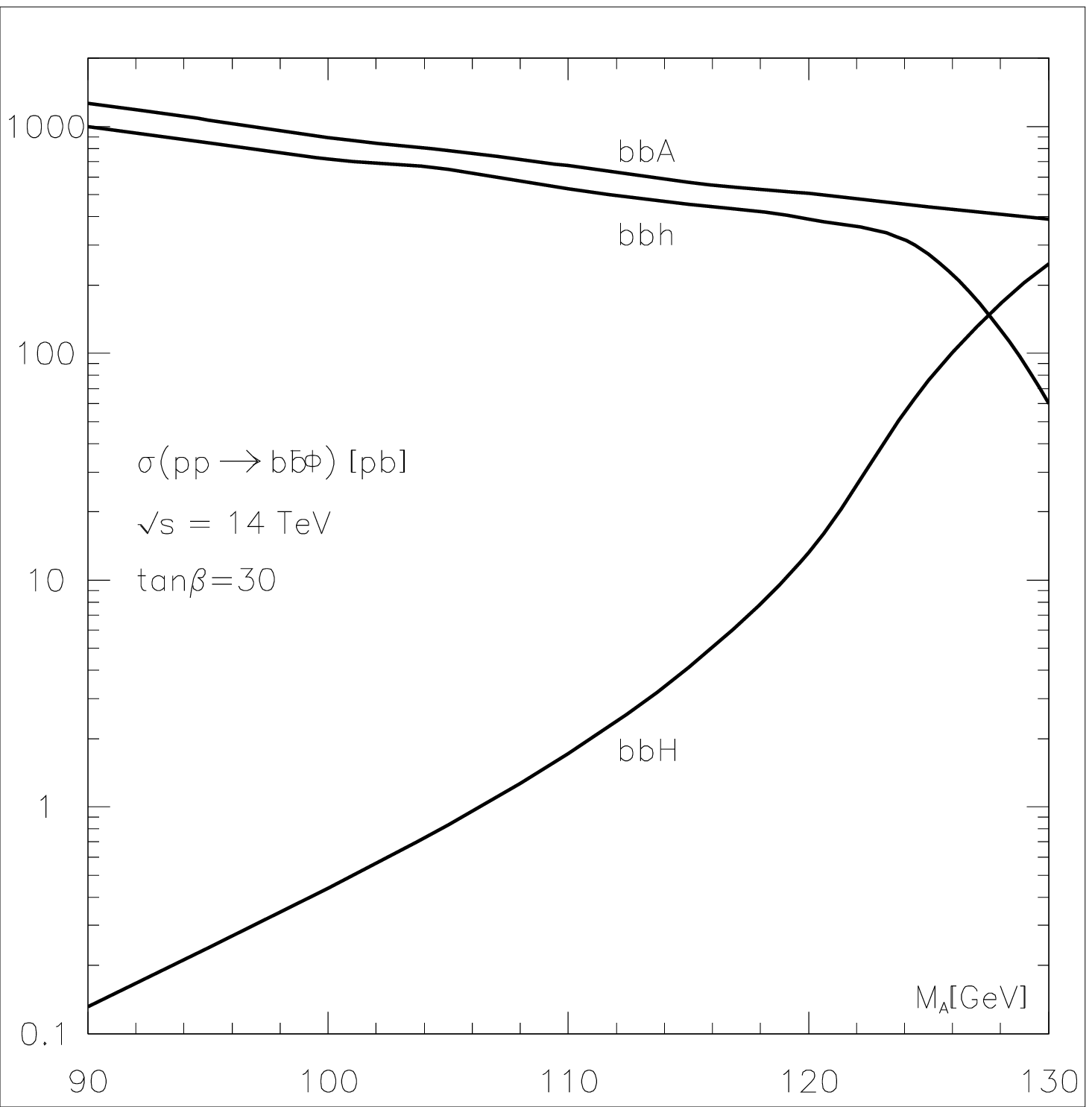,bbllx=3,bblly=3,bburx=418,bbury=420,height=6.5cm,width=7.5cm,clip=}\\[.1cm]
\epsfig{figure=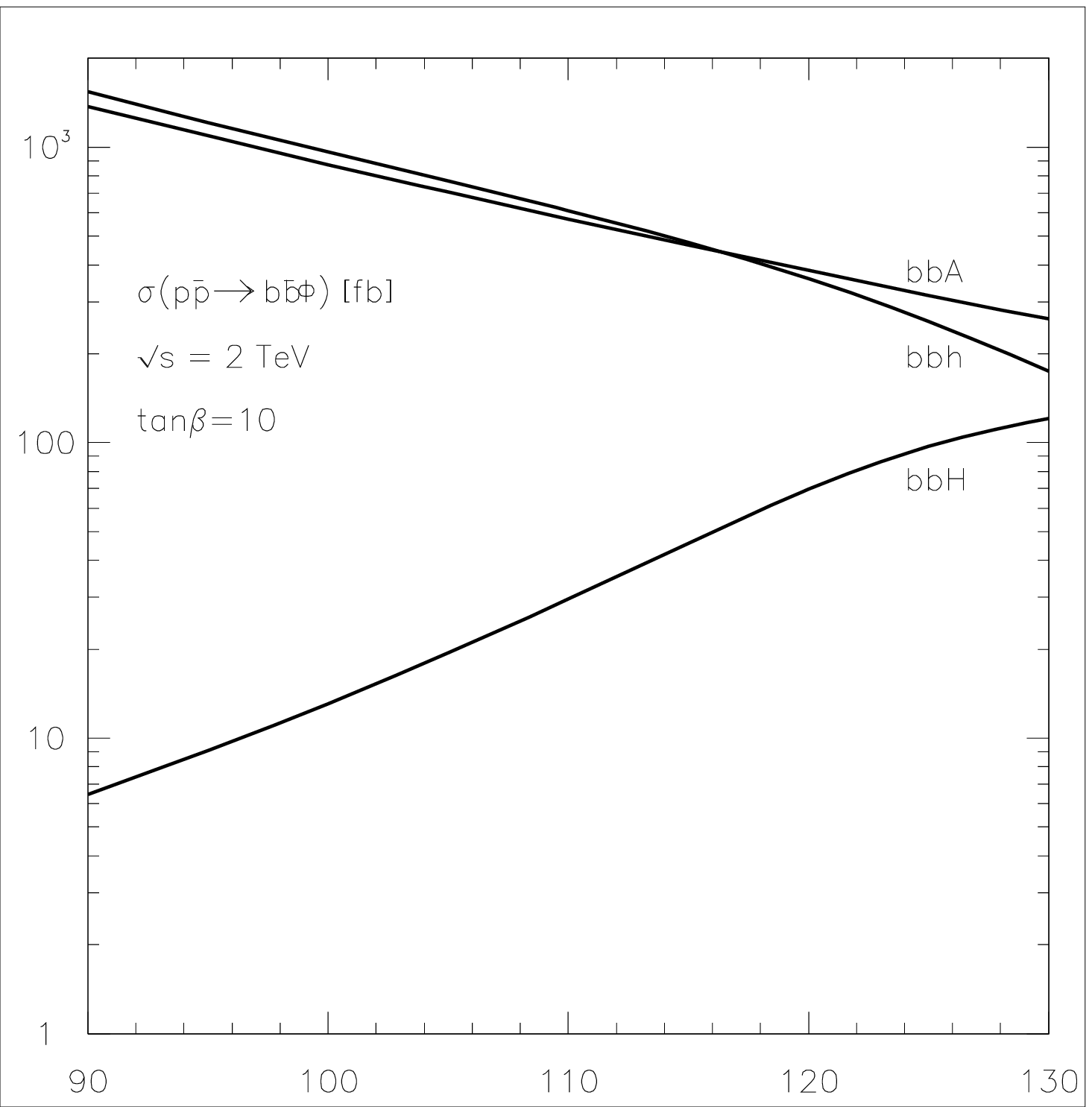,bbllx=3,bblly=3,bburx=418,bbury=420,height=6.5cm,width=7.5cm,clip=}\hspace{1cm}
\epsfig{figure=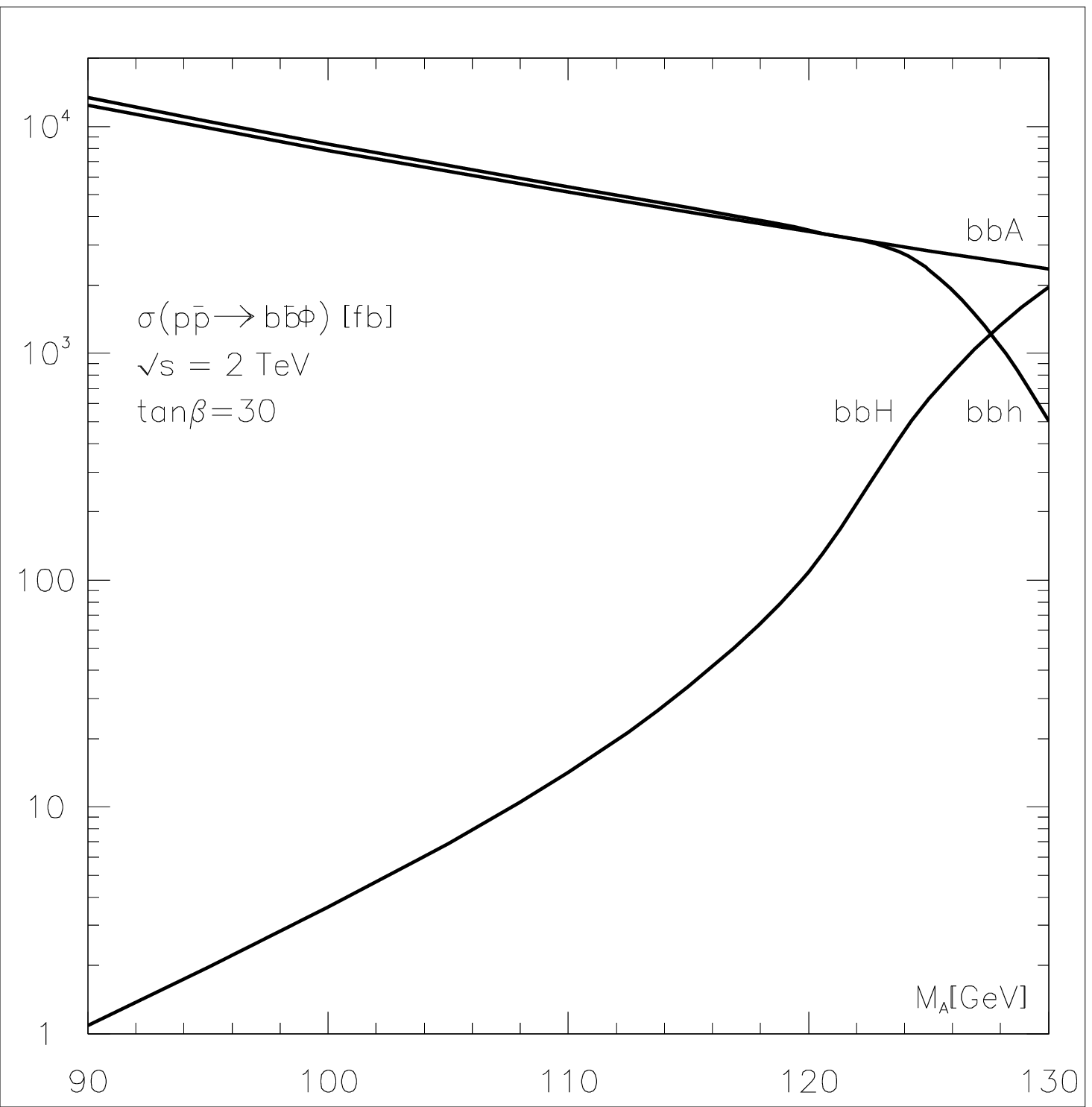,bbllx=3,bblly=3,bburx=418,bbury=420,height=6.5cm,width=7.5cm,clip=}\\[-.2cm]
\caption{\it Production cross sections at the LHC (upper panel) and at the 
Tevatron (lower panel) for Higgs bosons in association with $b$ quarks
as functions of  $M_A$ for $\tan \beta=10,30$. }
\end{center}
\vspace*{-.7cm}
\end{figure}

Because of the small number of events, the detection of the $\Phi_H$ boson in
this process is practically hopeless. In turn, the detection of the $A$ and
$\Phi_A$ bosons is more promising than in the $gg$ fusion process since we do
not have to rely on decays into photons or massive gauge bosons which here, are
absent or strongly suppressed.  Indeed, the additional two $b$--quarks in the
final state, which can be rather efficiently tagged using micro--vertex
detectors, will reduce dramatically the QCD backgrounds [especially since the
production cross sections can be very large] to the level where the final
states $A, \Phi_A \to \tau^+ \tau^-$ at the LHC and $A, \Phi_A \to \bar{b}b$ at
the Tevatron could be easily detectable for $\tan \beta \gsim 10$. \bigskip

\noindent {\bf c) Associated production with top quarks}

\begin{figure}[htbp]
\vspace*{-.5cm}
\begin{center}
\epsfig{figure=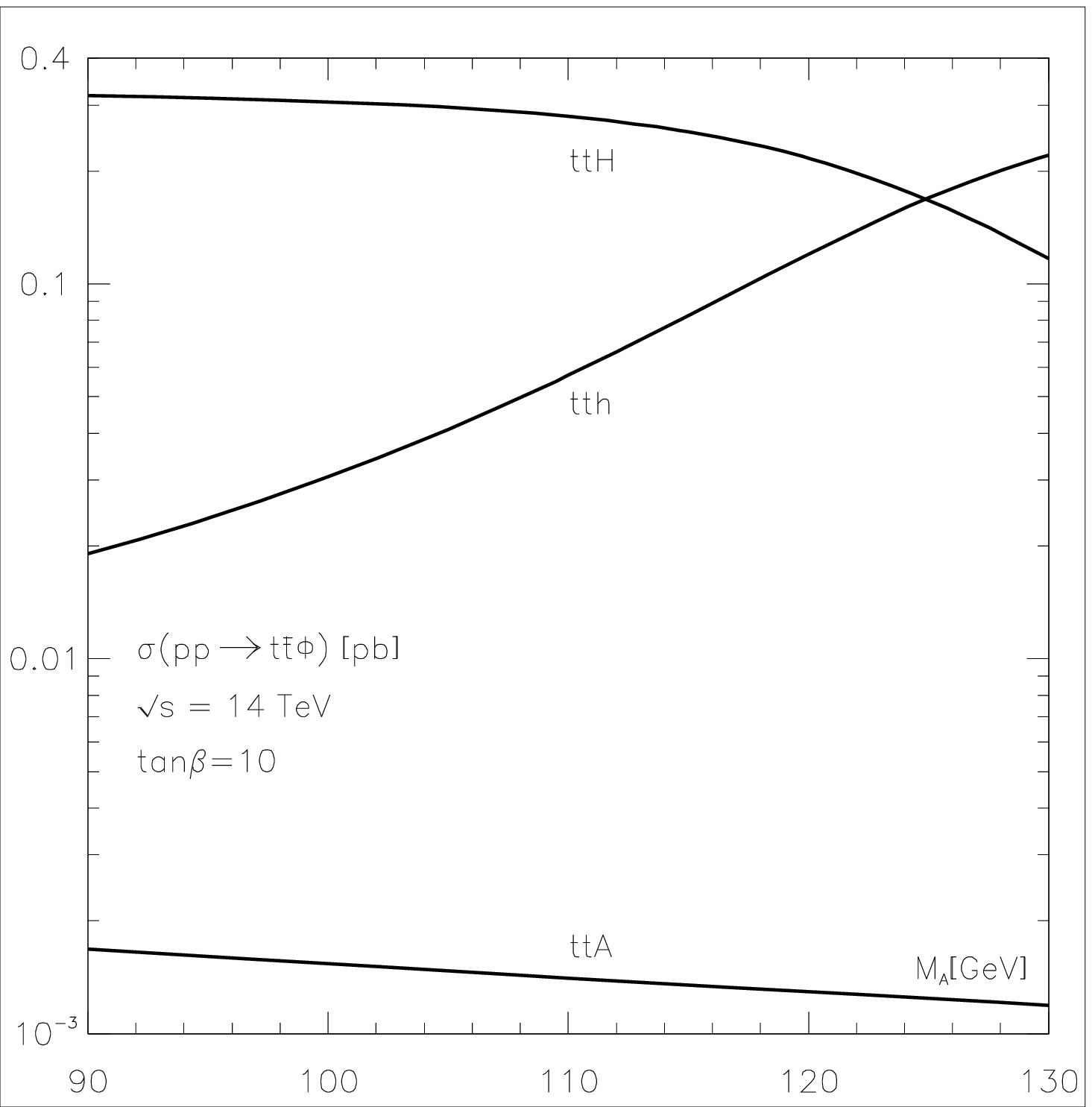,bbllx=3,bblly=3,bburx=418,bbury=420,height=6.5cm,width=7.5cm,clip=}\hspace{1cm}
\epsfig{figure=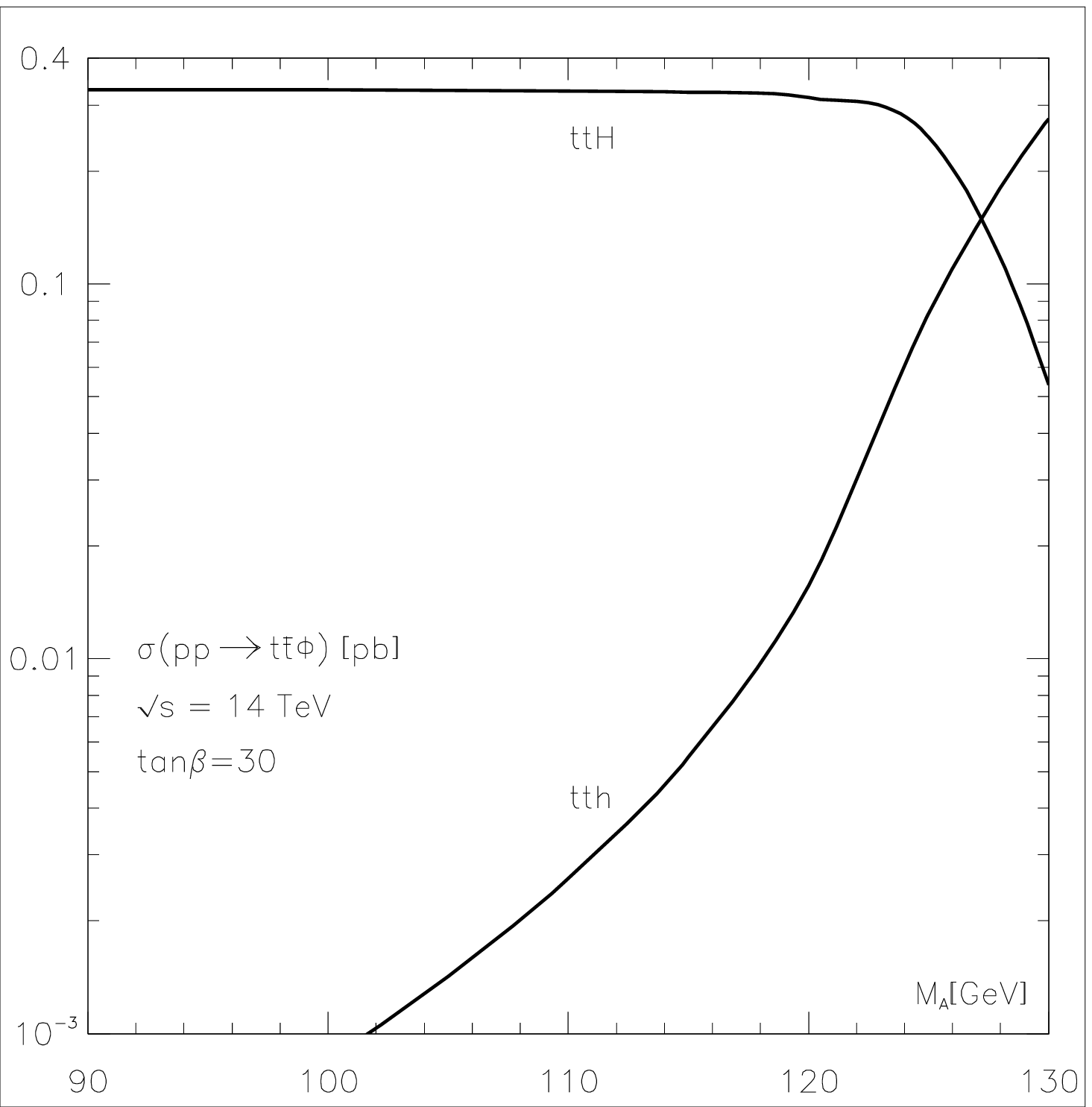,bbllx=3,bblly=3,bburx=418,bbury=420,height=6.5cm,width=7.5cm,clip=}\\[.1cm]
\epsfig{figure=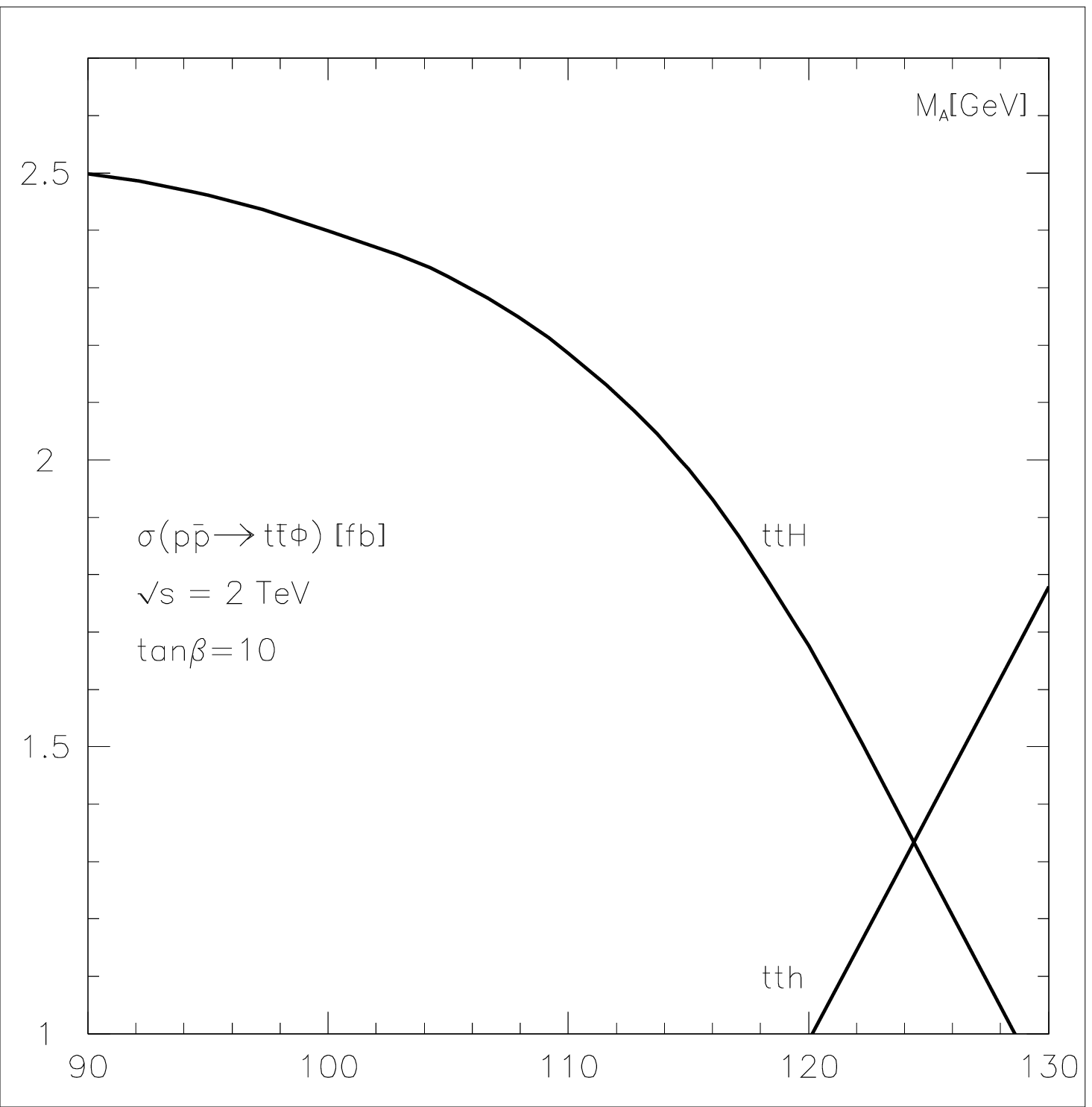,bbllx=3,bblly=3,bburx=418,bbury=420,height=6.5cm,width=7.5cm,clip=}\hspace{1cm}
\epsfig{figure=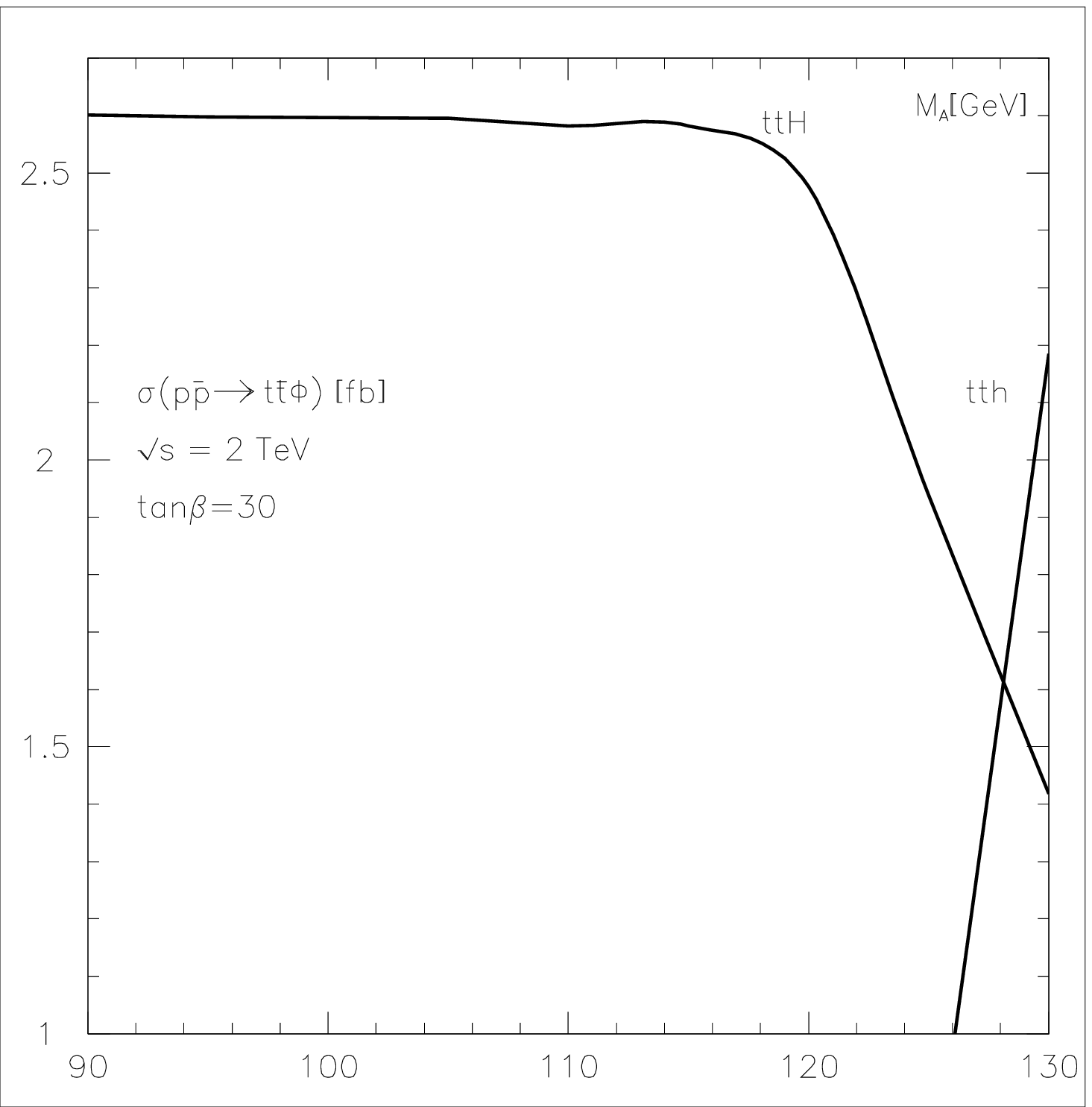,bbllx=3,bblly=3,bburx=418,bbury=420,height=6.5cm,width=7.5cm,clip=}\\[-.1cm]
\caption{\it Production cross sections at the LHC (upper panel) and at the 
Tevatron (lower panel) for Higgs bosons in association with top quarks
as functions of $M_A$ for $\tan \beta=10,30$. }
\end{center}
\vspace*{-.5cm}
\end{figure}

Here, the production cross sections are suppressed by the smaller phase space
[in particular at the Tevatron] compared to the previous case and by the fact
that the $t\bar{t}$--Higgs Yukawa coupling is not enhanced compared to the SM
case. To the contrary, the cross sections are strongly suppressed by $\tan^2
\beta$ factors for the production of the $A$ and $\Phi_A$ bosons. A reasonably
large cross section is obtained for the production of the $\Phi_H$ boson,
reaching the level of $\sim 0.3$ pb at the LHC and 2.5 fb at the Tevatron. In a
large part of the parameter space we are concerned with, i.e. for 90 $\lsim M_A
\lsim 130$ GeV and $\tan \beta \gsim 10$ [even around the turning point $ M_A
\sim 125$ GeV], the sum of the cross sections for $h$ and $H$ production  is
comparable to the one in the SM, since the sum of the squared couplings 
$g_{ht \bar{t}}^2+g_{Ht \bar{t}}^2 =1/\sin^2\beta \sim 1$ for $\tb \gsim 10$ 
and the Higgs masses are comparable. \s

The detection of the $\Phi_H$ boson in this process is in principle possible
either through its $\gamma \gamma$ or $\bar{b}b$ decay modes. In the $\gamma
\gamma$ channel, the detection is however more difficult in general than in the
SM, due to the reduced $\gamma \gamma$ branching ratio in most of the parameter
space as discussed previously. The $\Phi_H \to \bar{b} b$ final state gives a
more promising detection signal since the branching ratio is in general 
(slightly) larger than in the SM. \bigskip

\noindent {\bf d) WW/ZZ fusion processes} 

\begin{figure}[htbp]
\vspace*{-.5cm}
\begin{center}
\epsfig{figure=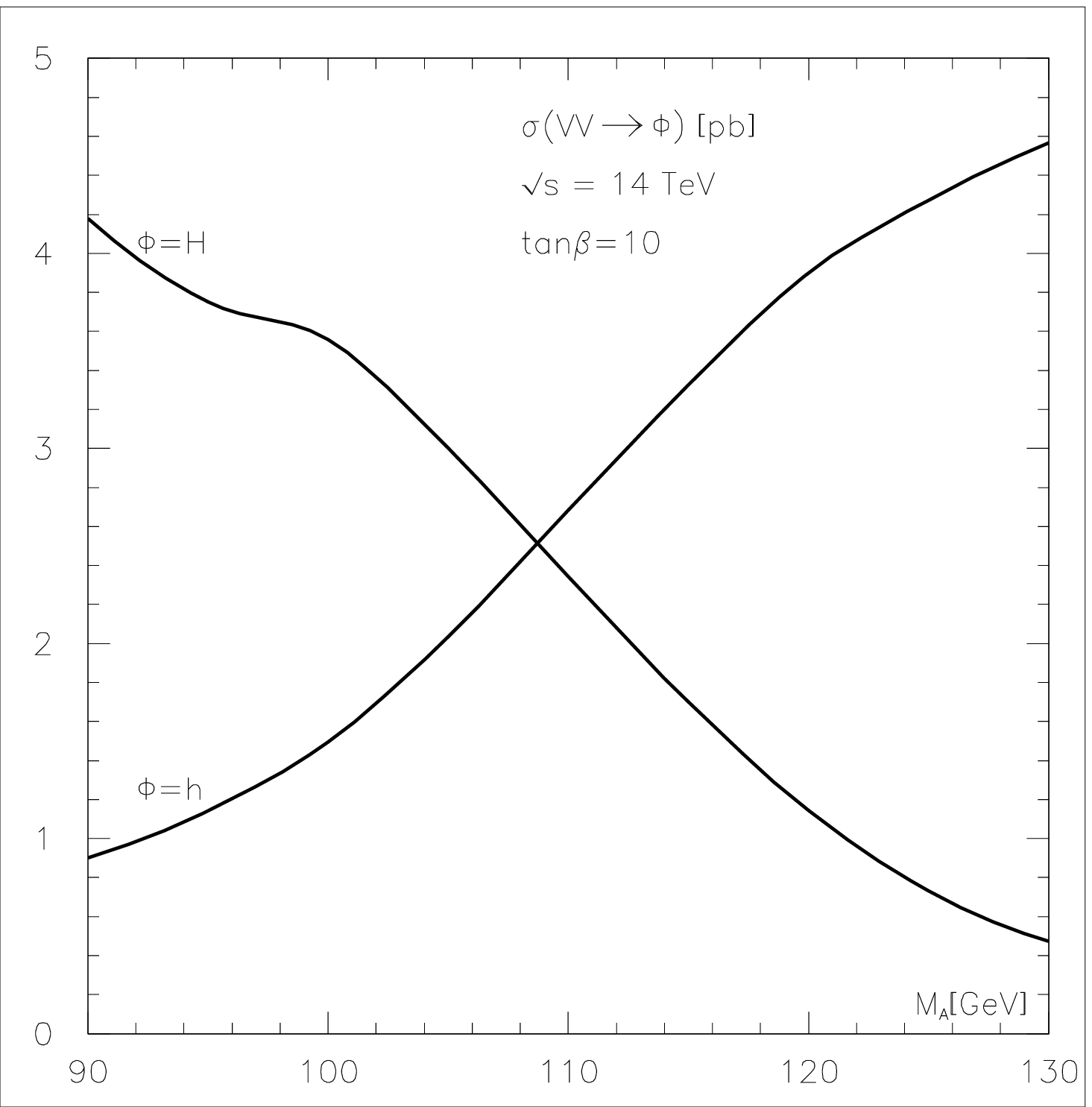,bbllx=3,bblly=3,bburx=418,bbury=420,height=6.5cm,width=7.5cm,clip=}\hspace{1cm}
\epsfig{figure=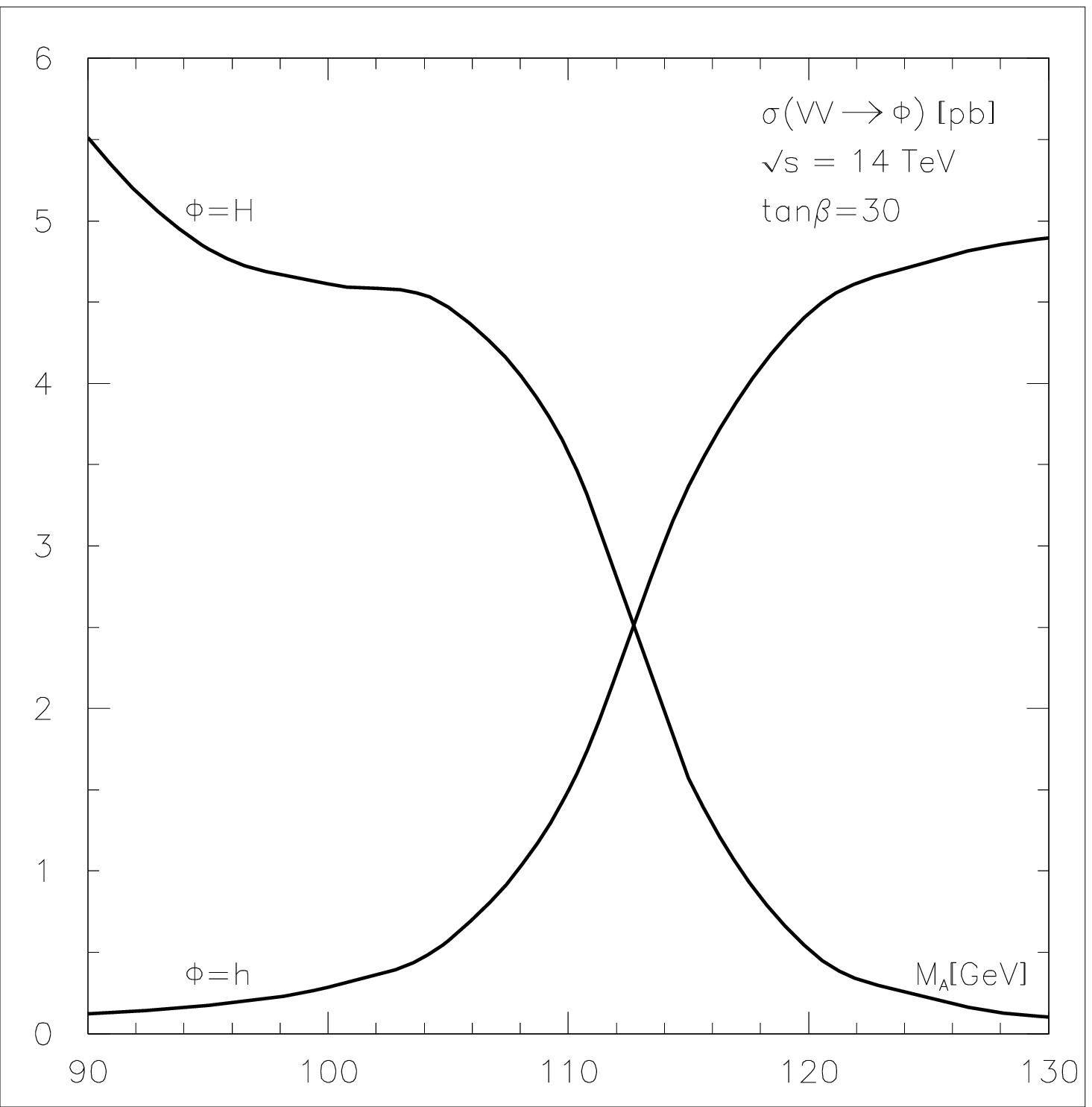,bbllx=3,bblly=3,bburx=418,bbury=420,height=6.5cm,width=7.5cm,clip=}\\[.1cm]
\epsfig{figure=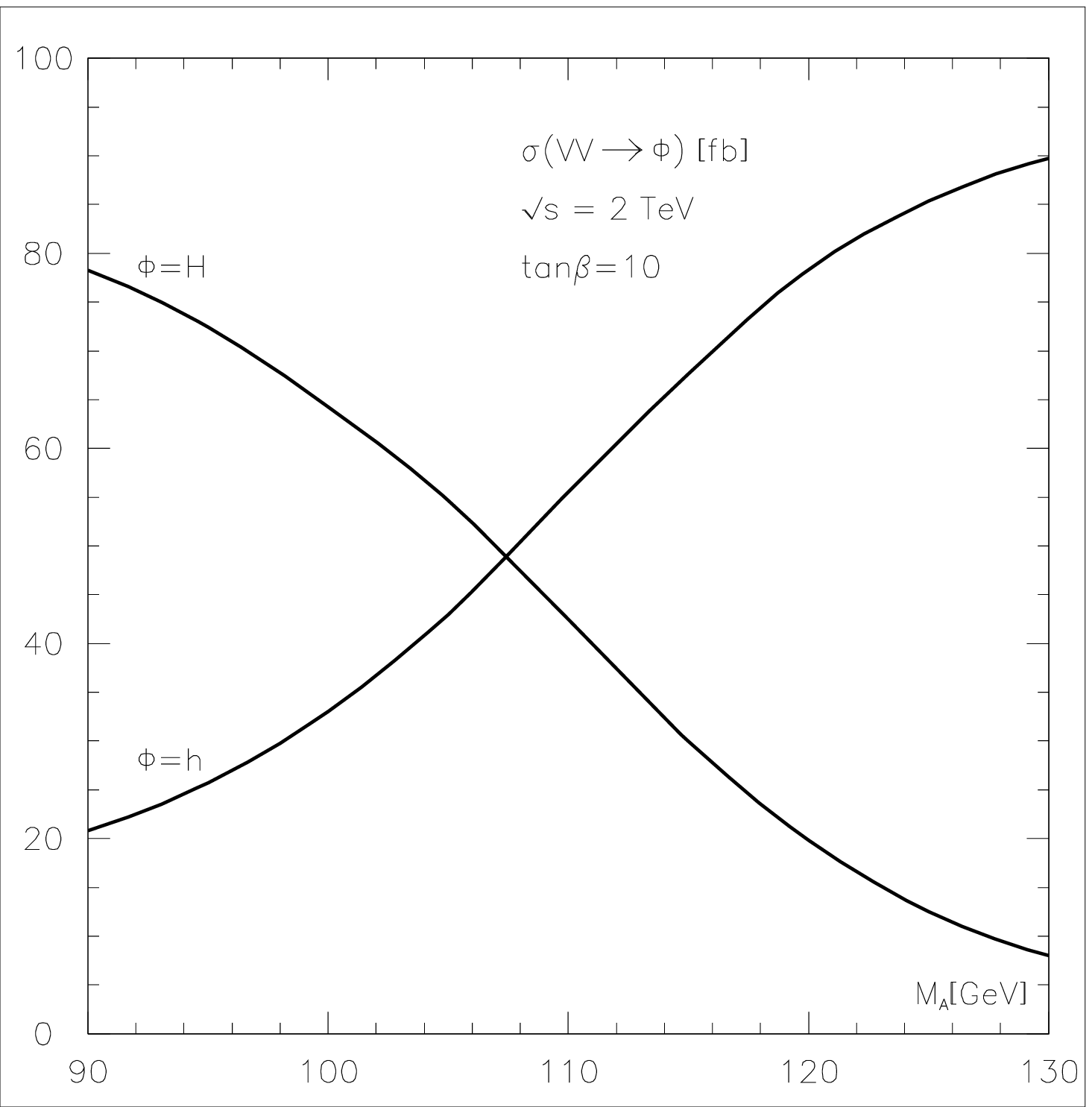,bbllx=3,bblly=3,bburx=418,bbury=420,height=6.5cm,width=7.5cm,clip=}\hspace{1cm}
\epsfig{figure=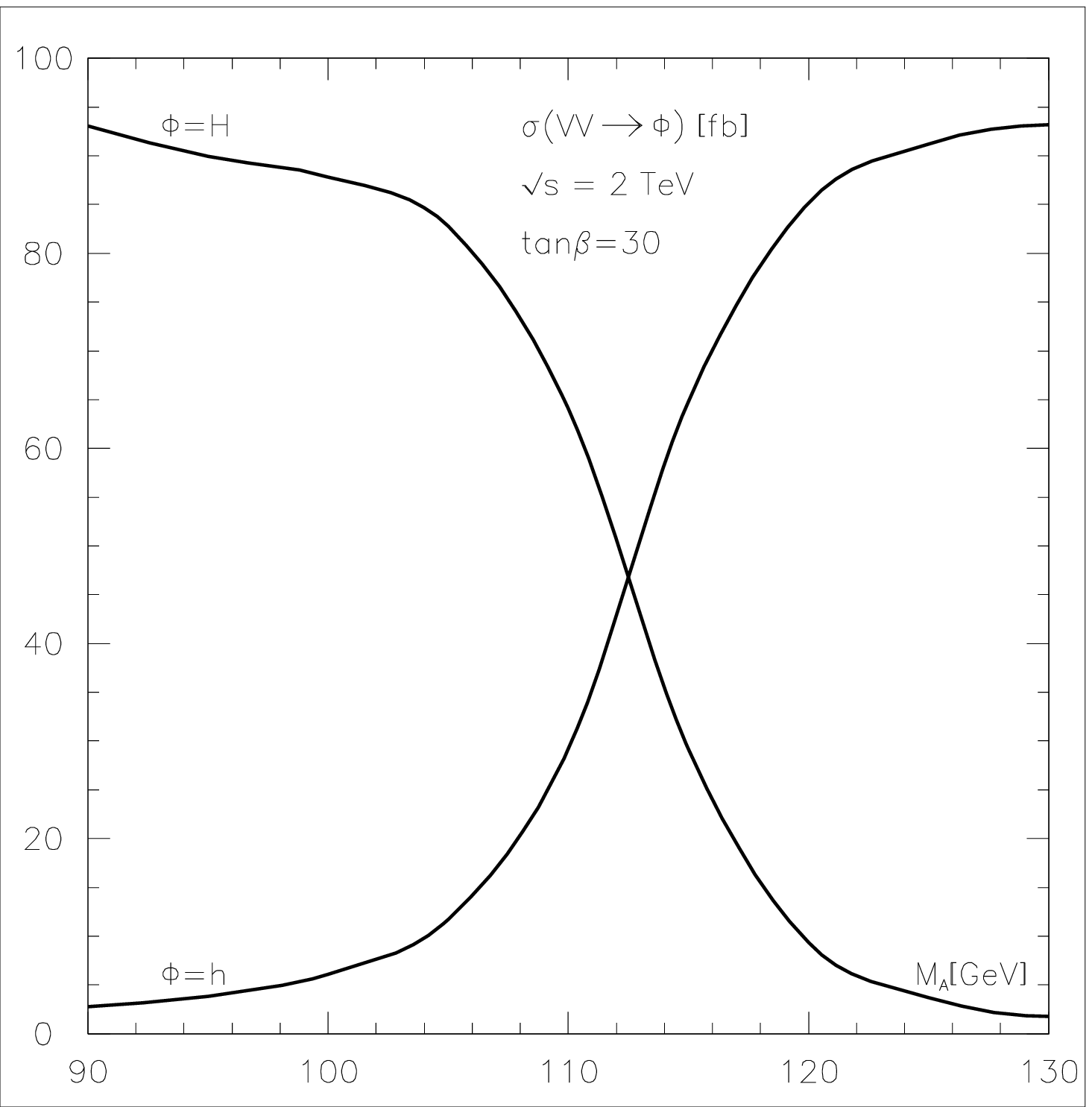,bbllx=3,bblly=3,bburx=418,bbury=420,height=6.5cm,width=7.5cm,clip=}\\[-.1cm]
\caption{\it Production cross sections at the LHC (upper panel) and at the 
Tevatron (lower panel) for the vector boson fusion mechanisms as functions of 
$M_A$ for $\tan \beta=10,30$. }
\end{center}
\vspace*{-.4cm}
\end{figure}

Here the cross sections are two to three orders of magnitude smaller than in
the $gg \to \Phi$ and $b\bar{b}\Phi$ processes. They only reach the level of 
$\sim 5$ pb at the LHC and $\sim 0.1$ pb at the Tevatron for a SM--like Higgs 
boson which has maximal coupling to gauge bosons. For the $\Phi_A$ boson, the 
rates are too small in particular for $\tan \beta \gsim 10$. \s

At the Tevatron, the cross sections are very small and this process is too difficult
to be used.  At the LHC, the decay into $\tau^+ \tau^-$ final states would allow for
the detection of the $\Phi_H$ particles, similarly as in the SM model, by taking
advantage of the energetic quark jets in the forward and backward directions which
allow for additional cuts to suppress the backgrounds. However, there are situations
where both products $\sigma (VV \to h,H) \times {\rm BR}( h,H \to
\tau^+\tau^-)$ are suppressed compared to the SM case as shown in Table 7 for several
values of $\tb$ and $M_A$. This occurs for large $\tb \gsim 30$ and 
small $M_A \sim 90$ GeV, where the ratio $\sigma (\Phi) \times {\rm BR}
(\tau \tau)|_{\rm MSSM}/ \sigma (H^0) \times {\rm BR}(\tau \tau)|_{\rm SM}$ with 
$\Phi=h(H)$ is strongly (slightly) suppressed. For $\tb \sim 10$ and $M_A \sim 130$ 
GeV, both ratios are slightly smaller than unity, 15\% for $\Phi=h$  and 30\% for 
$\Phi=H$, but in this case the sum of the ratios is larger than one. 
In addition, in the pathological regions where the $H$ boson couplings to $\tau$
leptons is suppressed [for instance $\tb\sim 50$ and $M_A\sim 110$ GeV], one can have
a very tiny cross section times branching ratio, since in this case, the lighter $h$
is $\Phi_A$ like and does not couple to vector bosons. \s

\begin{table}[htbp]
\renewcommand{\arraystretch}{1.8}
\begin{center}
\vspace*{-3mm}
\begin{tabular}{|c||c||c|c||c|c|} \hline
$\ \tb \ $ & $M_A$ & $M_h$ & $\frac{\sigma (h) \times {\rm
BR}(\tau \tau)|_{\rm MSSM} } {\sigma(H^0) \times {\rm BR}(\tau
\tau)|_{\rm SM} } $ & $M_H$ & $\frac{\sigma (H) \times {\rm BR}
(\tau \tau) |_{\rm MSSM} } {\sigma (H^0) \times {\rm BR}(\tau \tau)
|_{\rm SM} } $ \\ \hline
   & 90  &  84.9 & $6.6 \cdot 10^{-2}   $ & 125.1 & $1.16$ \\ 
10 & 110 &  103.9 & $0.2 $ & 126.8 & $1.24$ \\ 
   & 130 &  117.1 & $0.85$ & 134.0 & $0.68$
\\ \hline \hline
   & 90  &  85.9 & $3.1\cdot 10^{-3}   $ & 124.0 & $0.63$ \\ 
30 & 110 &  106.6 & $1.3\cdot 10^{-2} $ & 124.2 & $1.32$ \\ 
   & 130 & 123.2 & $1.2$ & 128.0 & $0.25$
\\ \hline \hline
   & 90  & 85.7 & $1.2\cdot 10^{-4}   $ & 124.3 & $0.49$ \\ 
50 & 110 & 106.5 & $4.8\cdot 10^{-4} $ & 124.4 & $6.3 \cdot 10^{-4} $ \\ 
   & 130 & 124.3 & $1.4 $ & 127.1 & $2.3 \cdot 10^{-2} $
\\ \hline \hline
\end{tabular}
\end{center}
\vspace*{-2mm}
\caption[]{\it The ratios $\sigma(VV \to \Phi) \times {\rm BR} (\Phi \to \tau 
\tau)$ in the MSSM with $\Phi=h,H$ relative to the SM case for a SM Higgs boson 
$H^0$ with the same mass, for three values of $\tb= 10, 30$ and 50 and three 
values of the pseudoscalar boson mass $M_A=90, 110$ and 130 GeV.}
\end{table}
\bigskip

\noindent {\bf e) Associated production with W/Z bosons} \s

At the LHC, these associated production processes would allow for the
measurement of the Higgs boson couplings to the massive gauge bosons [again in
conjunction with other production and decay processes].  The cross sections for
$W\Phi_H$ associated production are at the level of $\sim 1.5$ pb at the LHC
and $\sim 150$ fb at the Tevatron. They are of course slightly smaller than for
a SM Higgs boson with the same mass. The cross sections for $Z \Phi_H$
associated production are smaller by approximately a factor of 2. Because of
the fact that the gauge bosons are tagged through their leptonic decays,
$W$--boson final states are much more interesting since their leptonic
branching fractions are larger. \s

\begin{figure}[htbp]
\vspace*{-.7cm}
\begin{center}
\epsfig{figure=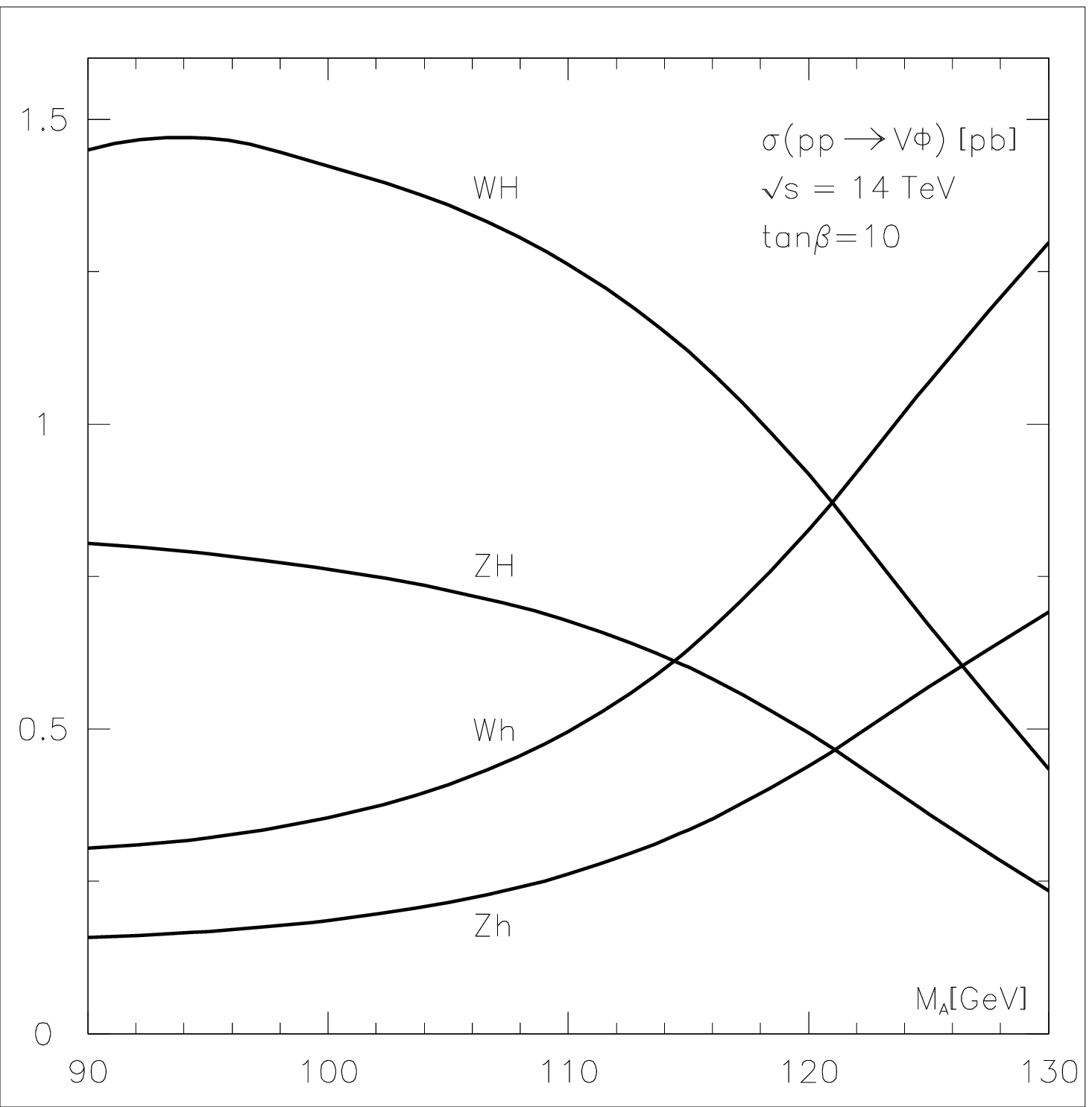,bbllx=3,bblly=3,bburx=418,bbury=420,height=6.5cm,width=7.5cm,clip=}\hspace{1cm}
\epsfig{figure=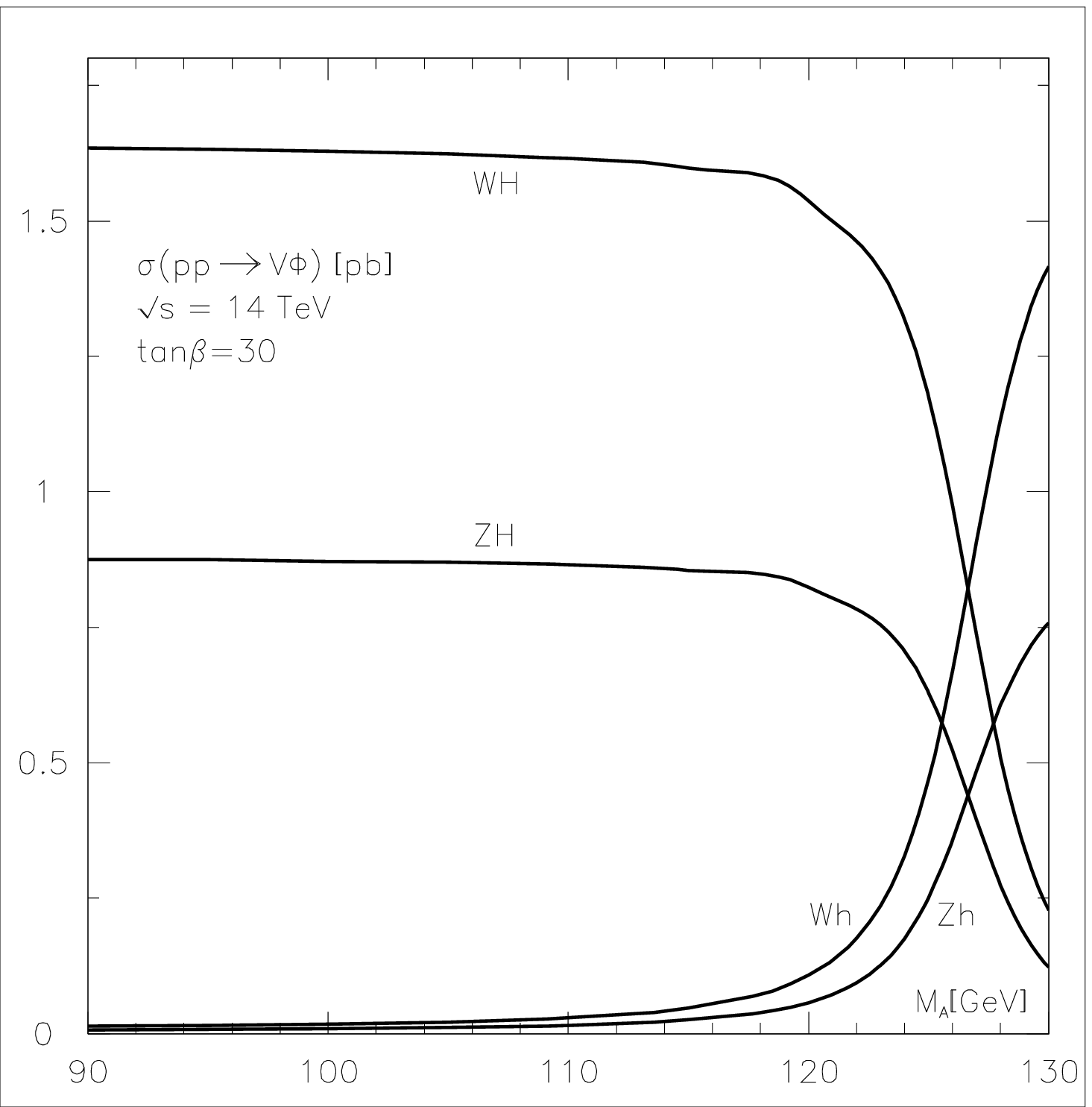,bbllx=3,bblly=3,bburx=418,bbury=420,height=6.5cm,width=7.5cm,clip=}\\[.1cm]
\epsfig{figure=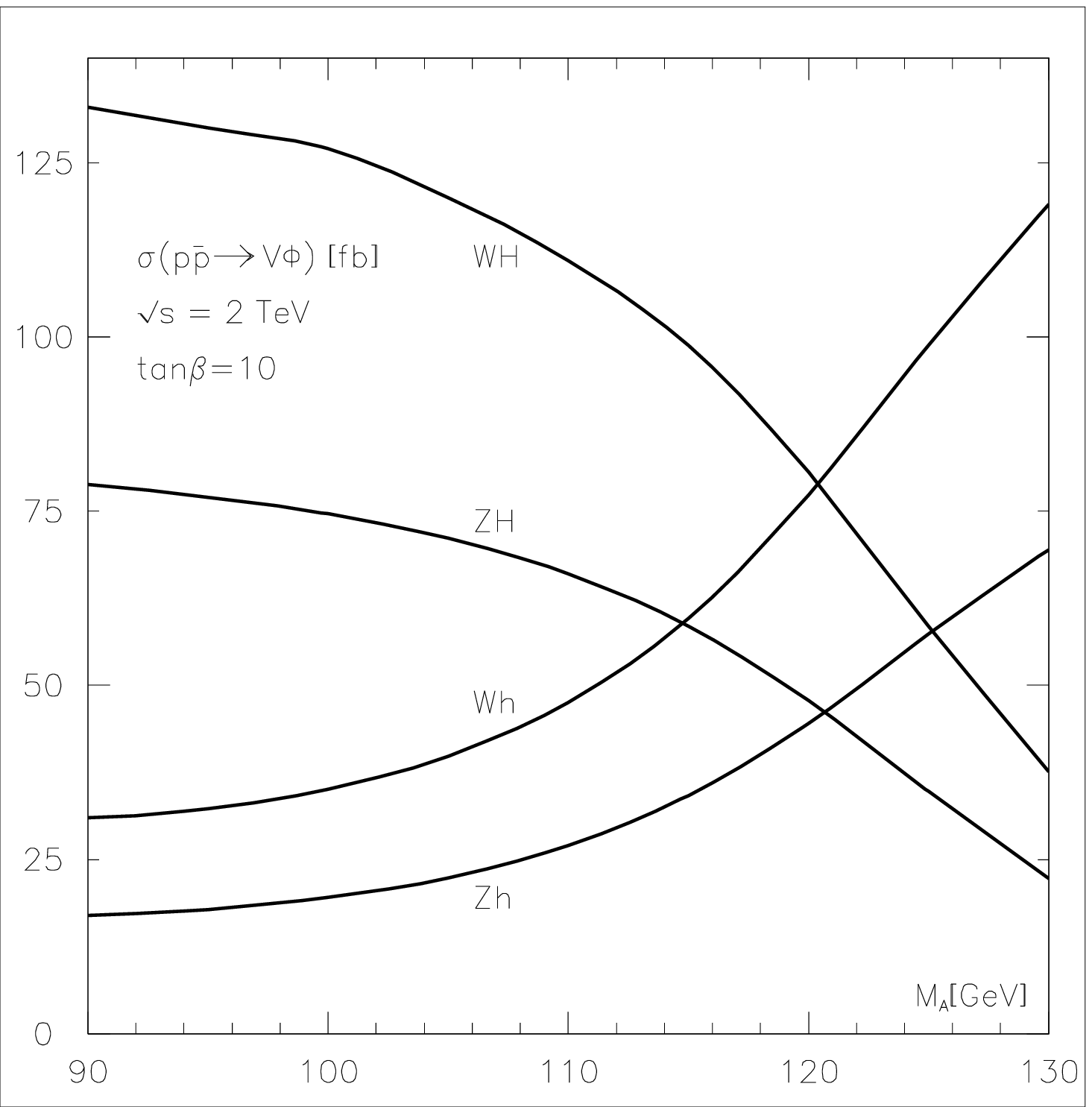,bbllx=3,bblly=3,bburx=418,bbury=420,height=6.5cm,width=7.5cm,clip=}\hspace{1cm}
\epsfig{figure=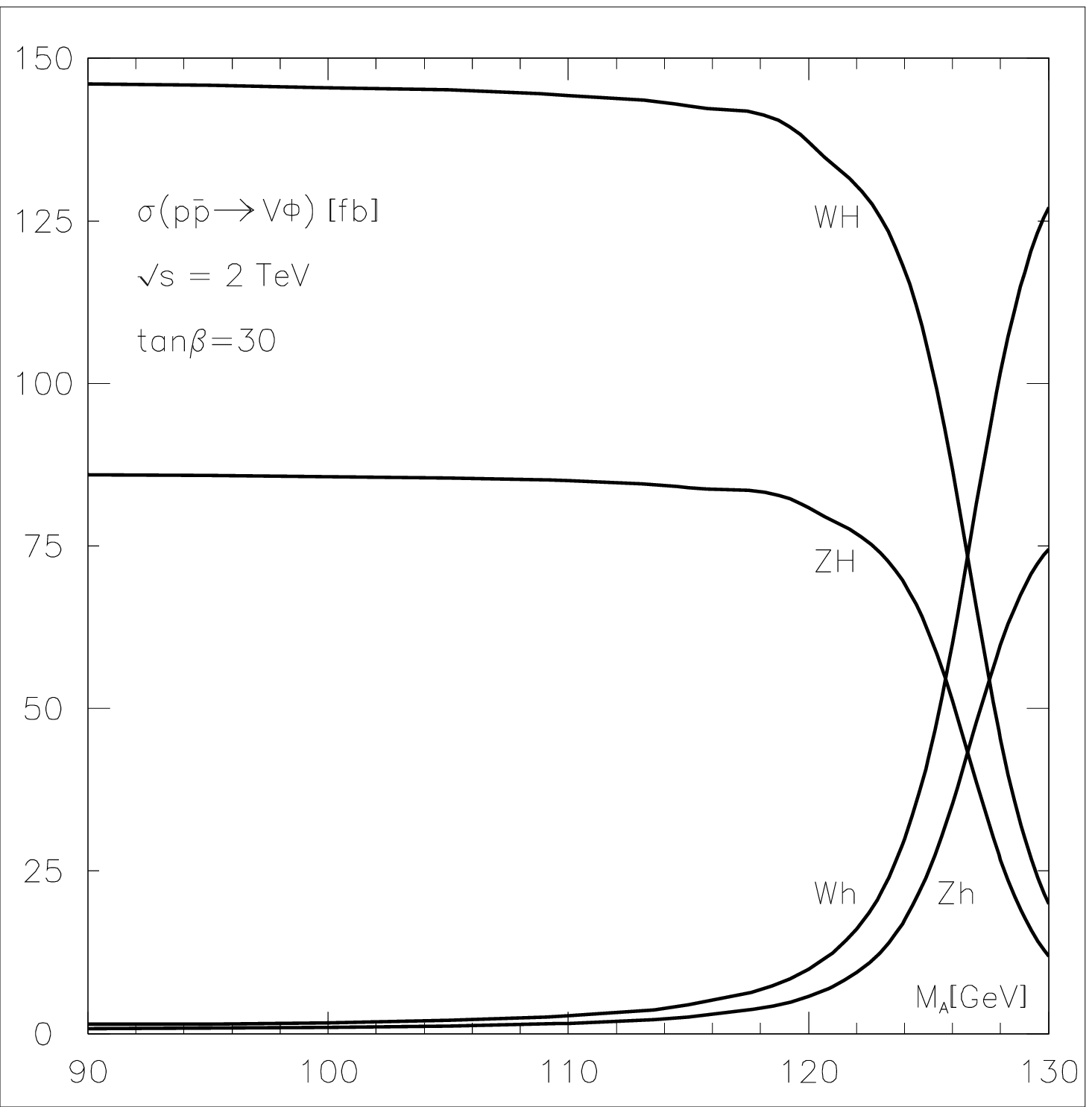,bbllx=3,bblly=3,bburx=418,bbury=420,height=6.5cm,width=7.5cm,clip=}\\[-.2cm]
\caption{\it Cross sections at the LHC (upper panel) and at the Tevatron (lower 
panel) for the associated Higgs production with gauge bosons as functions of 
$M_A$ for $\tan \beta=10,30$. }
\end{center}
\vspace*{-.6cm}
\end{figure}

At the Tevatron, these processes are the most promising ones to detect a
SM--like Higgs boson at the Tevatron \cite{HiggsTevatron}. Since the branching
fraction for the decay mode $\Phi_H \to b\bar{b}$ is always larger than in the
SM, except in the pathological cases discussed above, the situation is rather
favorable for the process $p\bar{p} \to W\Phi_H$ if the cross section is only
slightly suppressed compared to the SM case. In fact, the ratio of cross
sections times branching ratios into $b\bar{b}$, $\sigma (\Phi V) \times {\rm
BR}(bb)$, compared to the SM case, is the same as the one for the vector boson
fusion with the Higgs bosons decaying into $\tau^+ \tau^-$ final states [since
the cross sections are governed by the same $\Phi VV$ couplings and the ratio
BR$(\Phi \to \tau^+ \tau^-)$/BR($\Phi \to b\bar{b}) \sim m_\tau^2/3
\overline{m}_b^2$ is constant]. Therefore,  the situation is similar to that of
the previous case (d) and the possibility that the cross sections times
branching ratios are small for both $h$ and $H$ can occur. \s

The situation is more complicated at the LHC if the Higgs bosons have to be
detected through their $\gamma \gamma$ decay modes, since the $\gamma \gamma$ 
branching ratios are in general smaller than in the SM and the total decay 
widths of the states are much larger, as discussed previously [especially 
since already in the SM, the significance for this process is rather low]. 
\bigskip
 
\noindent {\bf f) Associated pair production} 

\begin{figure}[htbp]
\vspace*{-.5cm}
\begin{center}
\epsfig{figure=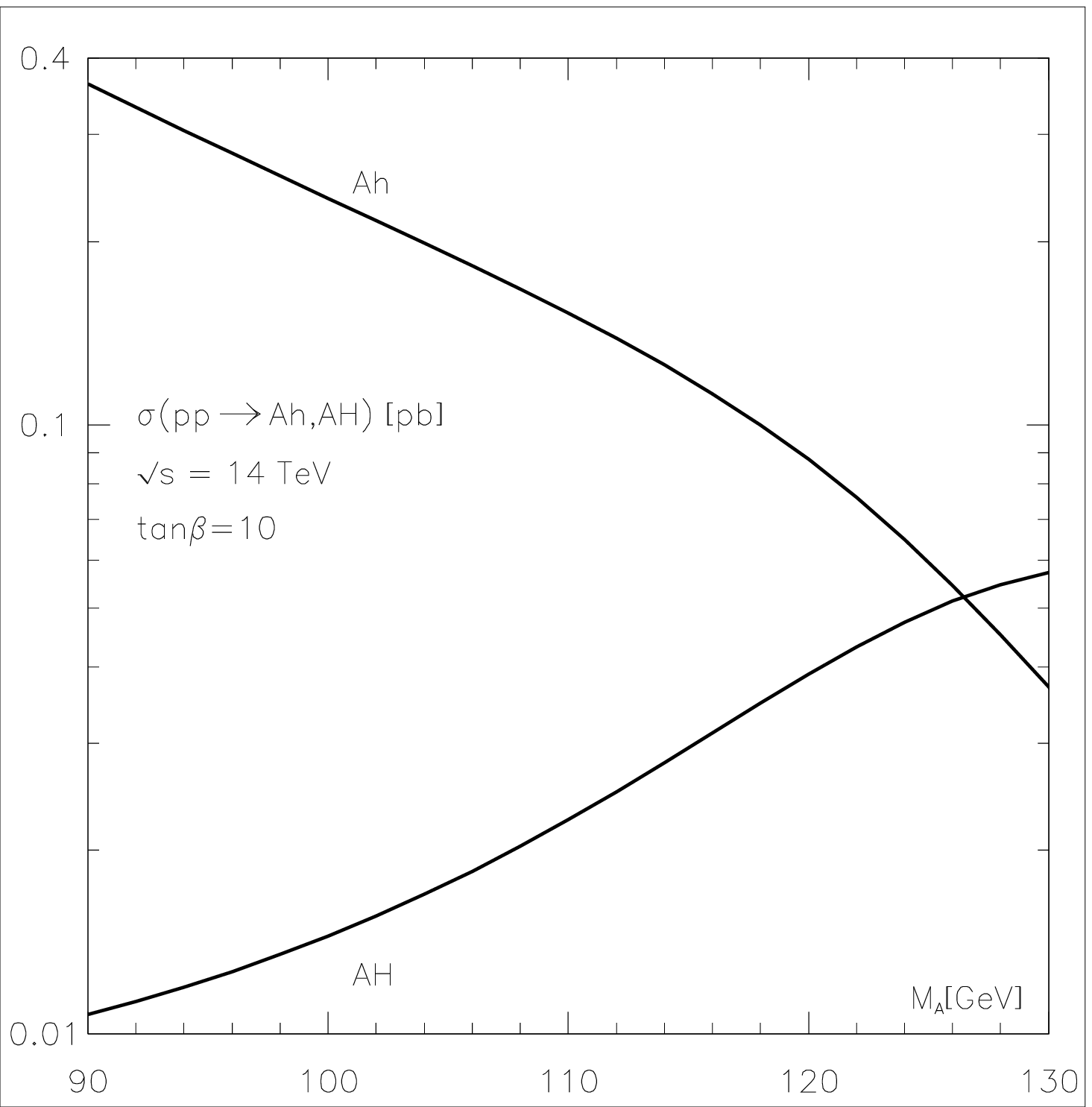,bbllx=3,bblly=3,bburx=418,bbury=420,height=6.5cm,width=7.5cm,clip=}\hspace{1cm}
\epsfig{figure=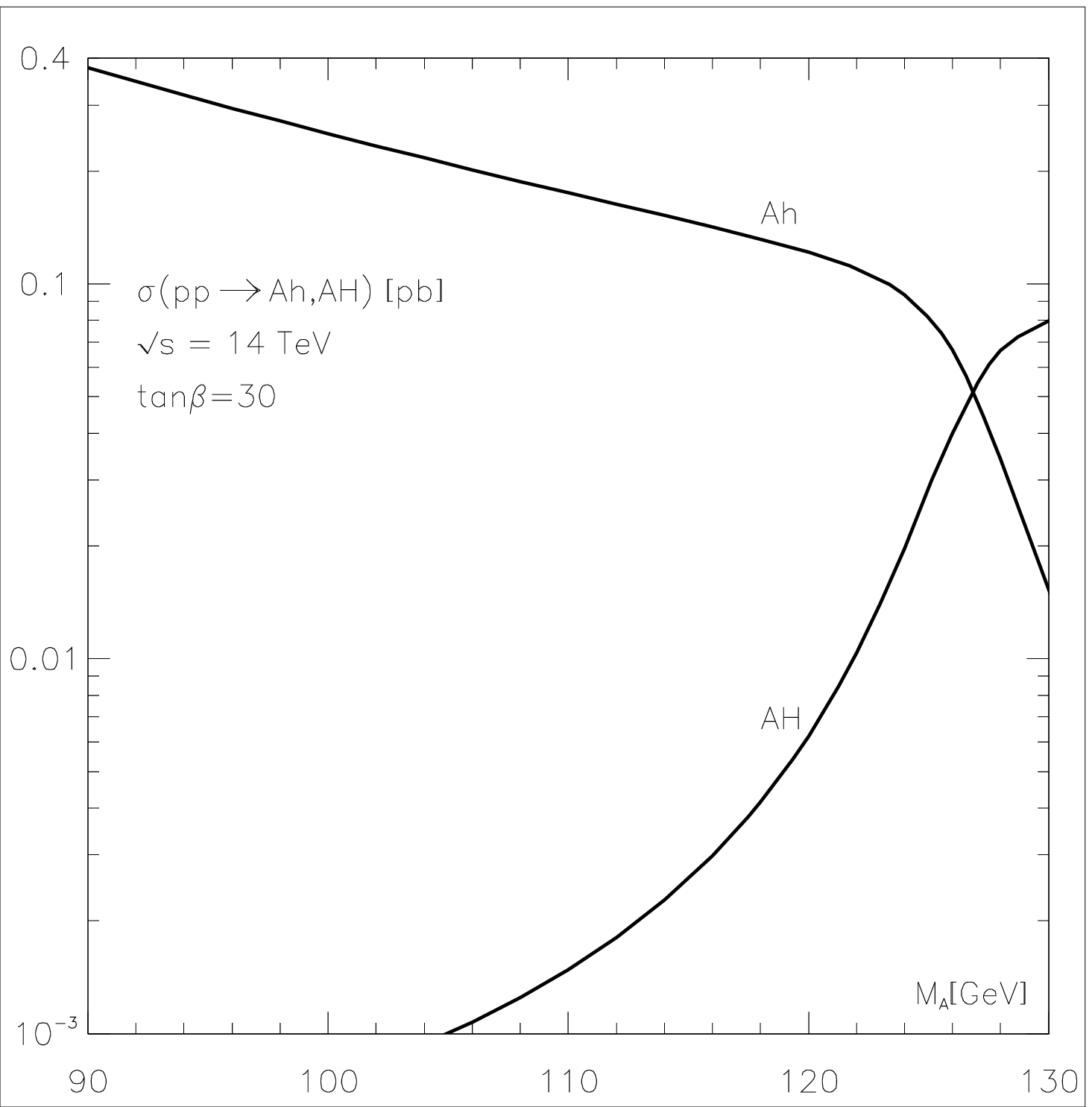,bbllx=3,bblly=3,bburx=418,bbury=420,height=6.5cm,width=7.5cm,clip=}\\[.1cm]
\epsfig{figure=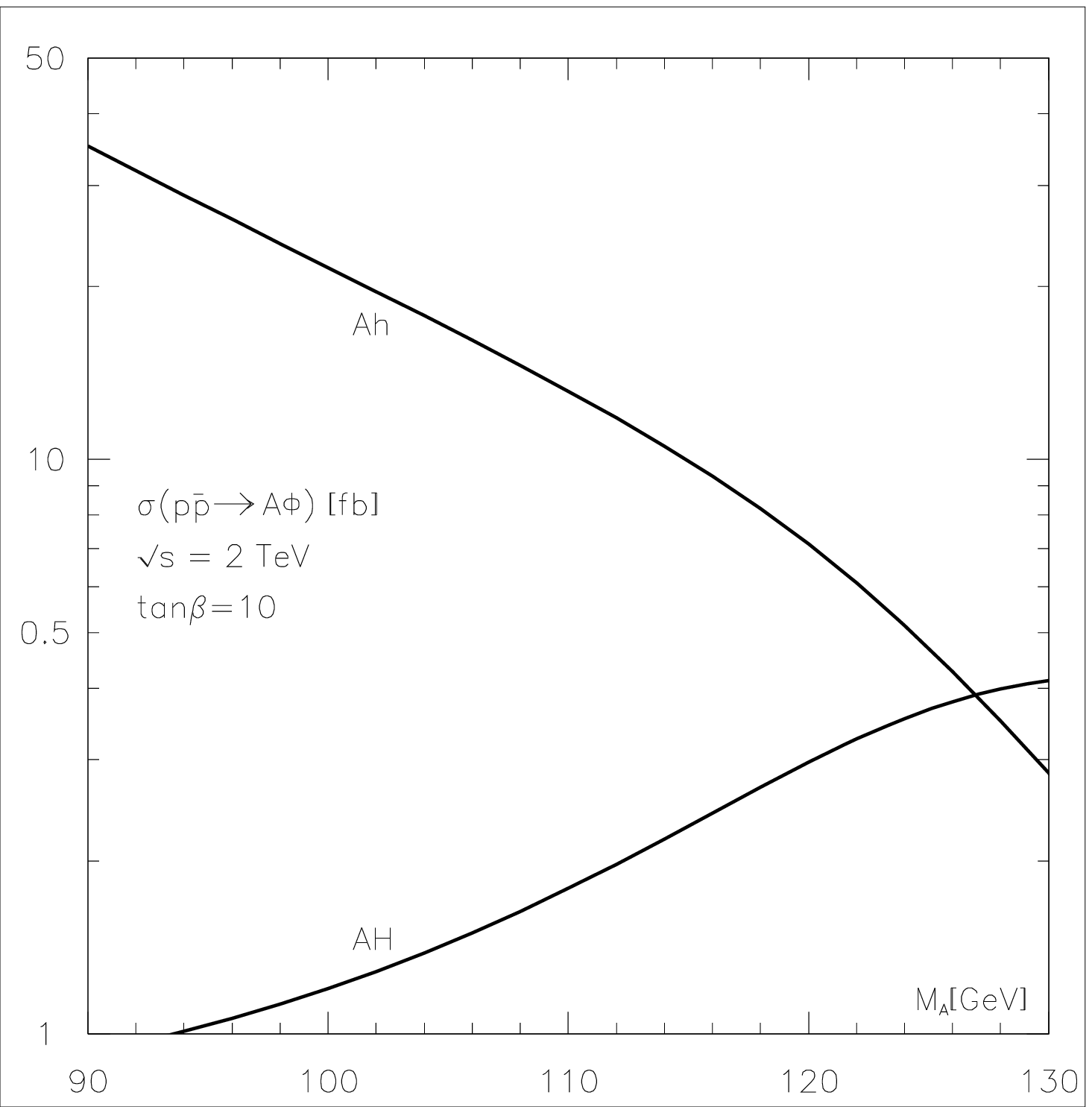,bbllx=3,bblly=3,bburx=418,bbury=420,height=6.5cm,width=7.5cm,clip=}\hspace{1cm}
\epsfig{figure=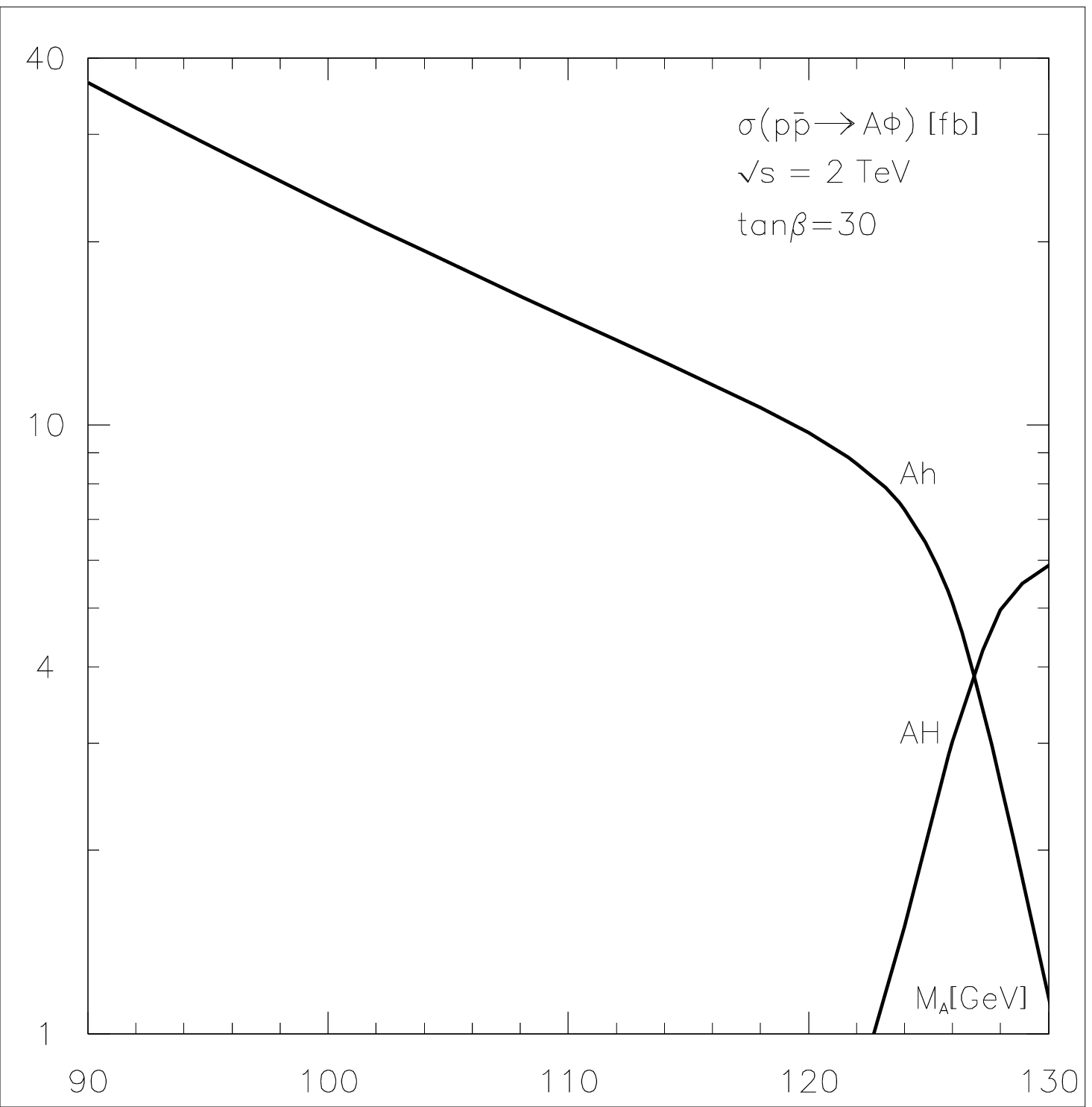,bbllx=3,bblly=3,bburx=418,bbury=420,height=6.5cm,width=7.5cm,clip=}\\[-.1cm]
\caption{\it Cross sections at the LHC (upper panel) and at the Tevatron 
(lower panel) for the associated Higgs boson pair production processes as 
functions of $M_A$ for $\tan \beta=10,30$. }
\end{center}
\vspace*{-.9cm}
\end{figure}

The associated pair production of CP--even and CP--odd Higgs bosons through the
$s$--channel exchange of a $Z$ boson has the lowest production cross section of
all processes. The event rates are reasonable only for the $q\bar{q} \to Ah$
process in the lower mass range of the $A$ boson, where the cross section is
above the level of 0.1 pb at the LHC and 10 fb at the Tevatron. In this range,
both the $A$ and $\Phi_A$ bosons will decay into $b\bar{b}$ pairs so that the
most advantageous final state [in terms of production rate] to consider is 4
$b$--jets, the rate for the $ b\bar{b}\tau^+ \tau^-$ final state being an order
of magnitude lower. \s

To our knowledge, the detection of MSSM Higgs bosons at the Tevatron and the 
LHC in these processes has not been subject to experimental scrutinity. This 
is probably due to the difficulty of extracting such a small 4 $b$--jet 
signal from the background. However, it might be interesting to look at such 
final states, not only for this particular process but also in Higgs boson 
pair production in higher order processes where the trilinear Higgs boson 
couplings are involved as will be discussed below. \bigskip

\noindent {\bf g) Pair production of Higgs bosons in higher order processes} \s

Pair production of Higgs bosons in the $gg$ fusion occurs through triangle loop
diagrams mediated by top and bottom quark exchange where a Higgs boson is
produced (off--shell in our case) and splits into two Higgs particles, and box
diagrams where the two Higgs particles are radiated from the internal top and
bottom quark lines. These processes are of higher order in the electroweak
coupling and therefore have small cross sections
\cite{DoubleHiggs1,DoubleHiggs2} compared to the dominant processes discussed
previously. However, they are interesting to consider since they involve the
trilinear Higgs boson couplings [in the vertex diagrams where a virtual Higgs
bosons splits into two Higgs bosons], the measurement of which is important in
order to reconstruct the MSSM Higgs potential. We will thus discuss shortly
these processes at the LHC and at the Tevatron. Additional processes such as
the double Higgs production in association with $W$ and $Z$ bosons, the double
Higgs production in the $WW/ZZ$ fusion processes and the triple Higgs
production in $q\bar{q}$ collisions, can also give access to the trilinear
Higgs boson couplings; however, the cross sections are in general smaller  than
in the $gg$ fusion mechanism and we will not discuss them here
\cite{DoubleHiggs2}.  \s

The production cross sections for the double production of Higgs bosons of any
kind, 
\beq (g) \ \ gg \to hh, Hh, HH, HA, hA, AA \nonumber 
\eeq
are shown in Figs.~13 for the LHC and Tevatron energies [for the latter case,
only for $\tan \beta=30$ and for those which are large enough].  At the LHC, the
cross sections are in general well below the level of 0.1 pb.  They can reach
the picobarn level only for very large $\tan \beta$ values, $\tan \beta \gsim
30$, small $A$ boson mass and only for the channels where only $A$ or $h$ are
involved, $gg \to hh$ and $AA$. In this regime, the Higgs bosons will decay
predominantly into $b\bar{b}$ pairs so that the final state will also be 4
$b$--quarks, as is the case for associated Higgs boson pair production which,
as discussed previously, has similar cross sections in the same $M_A$ range and
is a difficult process to detect experimentally.  In fact, when the cross
sections are sizable enough, the main contribution is due to the diagrams where
both Higgs bosons are emitted from the internal $b$--quark lines when the
Higgs--$\bar{b}b$ Yukawa coupling is strongly enhanced. The triangle diagram
which is sensitive to the trilinear Higgs boson coupling gives only a small
contribution, and these couplings can thus not be probed. At the Tevatron, the
cross sections are two orders of magnitude smaller than at the LHC because of
the reduced phase space and the much lower $gg$ luminosity. They are maximal
for the $gg \to hh$ and $AA$ processes where they reach the 10 fb level in the
same regime as at the LHC, i.e. for small $M_A$ and large $\tan \beta$ where
the contribution from Higgs emission from the internal $b$--quarks is enhanced
and the trilinear couplings play a minor role.  

\begin{figure}[htbp]
\begin{center}
\epsfig{figure=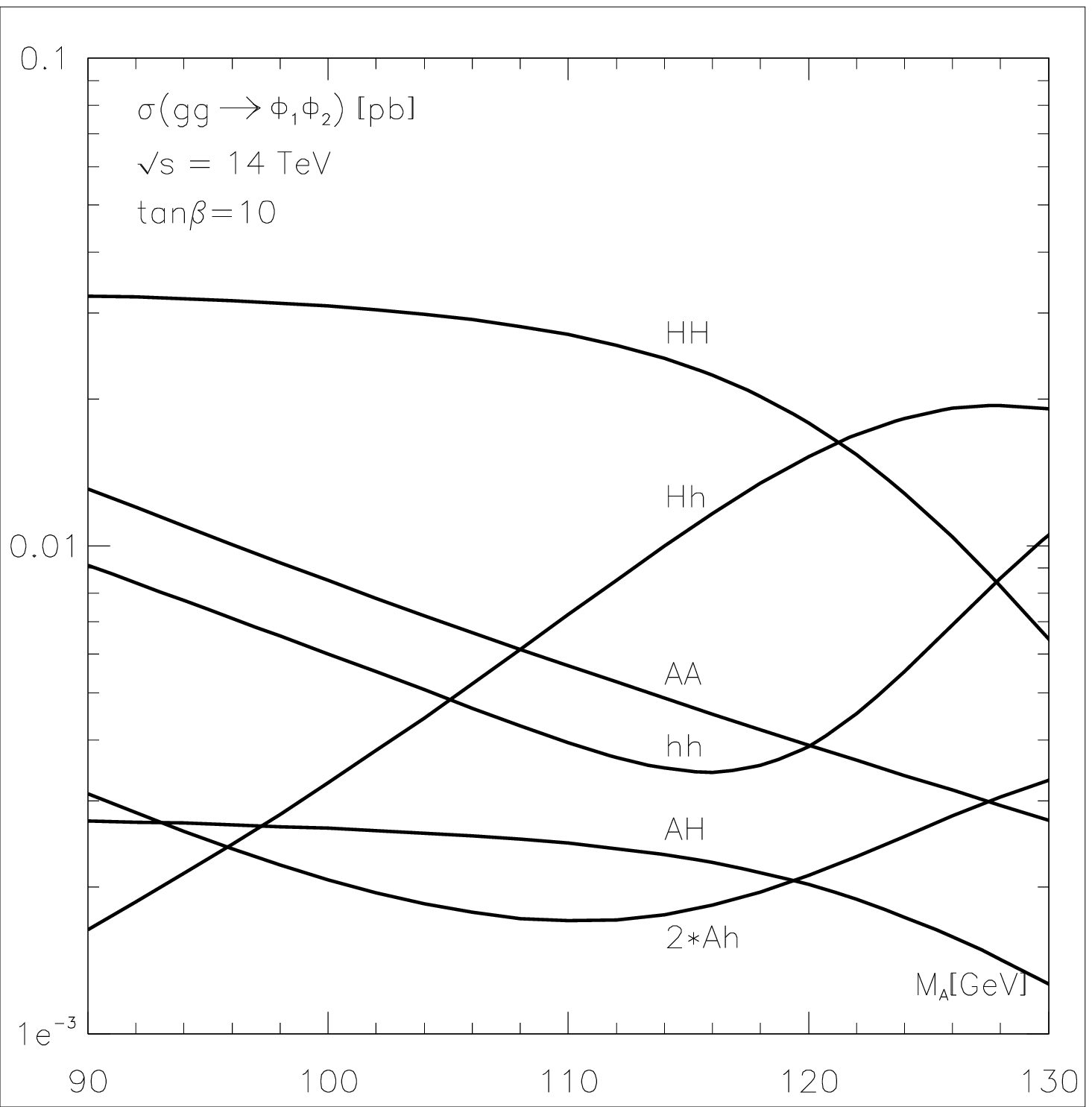,bbllx=3,bblly=3,bburx=410,bbury=420,height=6.5cm,width=7.5cm,clip=}
\hspace{1cm}
\epsfig{figure=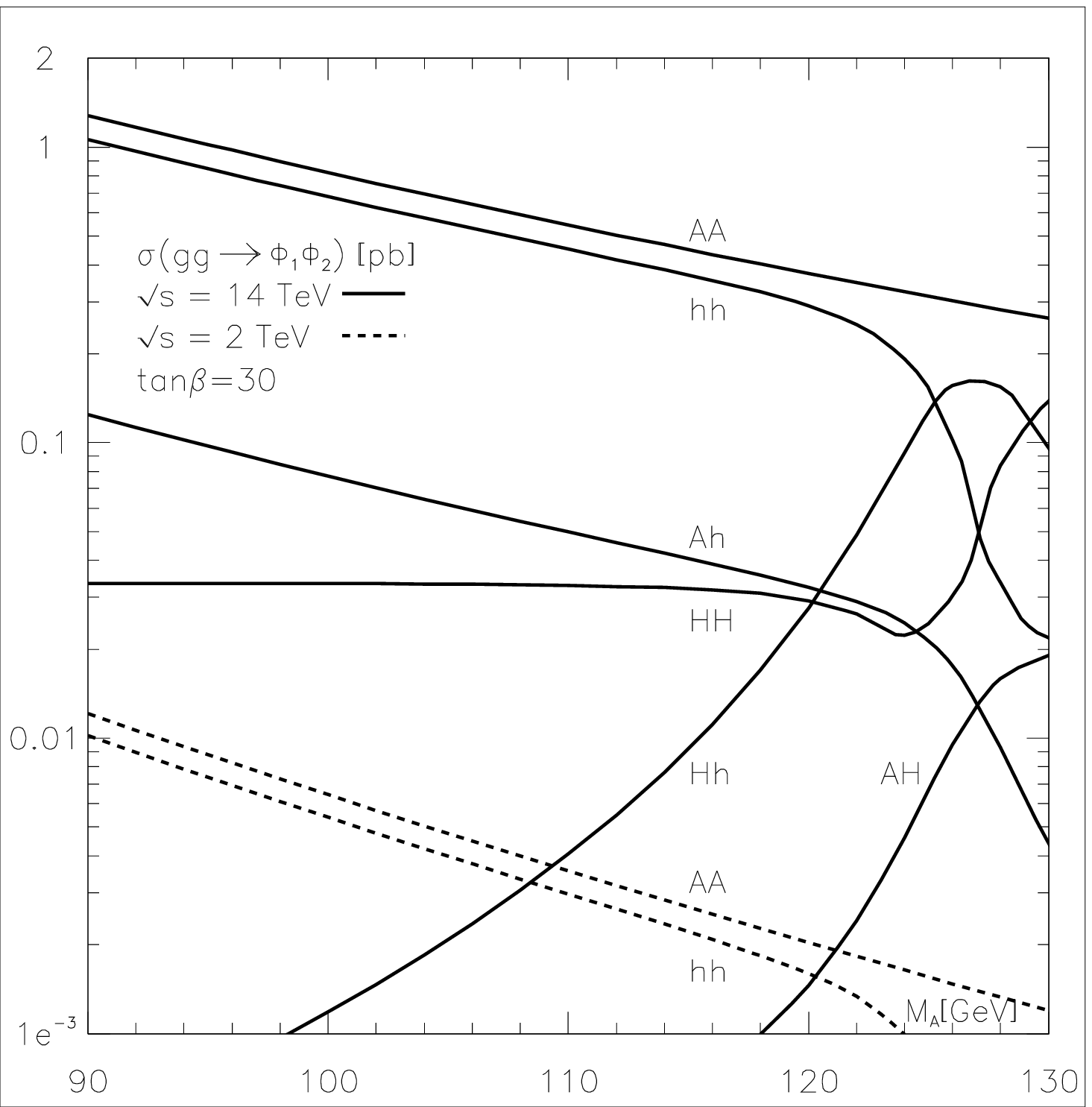,bbllx=3,bblly=3,bburx=410,bbury=420,height=6.5cm,width=7.5cm,clip=}
\\[.01cm]
\caption{\it Production cross sections at the LHC (full lines) and the Tevatron 
(dashed lines) for the double Higgs boson production processes $gg\to \Phi_1
\Phi_2$ as functions of $M_A$ for $\tan\beta=10$ (left panel) and $\tb=30$
(right panel).}
\end{center}
\vspace*{-1cm}
\end{figure}

\subsubsection*{5.2 Production at electron--positron colliders}

At $\ee$ linear colliders operating in the 500 GeV energy range,  the main 
production mechanisms for MSSM neutral Higgs particles in the $\ee$ option
are \cite{HHG,eexsections,eeHiggs}
\begin{eqnarray}
\begin{array}{lccl}
(a) & \ \ {\rm bremsstrahlung \ process} & \ \ \ee & \ra (Z^*) \ra Z+h,H \non \\
    & \ \ {\rm associated \ production \ process} & \ \ \ee & \ra (Z^*) \ra 
A+h,H \non \\
(b)  & \ \ WW \ {\rm fusion \ process} & \ \ \ee & \ra \bar{\nu}_e \ \nu_e 
\ W^*W^*  \ra \bar{\nu}_e \nu_e \ + h,H \non \\
     & \ \ ZZ \ {\rm fusion \ process} & \ \ \ee & \ra e^+ e^- Z^*Z^* \ra
e^+ e^- + h, H \non \\
(c)  & \ \ {\rm radiation~off~top~quarks} & \ \ \ee & \ra (\gamma^*,Z^*) \ra
t \bar{t}+h,H,A \non \\
   & \ \ {\rm radiation~off~bottom~quarks} & \ \ \ee & \ra (\gamma^*,Z^*) \ra
b \bar{b}+h,H,A \non
\end{array}
\end{eqnarray}

Again, the associated production of the pseudoscalar $A$ and $Z$ bosons, the
single production of the $A$ boson in the fusion processes as well as the pair
production of two CP--even or two CP--odd Higgs particles can only occur at higher
orders \cite{one-loop-Higgs} because of CP invariance. The production
cross sections are shown in Figs.~14--16 as functions of the pseudoscalar $A$
boson mass for a center of mass energy $\sqrt{s}=500$ GeV and for two values
$\tan \beta=10$ and 30.  Let us briefly discuss the cross sections of these
various processes. \bigskip

\noindent {\bf a) Higgs--strahlung and Higgs pair production} \s

These are the most interesting processes in this context.  The cross sections
for the bremsstrahlung and the pair production processes as well as the cross
sections for the production of the $h$ and $H$ bosons are mutually
complementary, coming either with a coefficient $\sin^2(\beta- \alpha)$ or
$\cos^2(\beta -\alpha)$. The cross section for $hZ$ production is large for
large values of $M_h$, i.e. close to the decoupling limit. By contrast, the
cross section for $HZ$ production is large for a light $A$ boson implying a
light $H$ boson with a large coupling to $Z$ bosons.  For the associated
production, the situation is opposite: the cross section for $Ah$ is large for
a light $h$ boson whereas $AH$ production is preferred in the complementary
region. \s

\begin{figure}[htbp]
\vspace*{-.5cm}
\begin{center}
\epsfig{figure=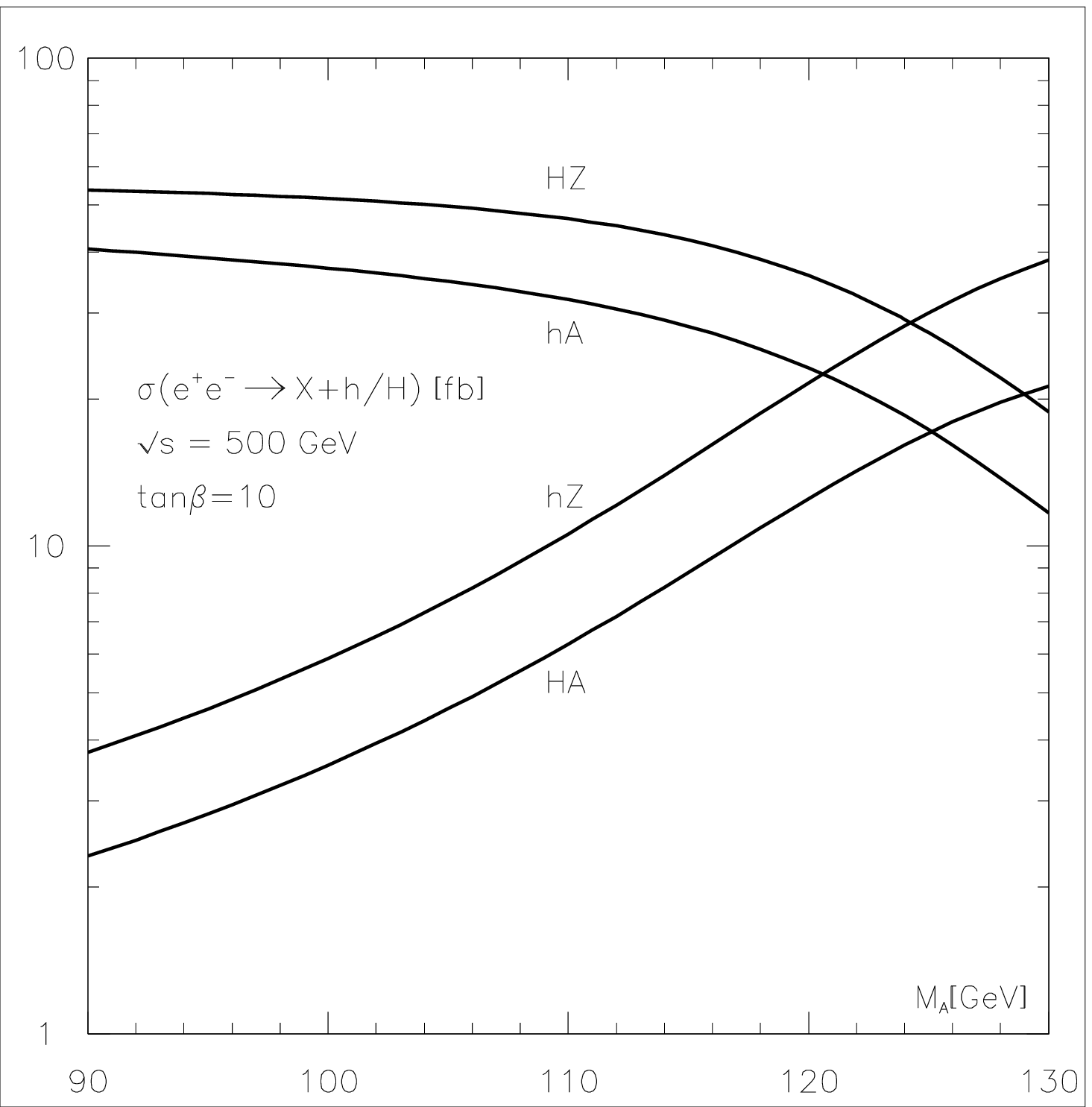,bbllx=3,bblly=3,bburx=418,bbury=420,height=6.5cm,width=7.5cm,clip=}
\hspace{1cm}
\epsfig{figure=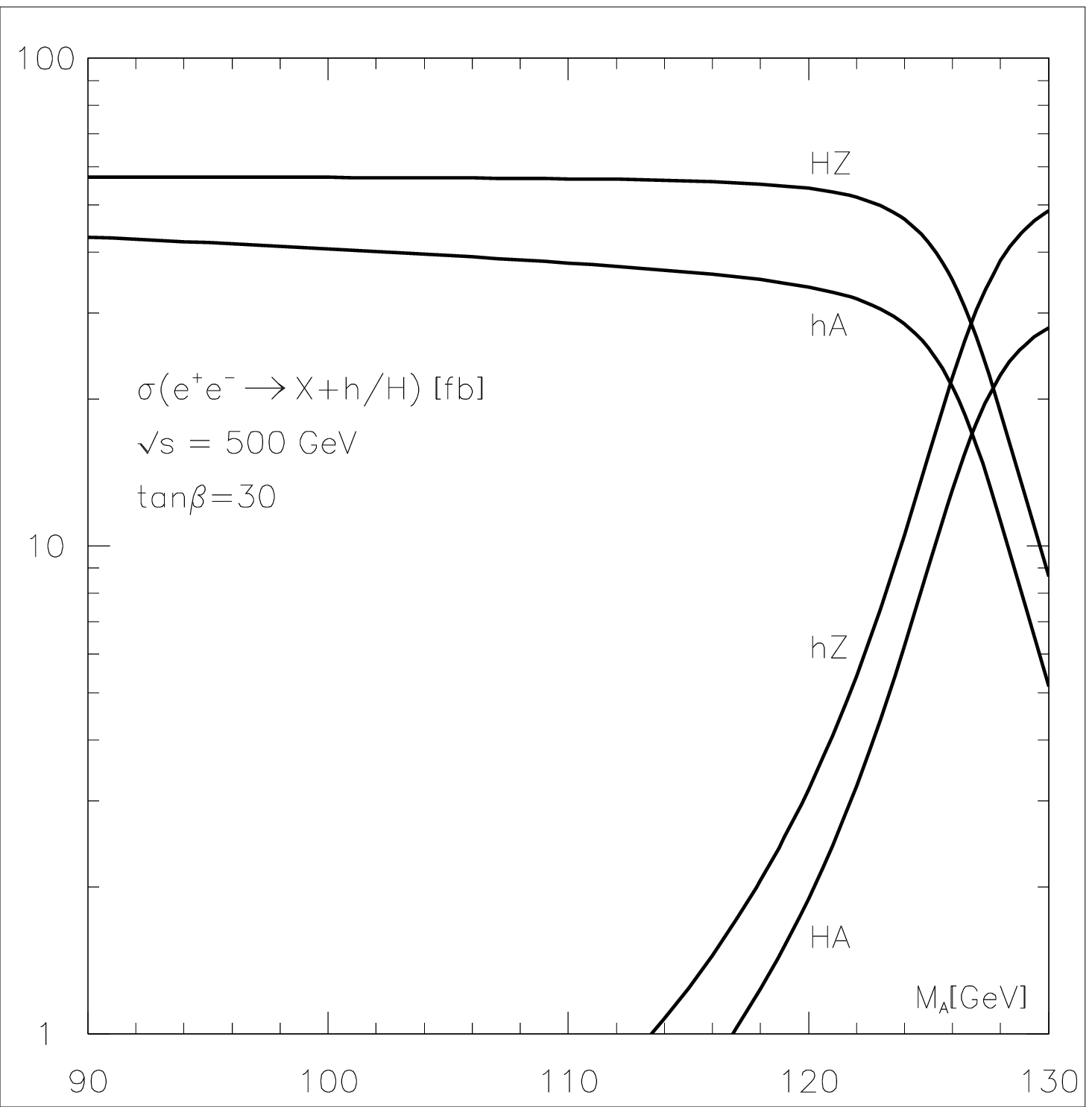,bbllx=3,bblly=3,bburx=418,bbury=420,height=6.5cm,width=7.5cm,clip=}\\[-.1cm]
\caption{\it Total cross sections at a 500 GeV $\ee$ collider in the Higgs
strahlung and associated Higgs production processes as functions of $M_A$ for 
$\tan \beta=10,30$. }
\end{center}
\vspace*{-.5cm}
\end{figure}

At $\sqrt{s}=500$ GeV, the sums of the cross sections $\sigma(\ee \to hZ+HZ)$
and $\sigma(\ee \to hA+HA)$ are, respectively, $\sim 60$ and $\sim 40$ fb in
the entire $M_A$ range. This means that with the integrated luminosity, $\int
{\cal L} \sim 500$ fb$^{-1}$, expected at the TESLA machine for instance,
approximately 30.000 and 20.000 events per year can be collected in these two
channels, respectively. In the Higgs--strahlung process, the signals consist mostly 
of a $Z$ boson and a $b\bar{b}$ pair, which is easy to
separate from the main background, $\ee \ra ZZ$ [for $M_h \simeq M_Z$,
efficient $b$ detection is needed]. For the associated Higgs boson production,
the signals consist mostly of four $b$ quarks in the final state, requiring
efficient $b$--quark tagging. \bigskip

\noindent {\bf b) Vector boson fusion processes} \s

The trend for the production cross sections of the $h$ and $H$ boson in vector
boson fusion  is the same as in the Higgs--strahlung processes since both are
proportional to the square of the Higgs--$VV$ couplings.  At $\sqrt{s}\sim 500$
GeV, the production rate for the $WW$ fusion mechanism is slightly larger than
the one for the Higgs--strahlung process, $\sigma(\ee \to H\nu \bar{\nu}+h \nu
\bar{\nu}) \sim 80$ fb compared to $\sigma(\ee \to hZ+HZ) \sim 60$ fb. A large
sample of events is therefore expected, 40.000 events per year with the
luminosity $\int {\cal L} \sim 500$ fb$^{-1}$, with a signature consisting
mainly of a $b\bar{b}$ or a $\tau^+ \tau^-$ pair and missing energy.  The cross
sections for the $ZZ$ fusion processes are one order of magnitude smaller than
the ones for $WW$ fusion, a consequence of the smaller neutral current
couplings compared to charged current couplings. The signal is however cleaner
due the additional $\ee$ pair in the final state. \s

Note that for other c.m.~energies, the Higgs--strahlung cross section, which 
scales as $1/s$, dominates at lower energies while the $WW$ fusion mechanism 
which has a cross section rising like $\log(s/M_H^2)$ dominates at higher 
energies. \s

\begin{figure}[htbp]
\vspace*{-.5cm}
\begin{center}
\epsfig{figure=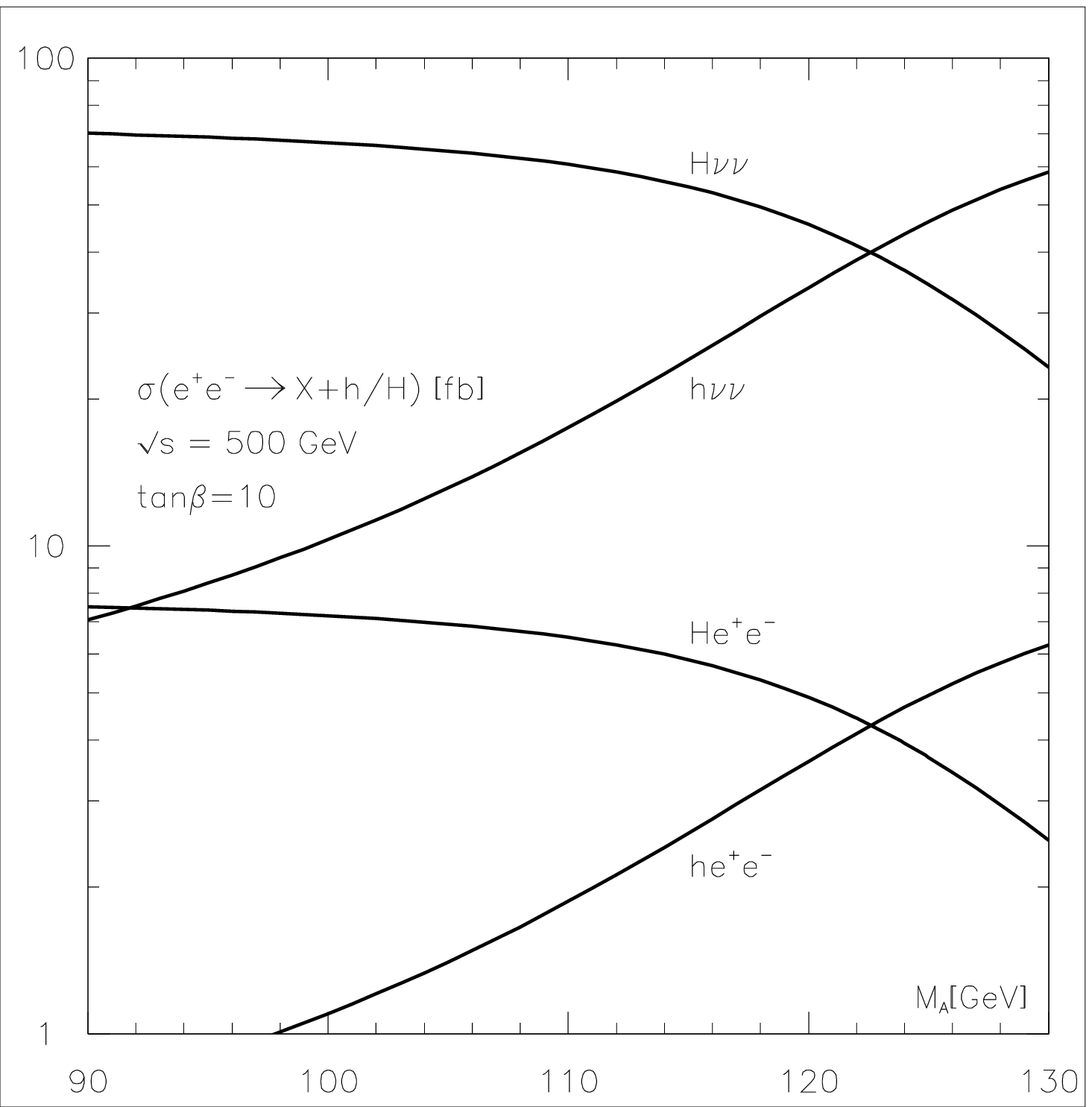,bbllx=3,bblly=3,bburx=418,bbury=420,height=6.5cm,width=7.5cm,clip=}
\hspace{1cm}
\epsfig{figure=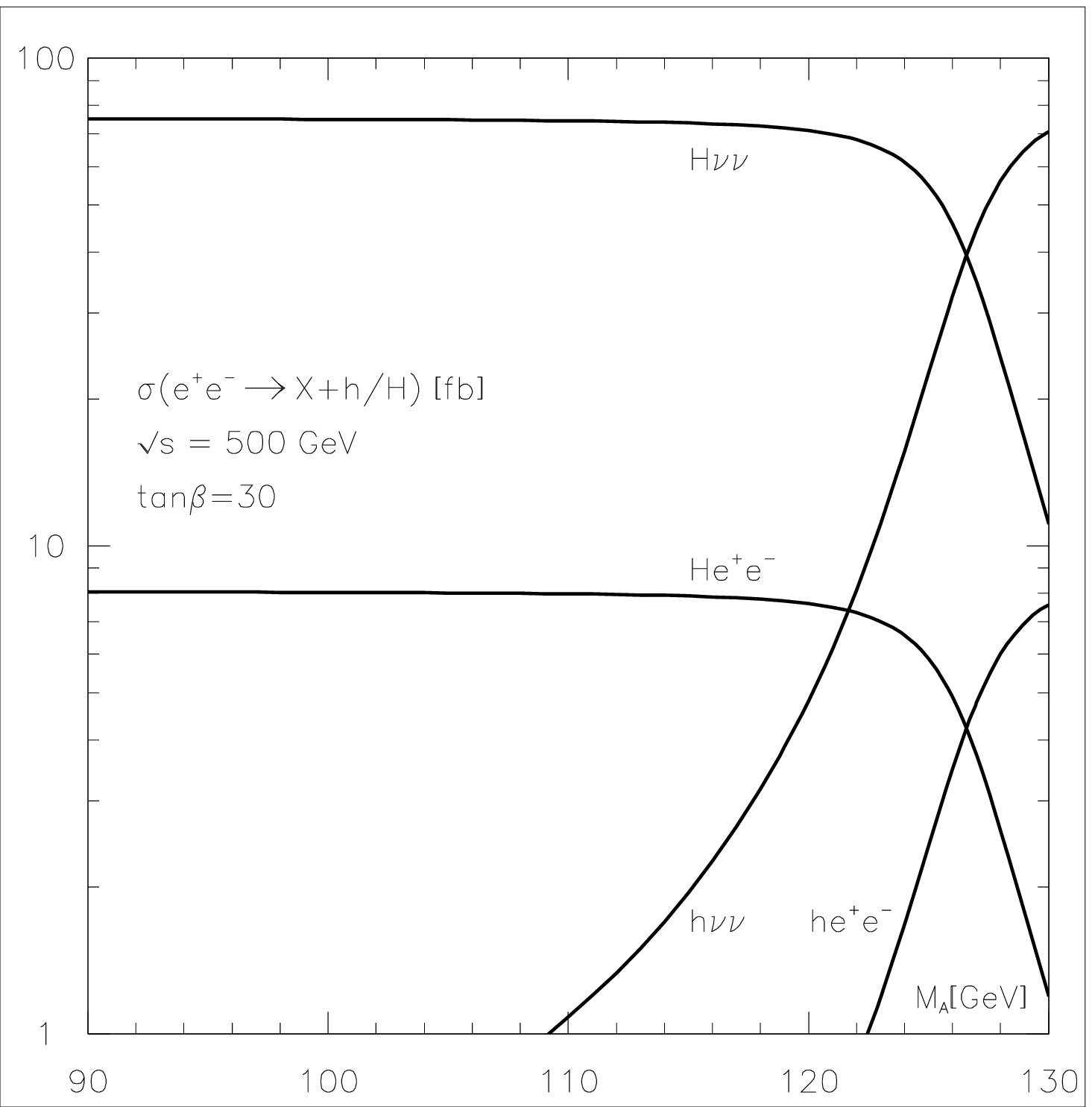,bbllx=3,bblly=3,bburx=418,bbury=420,height=6.5cm,width=7.5cm,clip=}\\[-.2cm]
\caption{\it Production cross sections at a 500 GeV $\ee$ collider in the 
vector boson fusion processes as functions of $M_A$ for $\tan \beta=10,30$. }
\end{center}
\vspace*{-.5cm}
\end{figure}

\noindent {\bf c) Higgs production with heavy quarks} \s

For Higgs boson production in association with top quarks, the production cross
sections are strongly suppressed for the $A$ and $\Phi_A$ boson for $\tan \beta
\gsim 10$ and are sizeable only for the $\Phi_H$ boson which has almost SM--like
couplings to the top quarks. Even in this case, they are however very small,
barely exceeding the level of 0.2 fb. This is due to the fact that at
$\sqrt{s}=500$ GeV, there is only a little amount of phase--space [$\sim 50$
GeV] available for this process. At higher energies, e.g. $\sqrt{s}=800$ GeV,
the cross sections can exceed the level of $\sim 1$ fb which would allow for the
measurement of the $\Phi_H t \bar{t}$ couplings since most of the cross section
is coming from the Higgs boson radiation off the top quarks. \s 

In the case of Higgs boson production in association with bottom quarks, we
have taken into account only the gauge invariant contribution coming from Higgs
boson radiation off the $b$--quark lines since a much larger contribution would
come from the associated production process $\ee \to Ah$ or $AH$, with one of
the Higgs boson decaying into $b\bar{b}$, or from the Higgs--strahlung process
$\ee \to Zh$ or $ZH$ with $Z\to b\bar{b}$. These resonant processes have been
discussed earlier and can be separated from the Higgs radiation off $b$--quarks
by demanding that the invariant mass of a $b\bar{b}$ pair does not coincide
with a $Z$ boson or another Higgs boson.  Because of the strong enhancement of
the $b\bar{b}$ Yukawa coupling, the cross sections can exceed the level of
$\sigma (\ee \to \bar{b}b A+ \bar{b}b \Phi_A) \gsim 0.7$ fb for $\tan
\beta\gsim 30$, allowing  for the possibility to measure directly $\tan \beta$
with a reasonable accuracy in this process \cite{tanbeta}. \bigskip

\begin{figure}[htbp]
\vspace*{-.5cm}
\begin{center}
\epsfig{figure=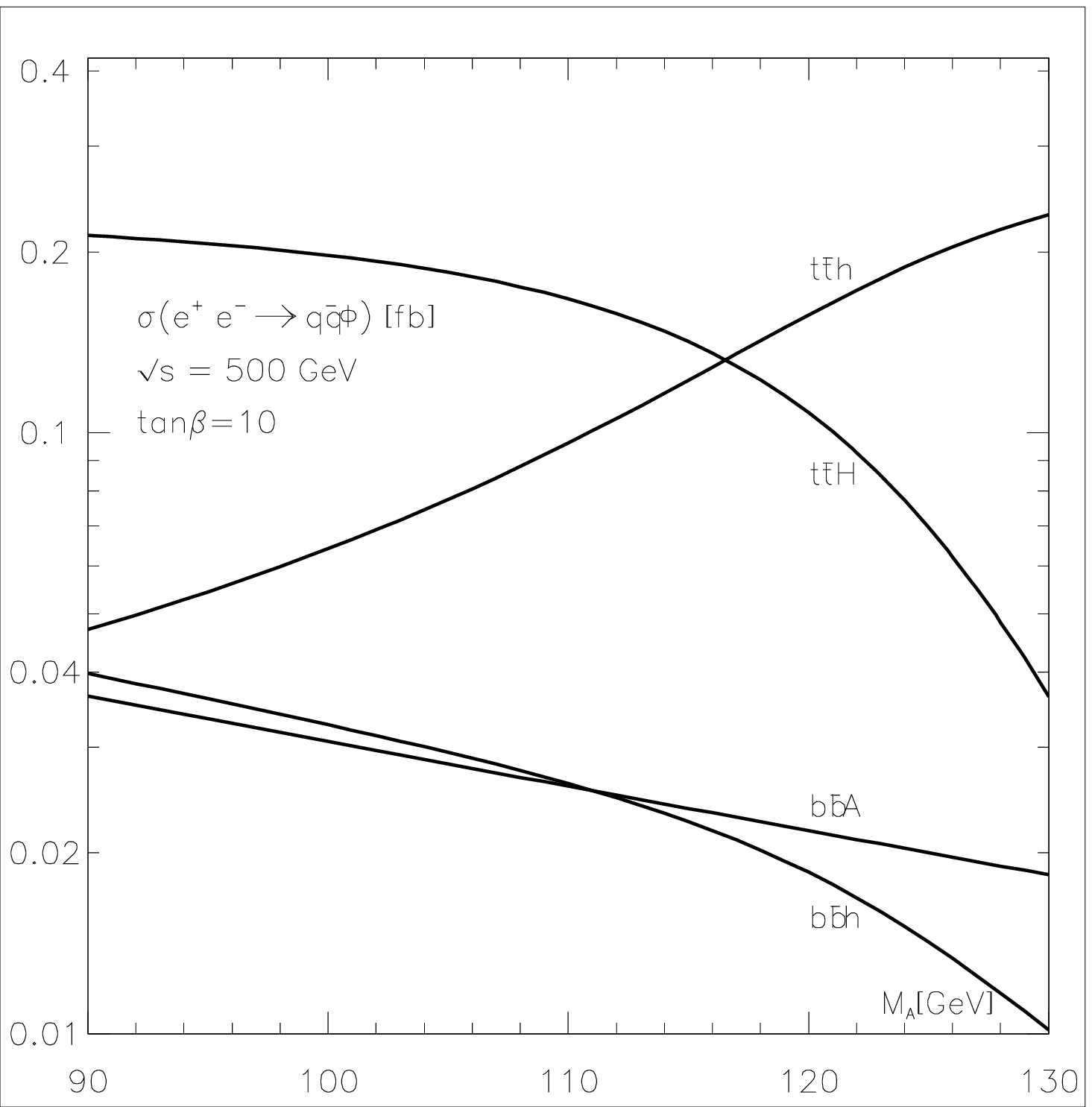,bbllx=3,bblly=3,bburx=418,bbury=420,height=6.5cm,width=7.5cm,clip=}
\hspace{1cm}
\epsfig{figure=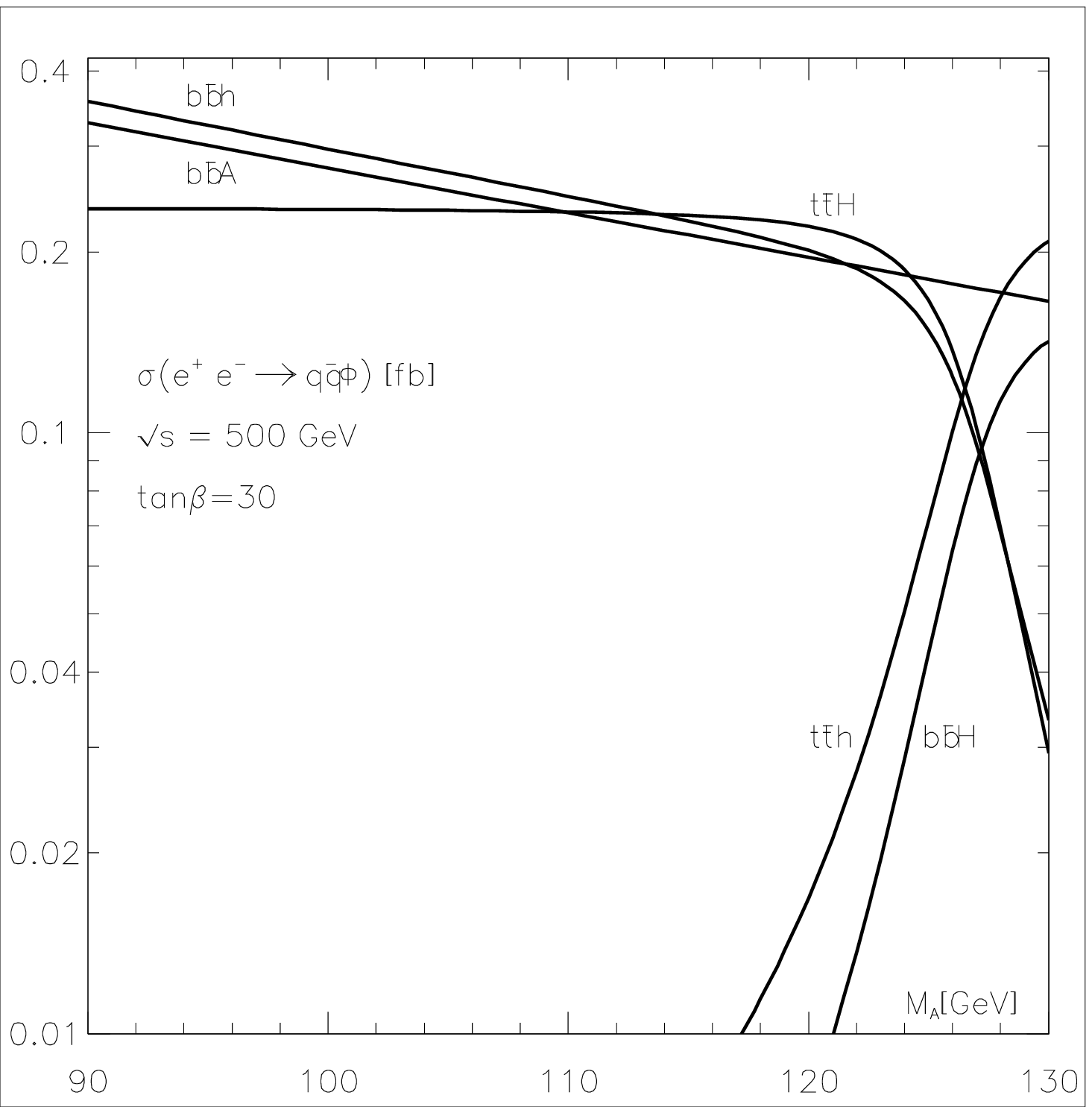,bbllx=3,bblly=3,bburx=418,bbury=420,height=6.5cm,width=7.5cm,clip=}\\[-.2cm]
\caption{\it Cross sections at a 500 GeV $\ee$ collider in associated Higgs
boson production with top and bottom quarks as functions of $M_A$ for $\tan 
\beta =10,30$. }
\end{center}
\vspace*{-.5cm}
\end{figure}

\noindent {\bf d) Double Higgs production in the strahlung process} \s

Finally, we also show the cross sections at $\sqrt{s}=500$ GeV and $\tan \beta=
10,30$ for the double Higgs boson production in the bremsstrahlung processes
\cite{DHiggsee} 
\beq
(d) \ \ {\rm Double~Higgs-strahlung} \ \ \ \ee \to Zhh, ZHh,ZHH, ZAA \non
\eeq
which are the largest [the other production processes with more than one Higgs
boson in the final state, such as $WW$ fusion or triple Higgs production via
$Z$ exchange, have too low rates at this energy]. In particular, $\sigma(\ee
\to Z \Phi_H \Phi_H)$ can be of the order of 0.1--0.2 fb [i.e.~$\sim 50-100$
events for $\int {\cal L} \sim 500$ fb$^{-1}$] which would allow a reasonable
measurement of the $\Phi_H^3$ trilinear coupling with some accuracy 
\cite{ZHHexp}.  The measurement of the other Higgs boson self--couplings is 
more challenging.  
 
\begin{figure}[htbp]
\vspace*{-.2cm}
\begin{center}
\epsfig{figure=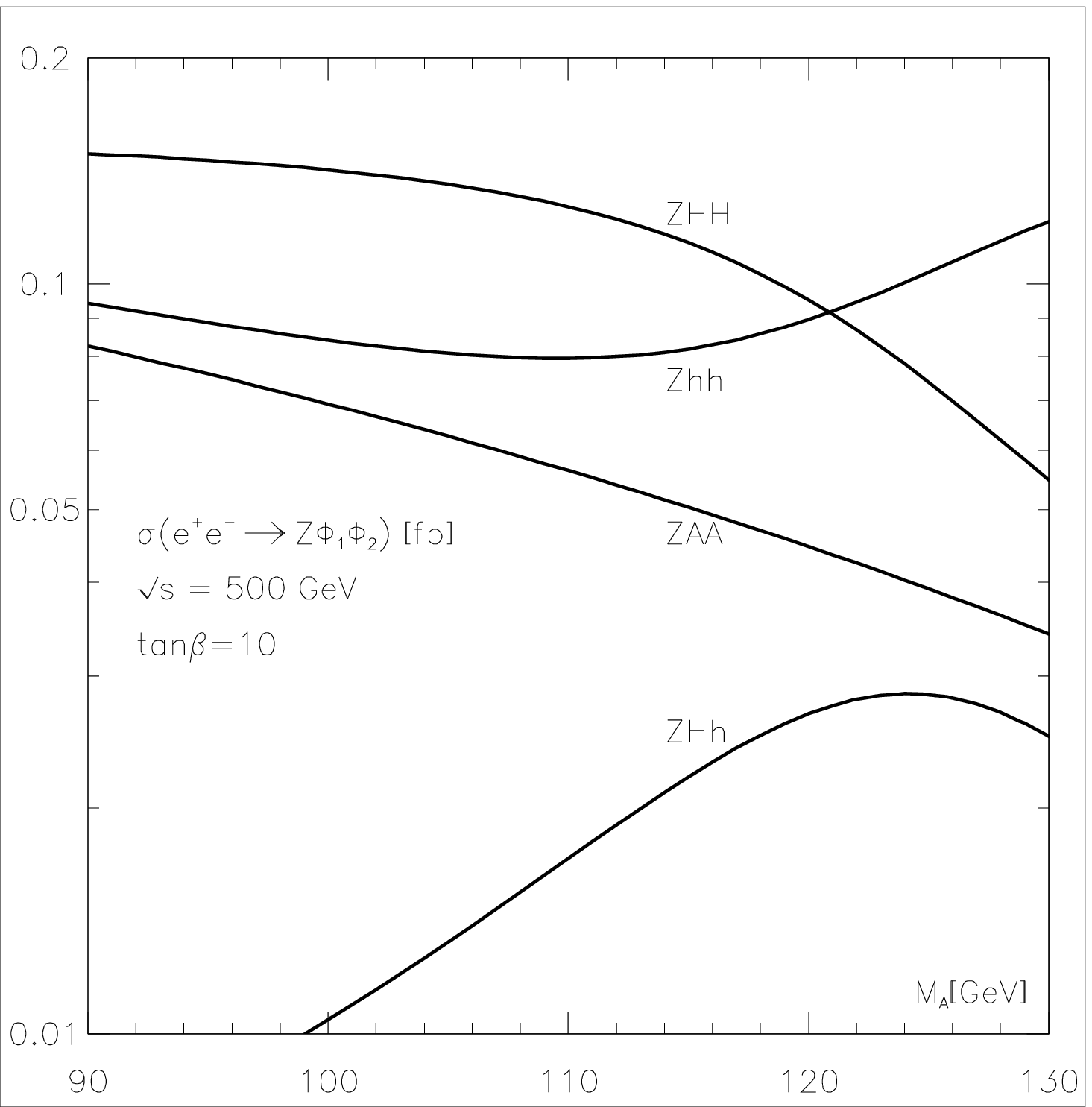,bbllx=3,bblly=3,bburx=410,bbury=420,height=6.5cm,width=7.5cm,clip=}
\hspace{1cm}
\epsfig{figure=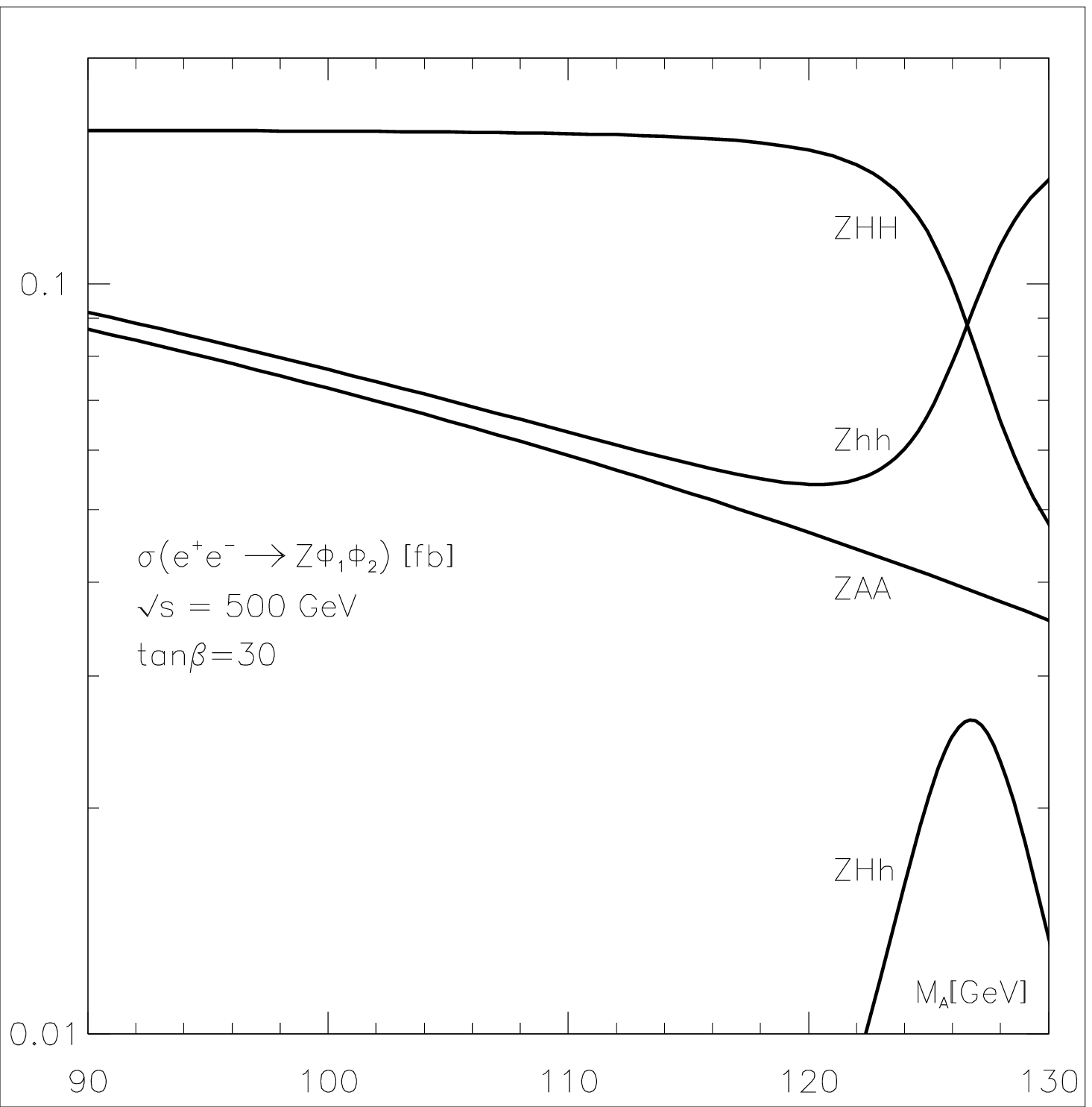,bbllx=3,bblly=3,bburx=410,bbury=420,height=6.5cm,width=7.5cm,clip=}
\\[.4cm]
\caption{\it Production cross sections at a 500 GeV $\ee$ collider in the 
double Higgs production processes $\ee \to Z \Phi_1 \Phi_2$ as functions of 
$M_A$ for $\tan\beta=10$ (left panel) and $\tan\beta=30$ (right panel).}
\end{center}
\vspace*{-.5cm}
\end{figure}

\subsubsection*{5.3 Resonant Higgs production at $\gamma \gamma$ colliders}

$\gamma \gamma$ colliders, in which the high--energy photon beams are generated
by Compton back--scattering of laser light \cite{laser}, provide useful
instruments to search for the neutral Higgs bosons and to test their properties
\cite{gamgam,gamgam2}.  Center of mass energies of the order of 80\% of the
$\ee$ collider energy and integrated luminosities $\int {\cal L} \sim 100$ fb,
as well as a high degree of longitudinal photon polarization can be reached at 
these colliders \cite{newgamma}. \s

Tuning the maximum of the $\gamma \gamma$ spectrum to the value of the Higgs 
boson mass, which is assumed to be already known from the $\ee$ option with
some accuracy, the Higgs particles can be formed as $s$--channel 
resonances, 
$$ 
 \ \ \ \gamma \gamma \ \to \ h, H , A
$$ 
decaying mostly into $b\bar{b}$ pairs. The main background, $\gamma \gamma \to
b\bar{b}$, can be suppressed by choosing proper helicities for the initial 
electron, positron and laser photons which  maximizes the signal cross 
section, and eliminating the gluon radiation by taking into account only the 
two--jet cross section. \s

Following the analysis of Ref.~\cite{gamgam2}, to which we refer for details,
we show in Fig.~18, the cross sections for resonant two--jet signal $\gamma
\gamma \to h,H, A \to b\bar{b}$ production. A cut on the scattering angle of
the $b$ quark has been applied, $|\cos\theta|<0.5$, which increases the ratio
of the signal cross section to the $\gamma \gamma \to b\bar{b}$ two--jet
background cross section.  [Note that the resummed NLO QCD corrections to both
the signal, background and the interference have been properly taken into
account in the 2--jet final state]. The Compton spectrum has been integrated
around the resonant Higgs mass in bins of $\Delta= \pm 3$~GeV in order to
account for the limited $b\bar{b}$ mass resolution. Clear signals can be
obtained for $\tan \beta=10$ and 30, in the 90 GeV $\lsim M_A \lsim$ 130 GeV
mass range, except for the pseudoscalar $A$ boson for $\tan \beta=10$ since in
this case, the $A \gamma \gamma$ coupling, mediated only by $b$ and $t$ quark 
loops, is suppressed.  

\begin{figure}[htbp]
\vspace*{-.5cm}
\begin{center}
\epsfig{figure=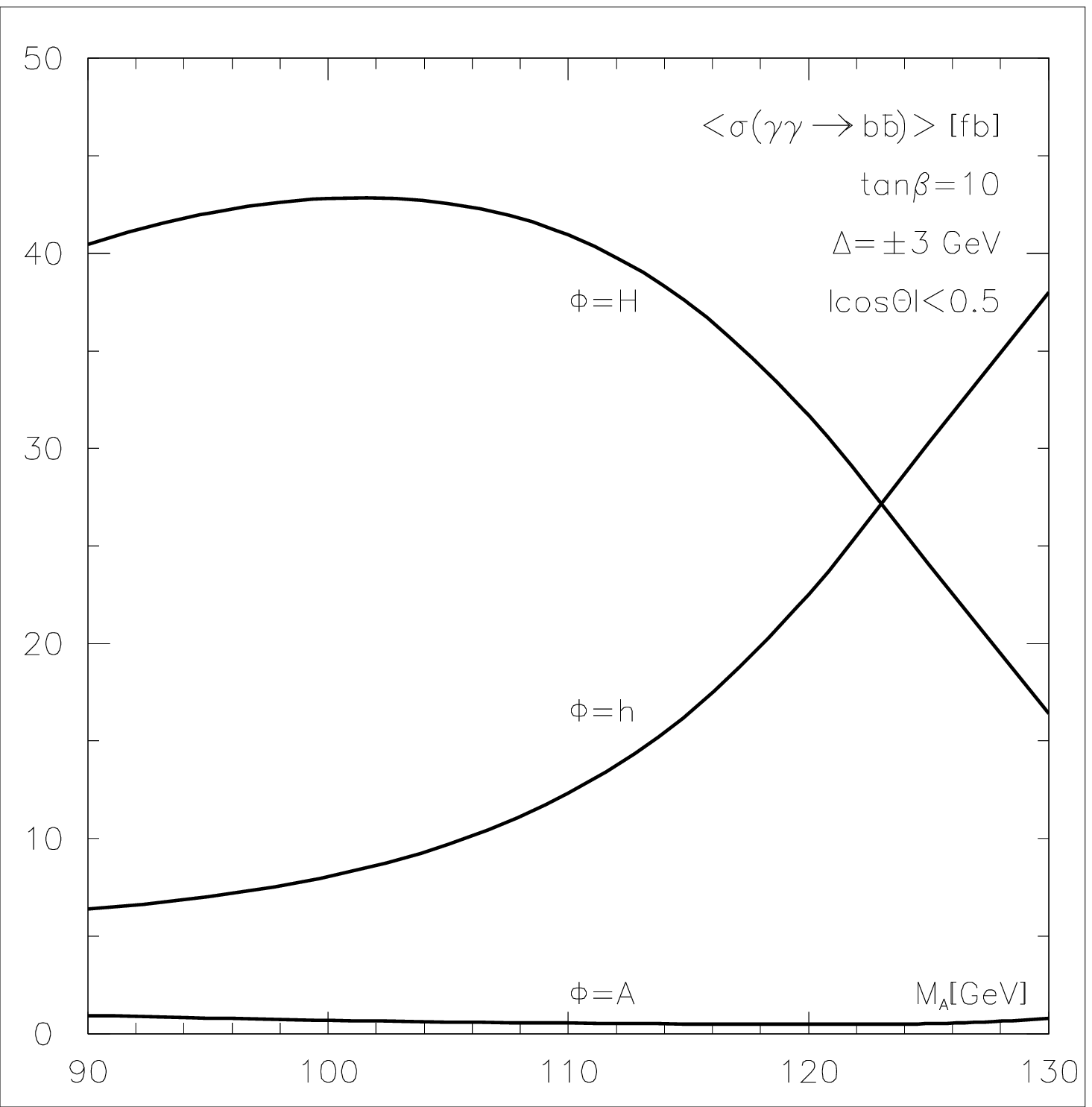,bbllx=3,bblly=3,bburx=410,bbury=420,height=6.5cm,width=7.5cm,clip=}
\hspace{1cm}
\epsfig{figure=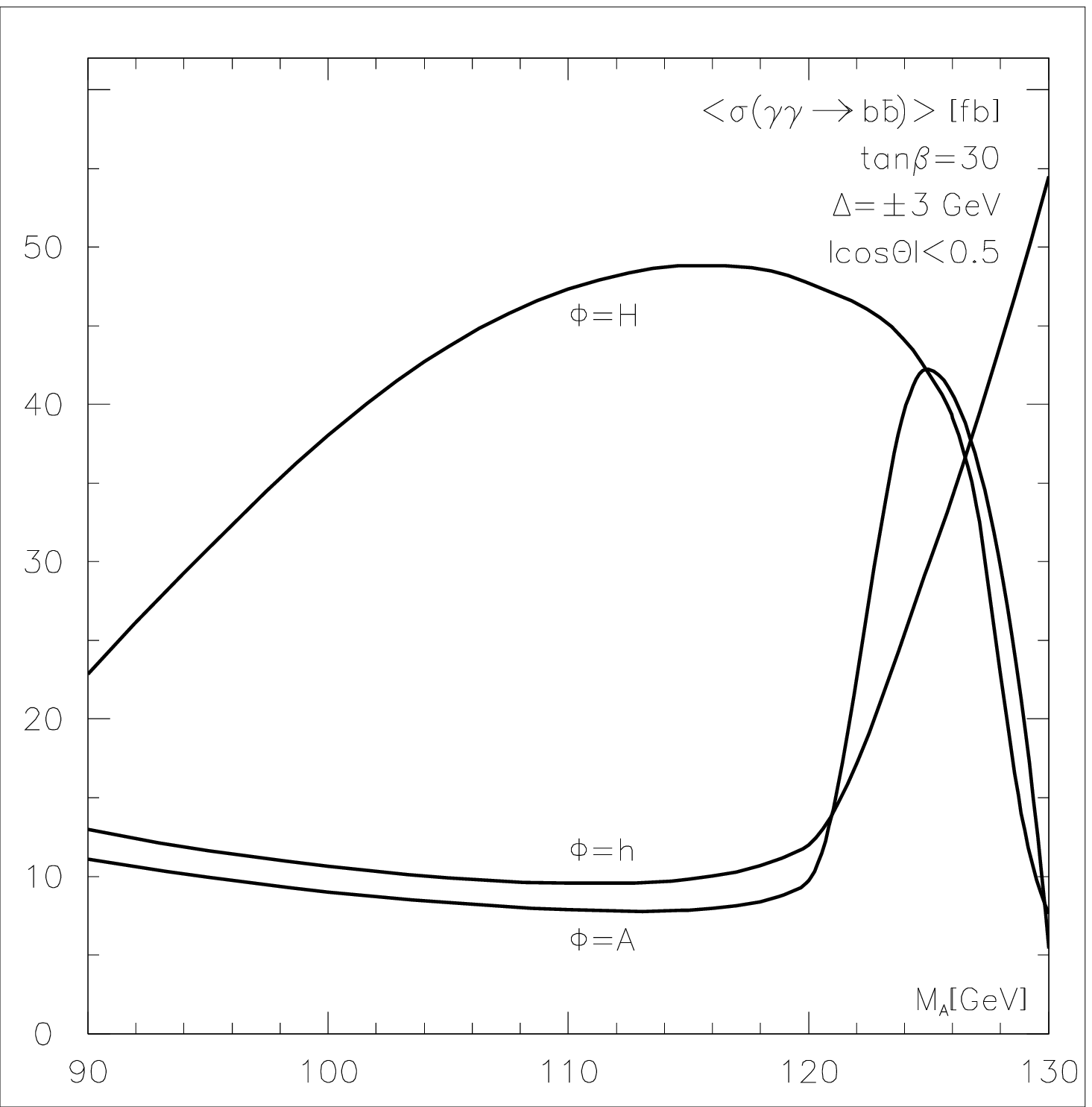,bbllx=3,bblly=3,bburx=410,bbury=420,height=6.5cm,width=7.5cm,clip=}
\\[-.1cm]
\caption{\it Total cross sections for the resonant Higgs production $\gamma 
\gamma \to h,H,A \to b\bar{b}$, in Compton backscattered $\gamma\gamma$ fusion 
for $\tb =10$ and $30$. The maximum of the $\gamma \gamma$ sub-energy is tuned 
to the mass of the $\Phi$ boson. }
\vspace*{-.5cm}
\end{center}
\end{figure}

\subsubsection*{5.4 Resonant Higgs production at $\mu^+ \mu^-$ colliders}

The ability of a future muon collider to investigate the Higgs sector of the 
SM and MSSM has been discussed in numerous papers; see for instance 
Refs.~\citer{adv,Barger2}. The main advantages of a muon collider, 
compared to an electron--positron machine, are due to the fact that the muon 
has a much larger mass than the  electron, which means that: \s

$(i)$ The couplings of Higgs bosons to $\mu^+ \mu^-$ pairs are much larger than 
the couplings to $e^+ e^-$ pairs, yielding significantly larger rates for
$s$--channel Higgs boson production at a muon collider [the production rate is 
negligible in $\ee$ collisions]. \s

$(ii)$  A very narrow beam energy spread can be achieved, which leads to only
a rather small loss of production rates at muon colliders. \s

However, there is a dependence between the small beam energy spread, $R=dE/E$,
and the delivered luminosity. Present estimates for yearly integrated
luminosities are $\int {\cal L} =0.1, 0.22, 1$ fb$^{-1}$ for beam energy
resolutions of $R = 0.003\%, 0.01\%, 0.1\%$, respectively \cite{Barger1}.

Due to these features, a muon collider can be considered as a potential factory
for the MSSM Higgs bosons, which can be produced as $s$--channel resonances: 
$$ 
 \ \ \ \mu^+ \mu^- \ \to \ h, \ H , \ A
$$
This would allow for the determination of the masses and the total decay widths
of the MSSM neutral Higgs bosons with a high degree of accuracy. \s

Following Ref.~\cite{Barger2}, we use a Gaussian spread $\sigma_{\sqrt{s}}$ 
for the center of mass energy, for which the central value is set at the Higgs 
boson mass, and calculate the effective $s$--channel production cross section 
for a Higgs boson $\Phi$ decaying into a final state $X$, using the formula: 
\begin{equation}
 \sigma(\sqrt{s}) = \frac{4\pi}{M^{2}_{\Phi}}\frac{{\rm BR} (\Phi \rightarrow 
\mu^+ \mu^-) {\rm BR} (\Phi \rightarrow X)} {\Bigl[ 1+ \frac{8} {\pi} \Bigl(
\sigma_{\sqrt{s}}/ \Gamma_{\rm tot} (\Phi) \Bigr)^2\Bigr]^{1/2}}
\label{mumuone}
\end{equation}
where $\sqrt{s}=M_{\Phi}$, $\Gamma_{\rm tot}(\Phi) $ is the Higgs boson total 
decay width and  $\sigma_{\sqrt{s}}$ is given by 
\begin{equation}
\sigma_{\sqrt{s}} = 0.002 \, {\rm GeV} \, \Bigl( \, \frac{R}{0.003\%}\Bigr) \, 
\Bigl(\frac{\sqrt{s}}{100 \, \rm {GeV} }\Bigr)
\end{equation}

In the discussed region of the MSSM parameter space for the intense--coupling
regime, the three neutral Higgs bosons $h,H$ and $A$ are broad enough. 
Therefore, one can use a resolution $R = 0.01\%$ without too much loss of
production rates. In such a case, the cross--sections are functions of the
Higgs branching fractions and Higgs masses and practically do not depend on the
resolution.  The obtained cross sections as functions of the pseudoscalar Higgs
boson mass $M_A$ for $\tb=10$ and 30 are shown in Fig.~19 [using
eq.~(\ref{mumuone}) with $R=0.01\%$]. As can be seen, they are large enough to
allow for the production of a significant number of Higgs bosons, so that 
detailed studies of the profile of these particles can be performed.  

\begin{figure}[htbp]
\vspace*{-.2cm}
\begin{center}
\epsfig{figure=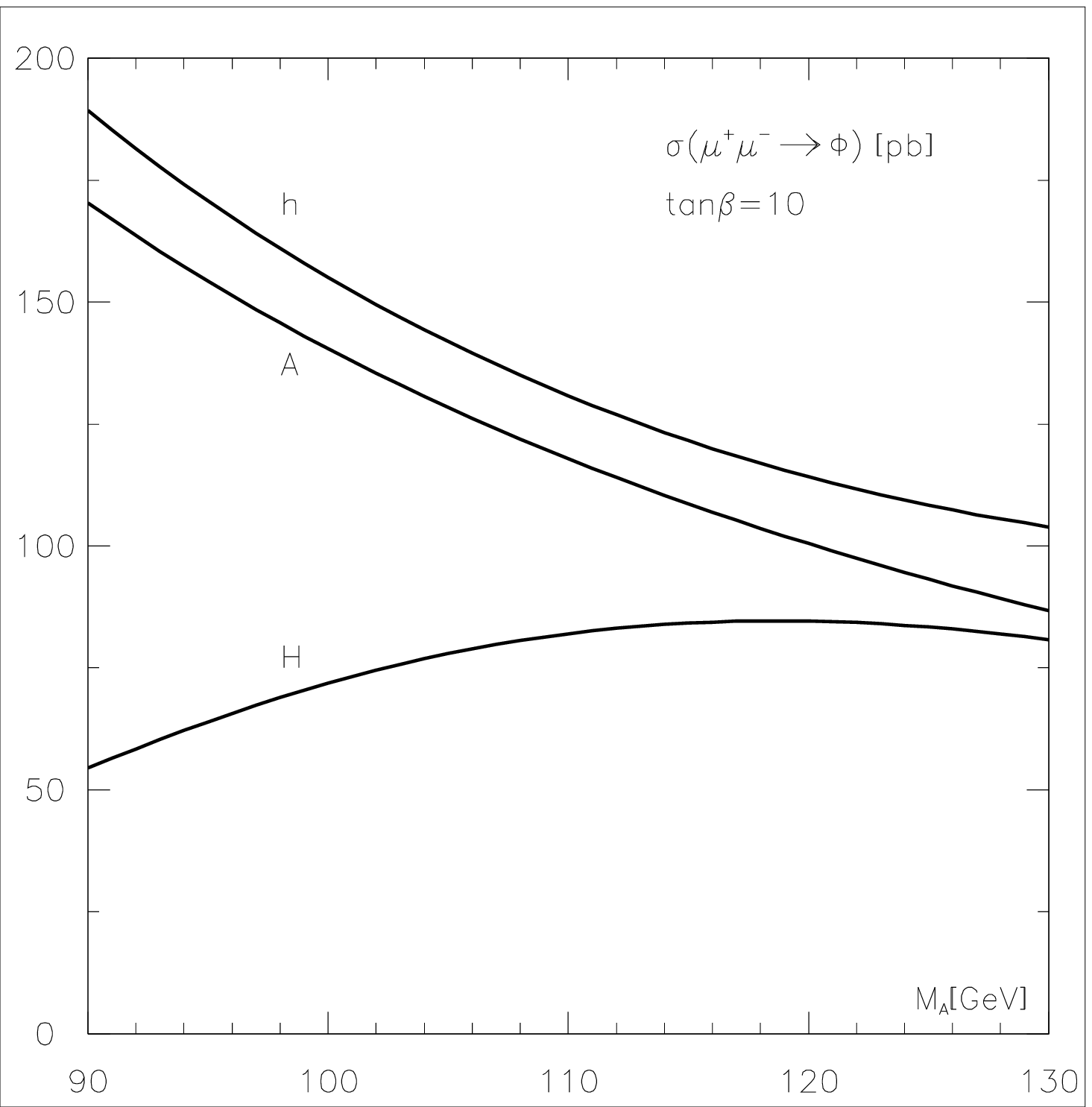,bbllx=3,bblly=3,bburx=410,bbury=420,height=6.5cm,width=7.5cm,clip=}
\hspace{1cm}
\epsfig{figure=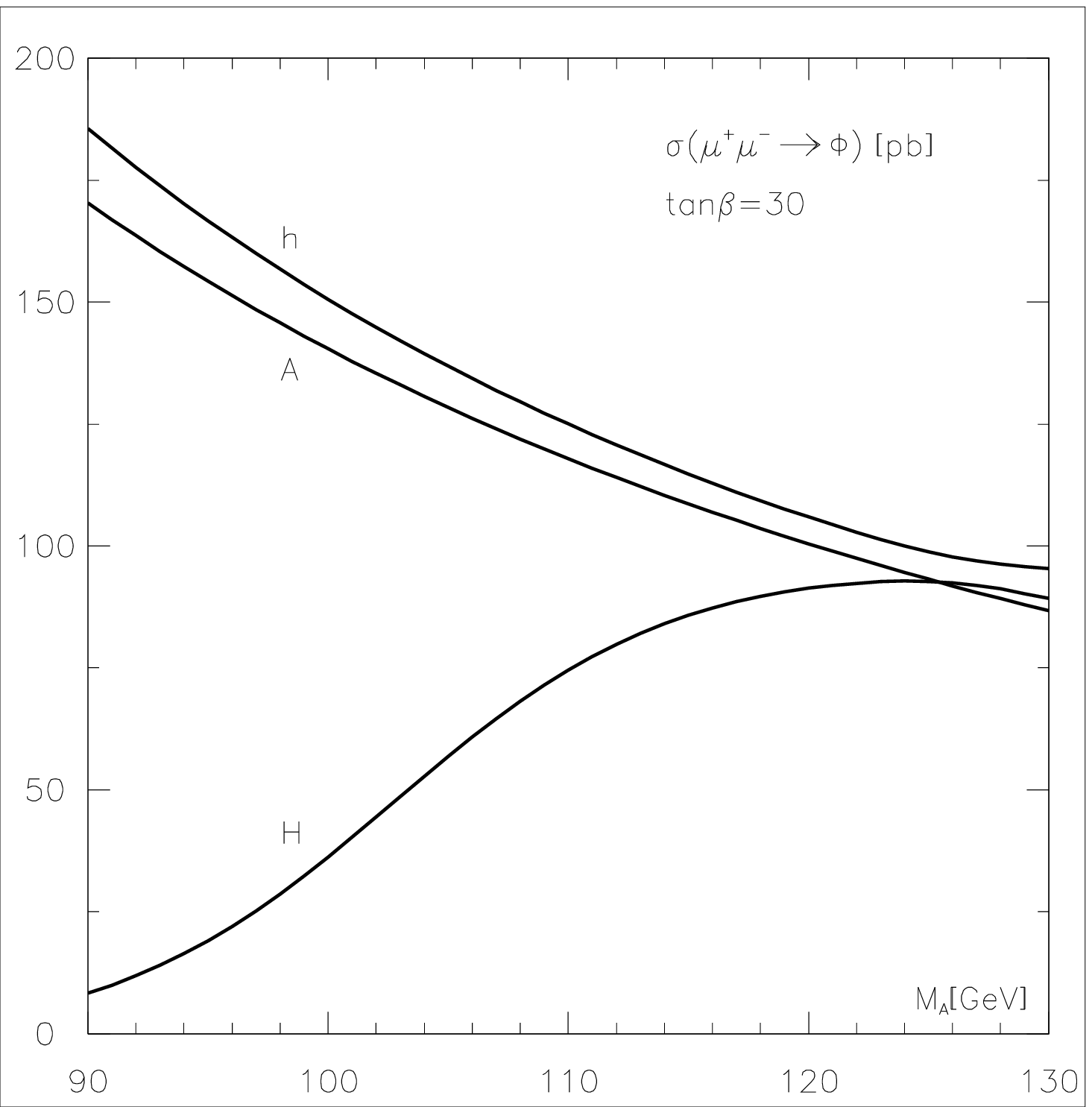,bbllx=3,bblly=3,bburx=410,bbury=420,height=6.5cm,width=7.5cm,clip=}
\\[-.1cm]
\end{center}
\caption{\label{fig:mumu}%
\it Cross sections for resonant MSSM Higgs boson ($h,H$ and $A$) production at 
a $\mu^+\mu^-$ collider as functions of $M_A$ for $\tb=10$ (left panel) and 
$30$ (right panel).}
\end{figure}

\subsection*{6. Conclusions} 

In this paper, we have performed a comprehensive analysis of the MSSM Higgs
sector in the intense--coupling regime, that is when the neutral Higgs
particles are rather light, 90 GeV $\lsim M_\Phi \lsim 130$ GeV, and the value
of the ratio of the vevs of the two MSSM Higgs doublet fields, $\tb$, is large.
In this scenario, one of the CP--even Higgs particles, $\Phi_H$, has SM--like
Higgs properties, i.e. couples almost maximally to massive gauge bosons and top
quark pairs, while the other CP--even Higgs particle, $\Phi_A$, has
pseudoscalar--like properties, i.e. couples strongly to bottom quarks and
$\tau$ leptons and has almost no couplings to massive gauge boson pairs;
however the $Z \Phi_A A$ coupling is almost maximal. \s 

We have first discussed the parameterization of the MSSM Higgs sector and
studied the radiative corrections due to the top/stop and bottom/sbottom
sectors to the Higgs boson masses, to their couplings to gauge bosons and
fermions [as well as to their self--couplings]. We have derived rather simple
expressions for these radiative corrections, which approximate rather
accurately the full corrections, and which allow to have a simple qualitative
understanding of the behaviour of these various parameters in this regime. \s

We have then analyzed the various experimental constraints on this 
intense--coupling regime, with large values of $\tb$ and relatively small Higgs
boson masses. We have shown that this scenario is still allowed by indirect
constraints from high--precision measurements such as the electroweak
parameters, i.e. the $W$ boson mass, the effective electroweak mixing angle
$\sin^2\theta_W$ and the $Z$ boson partial width and forward--backward
asymmetry  in $\bar{b}b$ final states, the radiative and loop induced
decay $b \to s\gamma$ and the anomalous magnetic moment of the muon,
$(g-2)_\mu$. We have also discussed the direct constraints from MSSM Higgs
boson searches at LEP2 and the Tevatron. From Tevatron searches for the charged
Higgs particle in top quark decays and the neutral Higgs bosons in the
associated production with $b\bar{b}$ pairs, $\tb$ values slightly above 50 are
still allowed. Imposing the LEP2 constraints, we have delineated the allowed
regions of the parameter space in $\tb$ and $M_A$ (and $M_h$) in various
scenarii for the mixing in the stop sector. We have also studied the
implications of the possible $\sim 2 \sigma$ evidence for a SM--like Higgs
boson with a mass around 115 GeV, and have shown that in the intense--coupling
regime, there are regions of the parameter space where this Higgs particle can
be either the lighter $h$ boson [for small to moderate mixing]  or the
heavier CP--even Higgs particle $H$ [in the no--mixing scenario]. \s 

For the decay modes, the pattern for the CP--even Higgs particles is in general
similar to the one for the pseudoscalar Higgs particle, i.e.  they decay
dominantly into $b\bar{b}$ [90\% of the time] and $\tau^+ \tau^-$ [10\% of the
time] pairs. There are three exceptions to this situation: \s

$(i)$ When the heavier $H$ boson mass is close to its minimal value. In this
case, it decays as the SM Higgs particle, i.e. the decays into $c\bar{c}, gg$
and $WW^*$ reach the few percent level.  \s

$(ii)$ When the lighter $h$ boson mass is close to its maximal allowed value,
the decays to other final states than $b\bar{b}$ and $\tau^+ \tau^-$ are also
sizeable. However, since the decoupling limit is not yet reached, the branching
ratios are in general smaller than for a SM Higgs boson.  \s

$(iii)$ There are regions of the parameter space where the $H$ boson couplings
to down--type fermions are strongly suppressed, thus enhancing the decays into
other final states.  \s

Thus there are regions of the parameter space where the decays of both the $h$
and $H$ bosons into the important channels $\gamma \gamma$ and $WW^*$ [and even
$\tau^+ \tau^-$ in some pathological situations] are suppressed compared to the
SM case. The total decay widths of the $A$ boson and one of the CP--even Higgs
boson, $\Phi_A$, are rather sizeable for large $\tb$ values [a few GeV], while 
the total decay width of the other CP--even Higgs particle, $\Phi_H$, is rather
small as for the SM Higgs boson [of the order of 100 MeV]. \s

A significant part of the paper was devoted to the analysis of the production of
the three neutral MSSM Higgs bosons at the up--coming colliders: the Tevatron
Run II, the LHC, a 500 GeV electron--positron linear collider in both its $\ee$
and $\gamma \gamma$ modes, as well as a future $\mu^+\mu^-$ collider. All the
neutral Higgs bosons [as well as the charged Higgs particle] will be
kinematically accessible even at the upgraded Tevatron and at an $e^+e^-$
collider with a c.m. energy $\sqrt{s} \gsim 300$ GeV. The Higgs particles can
be produced in various channels, allowing for a detailed study of the MSSM
Higgs sector and for the determination of the various parameters such as the
masses and the couplings. \s

At $\ee$ colliders, thanks to the clean environment, all Higgs particles can be
easily detected and the Higgs boson masses and couplings can be measured with a
relatively good precision. In particular, the couplings to photons, which could
be sensitive to new particles, can be accurately measured in the $\gamma \gamma$
option of the collider. All production channels will be effective: the
Higgs--strahlung and vector fusion processes for the SM--like $\Phi_H$ boson and
the associated Higgs pair production for the pseudoscalar $A$ and
pseudoscalar--like $\Phi_A$ particles. In addition, both the associated
production with top quarks [for the $\Phi_H$ boson] and associated production
with bottom quarks [for the $A$ and $\Phi_A$ particles] are possibly
accessible, allowing for the measurement of the Higgs boson Yukawa couplings 
to top and bottom quarks and the determination of the important parameter $\tb$.
The total decay widths of two Higgs particles would be large enough to be 
resolved experimentally. The total decay widths of all three Higgs bosons can 
be precisely measured at a $\mu^+ \mu^-$ collider. \s

At hadron colliders, the search for the MSSM Higgs particles in the
intense--coupling regime can also be performed in different channels. At the
Tevatron Run II with a very high luminosity, the SM--like $\Phi_H$ boson would
be accessible in the associated $p \bar{p} \to \Phi_H V$ [$V=W,Z$]
production process. In some cases, the production rates are larger than in 
the case of the SM Higgs boson; however, there are situations where the 
production rates are smaller for both the $h$ and $H$ bosons compared to 
the SM case [as is the case in the pathological situation discussed for 
the $H$ boson], making the detection more difficult. The $\Phi_A$ and $A$
particles can be produced in association with $b\bar{b}$ pairs with large rates
for large enough $\tb$ values and  the states can be detected in most of the
parameter space. \s

At the LHC, all Higgs particles can be produced with rather large cross
sections and many complementary production channels can also be effective.  In
particular, the production rates for the $A$ and $\Phi_A$ bosons in the $gg$
fusion mechanism and the associated production with $b\bar{b}$ pairs are 
strongly enhanced compared to the SM case, while the $\Phi_H$
particles can be accessed in all the processes, $gg, VV$ fusion and associated
production with top quarks and vector bosons. However, the experimental
detection of the particles can be slightly complicated for three reasons [this
also holds in some cases for the upgraded Tevatron]: \s

$(i)$ Since the three neutral Higgs particles are relatively close in mass, it 
might be difficult to separate between them, in particular in the case  of
the $A$ and $\Phi_A$ bosons which mainly decay into $b\bar{b}$ and $\tau^+
\tau^-$ final states which have a relatively poor resolution. \s 

$(ii)$ The distinction between these two [$A$ and $\Phi_A$] Higgs bosons can be
made slightly more difficult by the fact that their total decay widths can be 
rather large, a few GeV, making the signal peaks rather broad.\s 

$(iii)$ The clean $\gamma \gamma$ (as well as $WW^*$) final state signatures 
can be much less frequent than in the case of the SM Higgs particle, in 
particular when the lighter $h$ boson plays the role of $\Phi_H$. Again, 
the wider Higgs states would make the detection more difficult. \s

In conclusion: the phenomenology of the MSSM neutral Higgs sector in the 
intense--coupling regime is extremely rich. All Higgs states are kinematically
accessible at the next high--energy colliders, in various and complementary
production processes. In some cases, the techniques for searching these 
particles will be different  from the ones discussed in the case of the SM 
Higgs boson and even in the context of the MSSM Higgs bosons close to the 
decoupling limit. In this preliminary analysis, we have summarized the main 
features of these searches. However, detailed Monte--Carlo studies, including 
a more complete discussion of the various signals, a simulation of all the 
possible backgrounds, as well as a proper account of the experimental 
situation, will be needed to assess the potential of the colliders to discover 
these particles, and once discovered,  to measure their properties. 
\bigskip

\nn {\bf Acknowledgments}: \s

\nn We thank Aseshkrishna Datta for discussions. The work of
A.D. and M.M. is supported by the Euro--GDR Supersym\'etrie and by the 
European Union under contract HPRN-CT-2000-00149. The work of E.B. and A.V. 
is a partly supported by the INTAS 00-0679, CERN-INTAS 99-377 and RFBR 
01-02-16710 grants. E.B. thanks the Humboldt Foundation for the Bessel 
Research Award and DESY for the kind hospitality.

\end{document}